\documentclass[twocolumn,floatfix]{revtex4-2}

\usepackage[colorlinks,pdfusetitle,urlcolor=blue,citecolor=blue,linkcolor=blue,bookmarksnumbered,plainpages=false]{hyperref}
\usepackage{graphicx}
\usepackage{bm} 

\usepackage{amsmath}
\usepackage{amssymb}
\usepackage{amsfonts}

\usepackage{braket}
\usepackage{leftidx}
\usepackage{placeins}
\usepackage[percent]{overpic}

\usepackage[dvipsnames]{xcolor} 
 
\usepackage{mwe}

\usepackage{mathtools}
\usepackage{listings}

\lstdefinestyle{mystyle}{
    basicstyle=\ttfamily\footnotesize,
    breakatwhitespace=false,         
    breaklines=true,                 
    captionpos=b,                    
    keepspaces=true,                 
    numbersep=5pt,                  
    showspaces=false,                
    showstringspaces=false,
    showtabs=false,                  
    tabsize=2
}

\usepackage{macros}

\begin{document} 
\title{Dissipation across the ultrastrong-coupling regime of nanomechanical quantum Rabi systems}

\author{Janine C. Franz}
\email{janine.franz@u-bordeaux.fr} 
\affiliation{Université de Bordeaux, CNRS, LOMA, UMR 5798, F-33400 Talence, France}

\author{Fabio Pistolesi} 
\email{fabio.pistolesi@u-bordeaux.fr} 
\affiliation{Université de Bordeaux, CNRS, LOMA, UMR 5798, F-33400 Talence, France}

\date{\today}

\begin{abstract}
Mechanical resonators ultrastrongly coupled to quantum two-level systems provide a promising route towards mechanical qubits by introducing significant anharmonicity to the mechanical modes, particularly in the slow-oscillator regime. 
Although the resulting hybrid system is well described by the quantum Rabi model, a consistent treatment of dissipation remains challenging across the broad parameter space routinely probed in current nanotube electromechanical devices.
Here, we investigate dissipation in the open quantum Rabi model using a Born-Markov framework based on the slowly varying bath spectrum approximation, yielding a Lindblad master equation applicable far beyond conventional descriptions while recovering them in their respective limits. Using this framework, we analyze experimentally accessible observables across this parameter space.
As the secular approximation breaks down, phonon blockade progressively washes out. Our approach remains valid in this regime, enabling a quantitative description of the continuous evolution of phonon blockade with coupling strength and dissipation.
At finite temperature, we find a suppression of the temperature-induced increase of coherence decay rate for weak anharmonicity. 
Under driving, our approach remains applicable to substantially stronger perturbations than conventional dressed-state master equations and shows that an apparently classical observable can coexist with Wigner negativity. 
We further capture the weakly anharmonic regime arising from finite detuning in the double-quantum dot. 
These results establish a unified description of dissipation from weakly anharmonic operating regimes to the strongly anharmonic mechanical-qubit regime and provide experimentally relevant predictions for ultrastrong electromechanical systems.
\end{abstract}

\maketitle  
\section{Introduction}
Mechanical resonators provide an attractive platform for quantum science due to their long coherence times, small effective masses, and compatibility with a wide range of physical systems~\cite{poot_mechanical_2012, bachtold_mesoscopic_2022,engelsen_ultrahigh-quality-factor_2024}. 
Over the past two decades, advances in nanofabrication have enabled nanomechanical resonators to enter the quantum regime through ground-state cooling, coherent quantum control, and strong coupling to electromagnetic fields~\cite{oconnell_quantum_2010,teufel_sideband_2011,satzinger_quantum_2018,chan_laser_2011}. 
Among these systems, suspended carbon nanotubes are particularly promising, combining exceptionally low mass with mechanical quality factors exceeding $10^6$ in state-of-the-art devices~\cite{moser_nanotube_2014,baydin_carbon_2022}. Moreover, quantum dots can be defined directly within the suspended nanotube, such that the electronic and mechanical degrees of freedom are intrinsically co-localized~\cite{sapmaz_carbon_2003, steele_strong_2009, lassagne_coupling_2009,benyamini_real-space_2014,vigneau_ultrastrong_2022}. This unique architecture gives rise to exceptionally strong electromechanical coupling, reaching the ultrastrong-coupling (USC) regime in which the coupling strength becomes comparable to the mechanical frequency~\cite{khivrich_nanomechanical_2018,vigneau_ultrastrong_2022,samanta_nonlinear_2023}.

Building on these favorable mechanical properties, Ref.~\cite{pistolesi_proposal_2021} proposed encoding a mechanical qubit in a carbon nanotube coupled to a co-localized double quantum dot and has been realized in a related but ultimately different nano-mechanical set-up~\cite{yang_mechanical_2024}.
At sufficiently strong electromechanical coupling, the interaction induces an anharmonic spectrum. 
A good working point for the mechanical qubit is found when the mechanical resonance frequency is much smaller than the quantum dot transition frequency~\cite{pistolesi_proposal_2021}, 
this enables lower decoherence and an optimal frequency spectrum. 
In order to reach a sufficient anharmonicity of the spectrum to address the mechanical qubit the system needs to be then in the USC regime. 
Current experimental implementations operate with a ratio between these two frequencies at about an order of magnitude~\cite{moller_tunable_2026}.
For these reasons, it is important to fully elucidate this regime. 

The coupled nanotube--quantum-dot system  introduced above is naturally described by the quantum Rabi model (QRM) of a harmonic oscillator to a two-level system (TLS). As a paradigmatic model of light--matter interaction, the QRM has been extensively explored in order to understand cavity and circuit quantum electrodynamics, where experimental progress has enabled access to the USC and even deep-strong-coupling regimes~\cite{frisk_kockum_ultrastrong_2019,koch_quantum_2023,forn-diaz_ultrastrong_2019,qin_quantum_2024}.
While the closed quantum Rabi model has been solved exactly~\cite{braak_integrability_2011},
decoherence and dissipation play a crucial role for the quality of the mechanical qubit in the hybrid system and a consistent and practical description of dissipation remains an open challenge in a large parameter range~\cite{frisk_kockum_ultrastrong_2019}.
 
This problem is especially present in nanomechanical systems, where dissipation is typically dominated by the environment-coupling of the TLS, even when the relevant hybrid states are largely mechanical.  Capturing how these states inherit dissipation and decoherence from the TLS due to their coupling is therefore crucial.  
However, conventional approaches face significant limitations. 
The commonly used local Lindblad equation -- 
where dissipation is derived neglecting the interaction between the sub-systems
-- can lead to unphysical heating, whereas the full secular approximation (FSA) in the dressed basis avoids such artifacts, but imposes strong conditions on transition frequency separations~\cite{breuer_theory_2007,werlang_rabi_2008, beaudoin_dissipation_2011,frisk_kockum_ultrastrong_2019}. 
As a consequence, the secular approximation is restricted to regimes with sufficiently large anharmonicity.
Although the desired mechanical-qubit regime is sufficiently anharmonic for the secular approximation to provide an accurate description, current experiments necessarily explore a much broader parameter space. During device characterization and optimization, the system is routinely operated under conditions of stronger driving, increased dissipation, or reduced anharmonicity. 

In this work, we provide a unified description across the experimentally relevant parameter space, which lies largely beyond the reach of conventional Lindblad treatments. 
To this end, we employ a Born–Markov framework for dissipation in the quantum Rabi model, resulting in a Lindblad master equation that remains valid beyond the constraints of the full secular approximation. 
To achieve this, we build on the slowly varying spectrum (SVS) approximation~\cite{settineri_dissipation_2018,mccauley_accurate_2020,fernandez_de_la_pradilla_recovering_2024}. In particular, it was shown that the SVS approximation relies on the same conditions as  the underlying Born--Markov approximation at zero temperature~\cite{mccauley_accurate_2020}. Compared to earlier formulations, our approach combines a compact parametrization of dissipation leading to accurate thermalization at finite temperature and naturally incorporates decoherence while remaining practical for experimental modeling.
Since the underlying approximation is closely related to the Born--Markov approximation, the resulting SVS Lindblad equation is remarkably versatile. As we show, it reproduces both the secular approximated master equation and the local Lindblad master equation in their respective limits.

With this we can operate beyond the range of validity of conventional Lindblad treatments for nanomechanical quantum Rabi systems and 
we use the approach in this work to obtain quantitative predictions for experimentally accessible observables across the full parameter space relevant to nanotube devices.
We investigate several experimentally relevant signatures of dissipative dynamics in the nanotube–double-quantum-dot system: 
We first analyze the two-phonon correlation function, where we confirm that the onset of phonon blockade coincides with the parameter regime where the full secular approximation loses validity 
and the cross-over regime and beyond is inaccessible with conventional approaches. 
We then study the finite-temperature correlation spectrum and demonstrate systematic deviations from the results obtained by secular approximation, including modified decay rates and interaction-induced shifts of the mechanical resonance. Finally, we show that under finite driving the SVS description remains applicable over a substantially larger range of drive amplitudes than the full secular approximation, enabling predictions of driven dynamics beyond the validity of conventional approaches.

Taken together, our results provide the theoretical foundation needed to interpret and optimize nanotube-based mechanical qubits across the full range of conditions encountered in present-day experiments, not merely in the idealized regime for which existing treatments were designed.

In Sec.~\ref{sec:system}, 
we introduce the physical set-up that we are focussing on in detail and
we show that in the lower USC regime, the quantum Rabi model
with the TLS frequency much larger than the oscillator frequency
can be approximated by an anharmonic Kerr-oscillator 
via fourth-order perturbation theory, enabling us to use analytical methods in this perturbative regime, while we exploit numerical methods in the coupling regime beyond. 
Additionally, we show that the closed system is well described via a Born-Oppenheimer approximation for all coupling strengths from which we can conclude the emergence of an effective double well potential in the deep-strong coupling regime. 
In Sec.~\ref{sec:new_diss}, we introduce the Lindblad master equation used in this work and discuss its relation to related approaches. 
Subsequently, in Sec.~\ref{sec:diss_QRM} we apply the method to the considered system and obtain the SVS Lindblad master equation for the quantum Rabi model. 
We discuss its limitations and show its relation to conventional Lindblad equations which are recovered in their respective limit.  
We then study the two-phonon correlation function in presence of an infinitesimal drive as a function of coupling and dissipation in Sec.~\ref{sec:g2}, which characterizes the crossover from mechanical-TLS-like behavior in the regime of complete phonon blockade to that of a driven (an-)harmonic system. 
Beyond the regime of complete phonon blockade, conventional Lindblad descriptions become inadequate, motivating the revised approach presented in this work.
While the two-phonon correlation function has been measured previously~\cite{cohen_phonon_2015,hong_hanbury_2017}, it is not as easily accessible experimentally as its photonic counterpart. A more accessible observable is the spectrum of the mechanical displacement. 
In Sec.~\ref{sec:Sxx}, we hence calculate its thermal spectrum and show that it is indeed different from the one predicted by the conventional secular approximation when outside of its validity regime. 
Within the perturbative regime of USC, we show analytically that the secular approximation overestimates the actual decay of correlations at finite temperatures. We show that depending on the ratio between anharmonicity and decay rate, 
the transfer of coherence counteracts the temperature-induced increase of correlation decay and derive a temperature-dependent shift of the correlation function peak for finite anharmonicity. 
Similarly, we study the mean-square displacement when driving the system as a function of drive frequency in Sec.~\ref{sec:drive}. We show that when the driving amplitude exceeds the anharmonicity the undriven dissipator obtained via secular approximation is not appropriate to describe the system's dynamics even for sufficiently small dissipation rates, while the SVS Lindblad dissipator remains valid in its unperturbed form for a significantly larger range of driving amplitudes. 
We find that for sufficiently large dissipation and drive, the signal mimics then the one of a classical Duffing oscillator~\cite{lifshitz_nonlinear_2008}. 
However, by studying the Wigner function of the driven state and its negativity, we show that the system is nontheless in a non-classical state. 
  
\section{Closed system and anharmonicity}
\label{sec:system}
\begin{figure}  
\includegraphics[width= \linewidth]{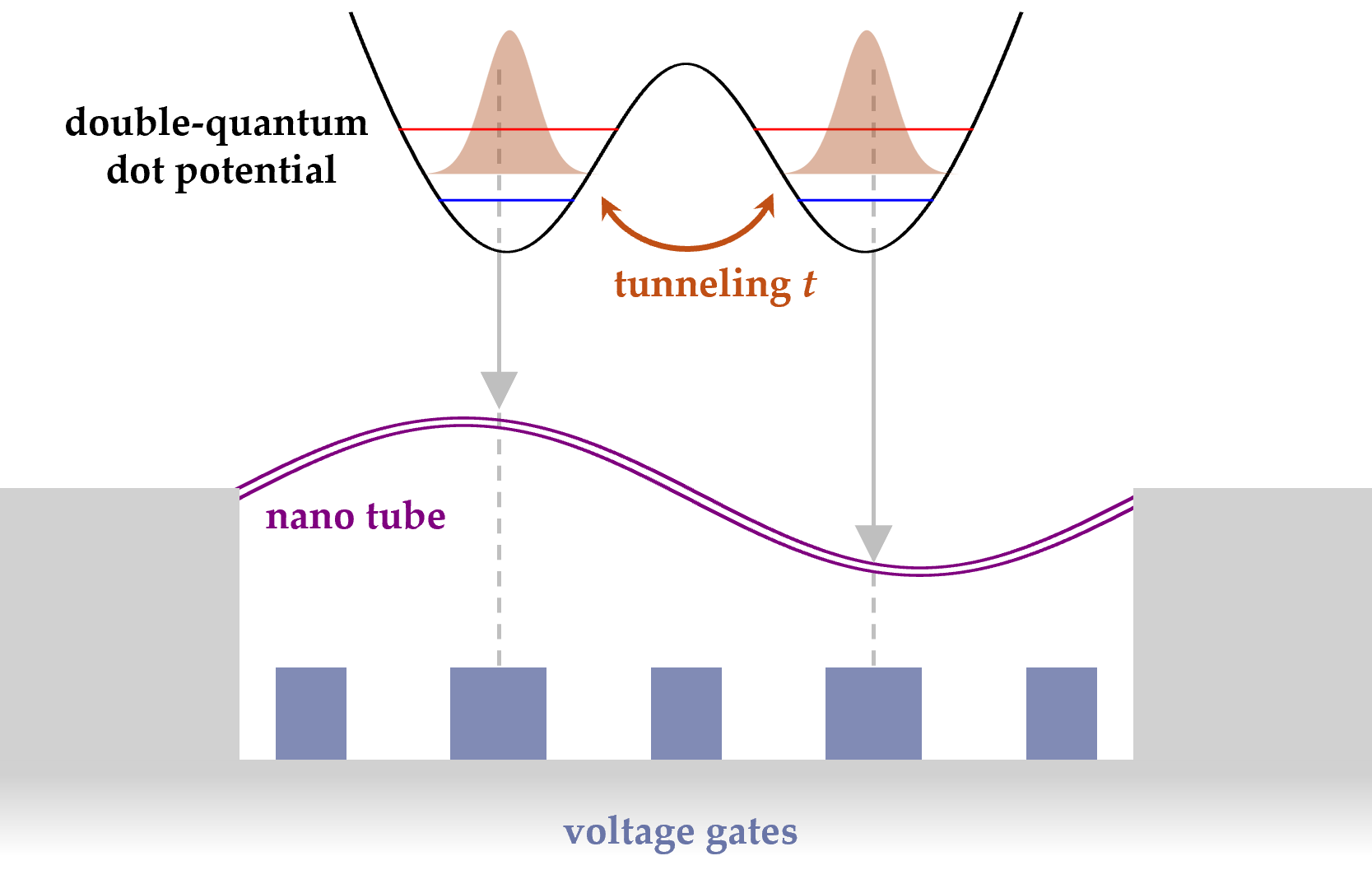} 
\caption{Scheme of the considered set-up. 
The second flexural mode of a suspended nanotube couples to the single-charge states of a double-quantum dot localized on the nanotube. For a symmetric quantum dot potential, due to
the tunneling strength between left and right dot,
the two eigenstates of the double dot are given by the anti-bonding and bonding state of a charge being localized in the left or right potential well. 
The energy splitting $\omega_\q$ of the dot states is then given by the tunneling strength $\omega_\q = 2 t$, here sketched as red and blue line inside the dot potential.
\label{fig:set-up}
 }
\end{figure}
We consider a suspended carbon nanotube
with an integrated double quantum dot which is realized on the nanotube via 
multiple voltage gates, see Fig.~\ref{fig:set-up}. 
With help of the voltage gates the electrostatic potential on the nanotube can be tuned to form a double-well potential for the charges, creating the double quantum dot, where we restrict ourselves to the case where two single-electron states are energetically accessible. 
Each of these states corresponds then to the electron populating the left or right dot and are coupled to each other via a tunneling strength $t$. 
By using a symmetric geometry of the double well potential, the coupling of the charge to the displacement of the second flexural mode of the nanotube is maximized and their physics may be modeled by the quantum Rabi Hamiltonian~\cite{pistolesi_proposal_2021}
\begin{align}
    H &= H_\m + H_\mathrm{tls} + g \sigma_x ( a + a^\dag) 
    \label{Hqrm} 
    \,, 
    \\
    H_\m &= \omega_\m a^\dag a
    \,,
    \\
    H_\mathrm{tls} &=  \omega_\q \sigma_z /2
    \,, 
\end{align} 
where we set $\hbar = 1$, $a$ and $a^\dag$ are the annihilation and creation operators of the mechanical excitations with frequency $\omega_\m$ and $\omega_\q = 2 t$ is the energy of the bare two-level system, given by the tunneling.  Here, the Pauli matrices are defined in the anti-bonding and bonding state basis $\ket{\pm}$, which are superpositions of the left- and right-dot-localised states $\ket{\mathrm{L/R}}$, \textit{i.e.}~$\ket{-} = ( \ket{\mathrm{L}} - \ket{\mathrm{R}})/\sqrt{2}$ and $\ket{+} = ( \ket{\mathrm{L}} + \ket{\mathrm{R}})/\sqrt{2}$.

We focus on the case of $\omega_\m \ll \omega_\q$, previously identified as an optimal working point for a mechanical qubit~\cite{pistolesi_proposal_2021} and realized in current experiments~\cite{moller_tunable_2026}. The parameters of this closely related experiment are $\omega_\m /(2 \pi)=0.8$~GHz, $\omega_\q /(2 \pi)=7.4$~GHz, an electromechanical coupling $g /(2 \pi)=0.5$~GHz, and a mechanical quality factor of $Q>10^5$. Since $\omega_\m \ll \omega_\q$, the Hamiltonian cannot be approximated by the Jaynes--Cummings model even for small couplings $g$: the counter-rotating terms are always of comparable magnitude to the co-rotating terms.

In this section, we analyze the eigenspectrum of the QRM in the ultrastrong, $0.1 \omega_\m < g < \omega_\m$, to deep-strong coupling regime, $\omega_\m < g$~\cite{frisk_kockum_ultrastrong_2019}, presenting two analytical approximations that let us understand and study the influence of the coupling on the resulting hybrid system. 
The first one is applicable in the lower ultrastrong coupling regime, where $g/\omega_\m \ll1$ can still be treated as a perturbative parameter and we approximately diagonalize the coupled Hamiltonian in fourth order of $g$ and map it onto a Kerr-oscillator, we call this regime the perturbative Kerr regime (PKR).
As we show and discuss, the mapping is quantitatively in good agreement with the full model with a surprisingly excellent accuracy for $g/\omega_\m < 0.5$ where $\omega_\m / \omega_\q = 10$.

The second is applicable for well separated time-scales $\omega_\q \gg \omega_\m$ in the spirit of a Born-Oppenheimer approximation and shows that the coupled system can be described by a double-well potential for coupling strengths $g$ exceeding a critical value in the deep-strong coupling regime~\cite{pistolesi_proposal_2021}.  
\subsection{Perturbative Kerr regime}
To analyze the system in the perturbative ultrastrong coupling regime, we may apply time-independent perturbation theory to approximately diagonalize the Hamiltonian up to fourth order in the coupling, 
\begin{multline}
    H =   \tilde{\omega}_\q \tau_z/2
    +
    (\tilde{\omega}_\m + \Delta \omega_\m/2 ) b^\dag b 
    \\
    + 
    (\Delta \omega_\m/2 ) \tau_z b^\dag b 
    - 
    (\chi/2 )\tau_z b^\dag b^\dag b b  
    \,,
\label{H_4th}
\end{multline}
where $b^{(\dag)}$ and $\tau_i$ act on the approximated eigenstates $\ket{\Psi_{n,\sigma}}$ as annihilation (creation) operator and Pauli matrices
, $b^\dag b \ket{\Psi_{n,\pm}} = n \ket{\Psi_{n,\pm}}$ and $\tau_z \ket{\Psi_{n,\pm}} = \pm \ket{\Psi_{n,\pm}}$ and they obey the usual commutation relations with each other. 
The introduced energies of the diagonalised Hamiltonian are given 
in the appendix, App.~\ref{app:4th_order_H}. A quantity of special interest is the Kerr-non-linearity which is given 
by 
\begin{align}
\chi 
&= 
\frac{4 g^4 \omega_\q \left(\omega_\m^2+3 \omega_\q^2\right)}{\left(\omega_\q^2-\omega_\m^2\right){}^3}
\label{chi}
\,,
\end{align}
and appears in fourth order of $g$. It can be approximated by $\chi \sim 12 g^4 / \omega_\q^3$ for $\omega_\q \gg \omega_\m$. 
The coefficients relating the approximated eigenbasis $\ket{\Psi_{n,\pm}}$ to the uncoupled basis $\ket{n \pm}$ defined  by  
$\ket{\Psi_{n \pm }}
  = 
\sum_{m, \sigma \in\{  \pm 1 \}}  \braket{ \Psi_{n \pm } |m \sigma } \, \ket{m \sigma} $
are given in the appendix up to fourth order in $g$ for the lower band $\ket{\Psi_{n-}}$, see App.~\ref{app:4th_order_H}.

A consequence of $\omega_\m \ll \omega_\q$ is that the upper band of the spectrum $\ket{\Psi_{n +}}$ is energetically separated from the lower band $\ket{\Psi_{n -}}$ and it is possible to only operate in said lower band.
Focussing on the lower band only, we can reduce the Hilbert space by projection $H \rightarrow \sum_{nm} \braket{\Psi_{n-} |H|\Psi_{m-} } \ket{\Psi_{n-}}\bra{\Psi_{m-}}$ and map the system on a single Kerr oscillator, 
\begin{align}
    H_\mathrm{Kerr} 
    = \tilde \omega_\m b^\dag b + (\chi/2) b^\dag b^\dag b b \,, 
    \label{HKerr}
\end{align}
with the transition frequencies between the dressed bosonic modes of the lower band ${\omega_{n,n+1} = E_{n+1, -} - E_{n,-}}$ (where $E_{n -}$ is the eigenenergy of $\ket{\Psi_{n-}}$) given by 
\begin{align}
    \omega_{n,n+1}
    = \tilde \omega_\m + n \chi 
    \,.
\end{align} 

Treating $g V = g \sx (a + a^\dag)$ as the perturbation is in general valid if $g \ll \omega_\m$ and $g \ll \omega_\q$, the energies of the bare Hamiltonian $H_0 = H_\m + H_\mathrm{tls}$.  
Because  $\omega_\q \gg \omega_\m$, the expansion parameter can range from  
$g/\omega_\q$ to $g /\omega_\m \gg g/\omega_\q$.
We can show that due to the transverse character of the coupling $V$, the correction terms remain suppressed by orders of $g/\omega_\q$ even when $g \lesssim \omega_\m$.

The energy correction $E_n^{(r)}$ of order $r$ is given by
\begin{align}
    E^{(r)}_n
    &= 
    g^r \tr \sum_{\{k_i\}} S_{k_1} V S_{k_2} V \dots V S_{k_{r+1}}
    \\
    S_{k} &= \sum_{m \neq n } 
    \frac{\ket{m}\bra{m}}{(E_n^{(0)} - E_m^{(0)})^k} 
    \quad \text{for } k > 0
\end{align}
and $S_0 = - \ket{n}\bra{n}$ with $\sum_i k_i = n -1$, and where $E_n^{(0)}$ is the $n$th eigenenergy of $H_0$.
To simplify the following discussion, let us focus on the contribution $e_n^{(r)}$ to the $r$th energy correction $E^{(r)}_n$ which depends on the most intermediate states,
\begin{align}
e_n^{(r)}
& =
g^r
\sum_{ {m_i } \neq n }
\frac{\braket{n| V|m_1} \braket{m_1| V|m_2} \dots \braket{m_{r}|V|n}}{(E_n^{(0)} - E_{m_1}^{(0)})\dots(E_n^{(0)} - E_{m_r}^{(0)})}
\end{align}
which depends on $r$ intermediate states $\ket{m_i}$.
Due to the transverse character of $V$, each intermediate state then differs from its predecessor by $\pm \omega_\q$ and $\pm \omega_\m$ and $r$ must be an even number to obtain a non-vanishing contribution.
Because of this, the largest expansion factor $g^r/\prod_i ( E_n^{(0)} - E_{m_i}^{(0)})$ found in $e_n^{(r)}$ for $\omega_\q \gg \omega_\m$ can be approximated by
\begin{align}
\frac{g^r}{\omega_\m^{r/2}\omega_\q^{r/2}}
=
\left(\frac{g}{\omega_\m}\right)^{r/2}
\left(\frac{g}{\omega_\q}\right)^{r/2}.
\end{align}
Hence, even for $g \lesssim \omega_\m \ll \omega_\q$, the perturbative expansion is expected to work well as long as $g/\omega_\q \ll 1$. 
Note, however, that this constitutes a bookkeeping issue when relating the formal perturbative order to the order in $g/\omega_\q$: the formal perturbative order $r$ contains terms ranging from $(g/\omega_\q)^r$ to $(g/\omega_\q)^{r/2}$.  

Another way to obtain the Kerr Hamiltonian and see its validity for $g \lesssim \omega_\m$, is given by first doing a dispersive expansion for $g \ll \omega_\q - \omega_\m$ as presented in Ref.~\cite{pistolesi_proposal_2021}, yielding the lower band Hamiltonian 
\begin{align}
    H_\mathrm{disp} &= \frac{\tilde \omega_\m -  \chi}{4}  ( x^2 + p^2 ) + \frac{\chi}{12} x^4 \,,
    \nn
    &= 
    \tilde \omega_\m a^\dag a  + \frac{\chi}{2} a^\dag a^\dag a a
    + \frac{\chi}{12} 
    \bigg( 
    a^4 + [a^\dag]^4  
    \nn
    & \qquad \qquad
    + 4 ( [a^\dag]^3 a + a^4 a^\dag ) 
    + 6 ( a^2 + [ a^\dag]^2 ) 
    \bigg)
\end{align}
with dimensionless $x = a + a^\dag$ and $p = - \mi ( a - a^\dag)$
and then apply an additional perturbative expansion in $0$-order to the remaining non-diagonal part of the dispersive Hamiltonian, \textit{i.e.}~approximating $\chi \ll \tilde \omega_\m$. 
 
\begin{figure*}[htbp]
\begin{overpic}[width=0.7\linewidth]{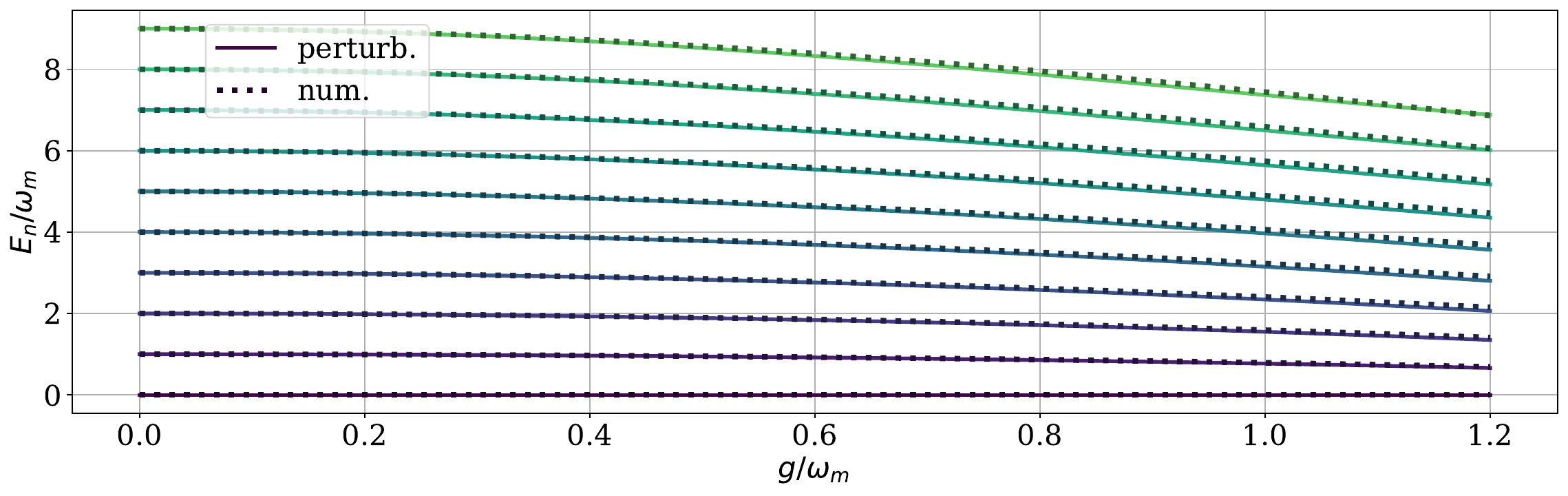}
\put(1,30){a)}
\end{overpic} 

\begin{minipage}{0.25\textwidth}
\begin{overpic}[width=\linewidth]{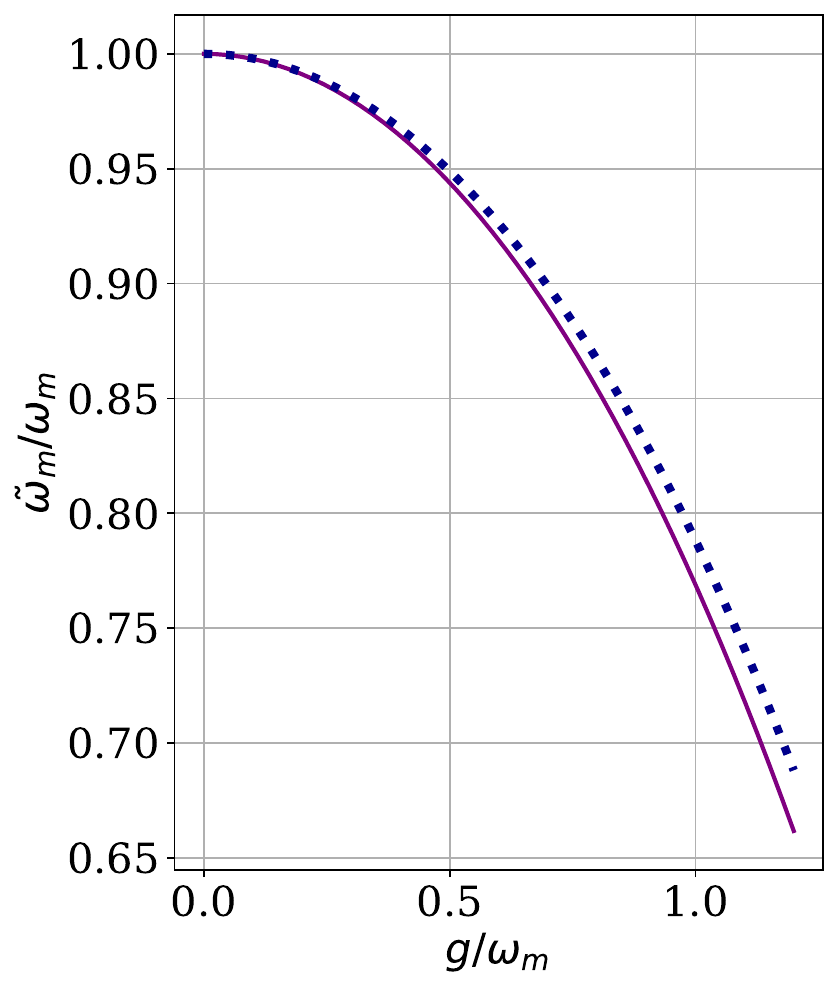}
\put(1,95){b)}
\end{overpic} 
\end{minipage}
\begin{minipage}{0.25\textwidth} 
    \begin{overpic}[width=\linewidth]{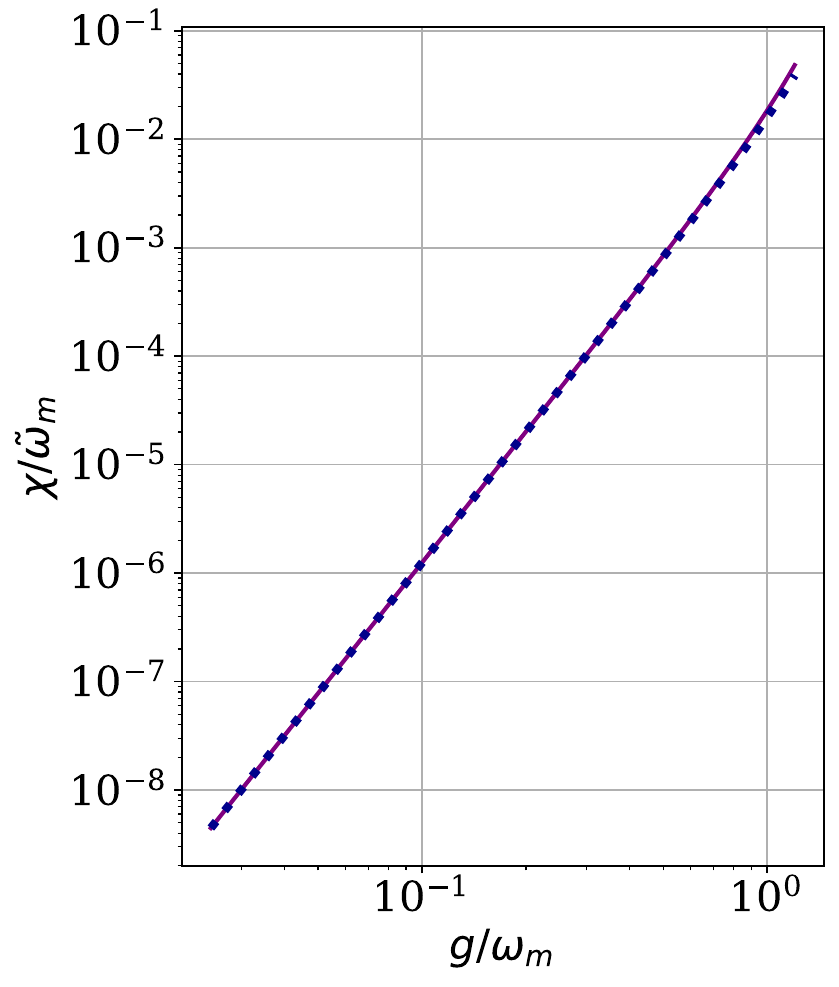}
\put(1,95){c)}
\end{overpic} 
\end{minipage}
\begin{minipage}{0.25\textwidth} 
        \begin{overpic}[width=\linewidth]{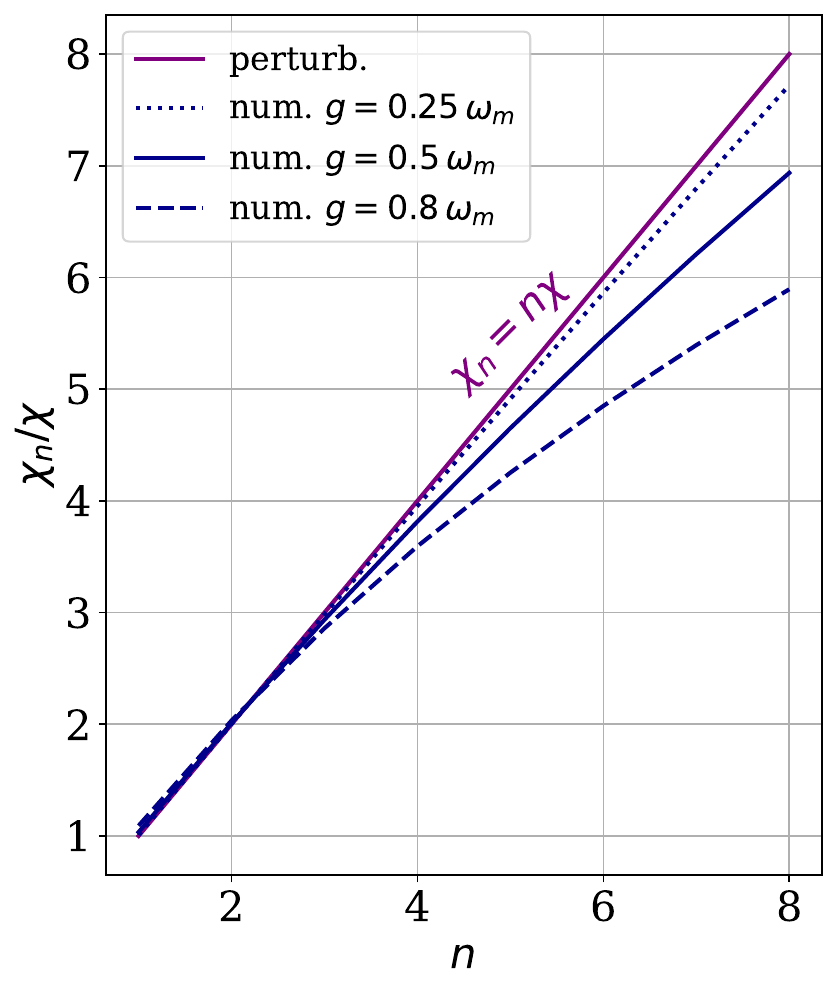}
\put(1,95){d)}
\end{overpic}  
\end{minipage}
\caption{
Comparison between perturbative diagonalization of the eigenvalue problem valid in the ultrastrong coupling regime and the numerical exact eigenenergies.
a) The 10 lowest eigenenergies from the perturbative approximation which yields a Kerr oscillator, see Eq.~\eqref{HKerr} compared to the numerical ones as a function of $g/\omega_\m$ for $\omega_\q = 10 \omega_\m$. 
b) The harmonic frequency $\tilde \omega_\m$ of the Kerr oscillator compared to the analogue numerical quantity $\tilde \omega_\m^\mathrm{num} = E_{1} - E_{0}$, \textit{i.e.}~the lowest-lying transition frequency as a function of coupling strength $g$. 
c) The anharmonicity of the Kerr oscillator compared to the numerical analogue quantity $\chi^\mathrm{num} = \omega_{21} - \omega_{10}$, where $\omega_{nm} = E_m - E_n$ and $E_n$ is the eigenenergy of the $n$th eigenstate. 
d) The difference between the $(n+1)$th transition frequency and the first transition frequency $\chi_n = \omega_{n+2,n+1} - \omega_{10}$ from the perturbative approximation and numerical diagonalisation for the first 10 transitions. In case of a Kerr oscillator this yields simply a linear dependency, $\chi_n = n \chi$ while the numerical diagonalisation shows deviations from the Kerr-oscillator behaviour with increasing $n$ and $g$. 
\label{fig:4th-order_comp}}
\end{figure*}

A comparison of the numerically exact eigenenergies and transitions with the fourth-order perturbative approximation is given in Fig.~\ref{fig:4th-order_comp}.
While the absolute value of the eigenenergies from the approximation fit well with the numerical exact ones even for large couplings $g \sim \omega_\m$, comparing the anharmonic behaviour shows that the actual eigenstructure of Eq.~\eqref{Hqrm} deviates from that of a perfect Kerr oscillator with increasing excitation number and coupling constant $g$. 
In a perfect Kerr oscillator the level-spacing behaves linear with excitation number $n$ and proportionality $\chi$, while the numerical exact solution shows increasingly non-linear behaviour in $n$ with increasing coupling $g$. 

\subsection{Born-Oppenheimer approximation}
In the deep strong coupling regime $g \ge \omega_\m$, we cannot approximately diagonalise the system in a similar manner. 
However, by using that the mechanical oscillator is slow with respect to the TLS, $\omega_\m \ll \omega_\q$, we can approximately map the Hamiltonian onto a 
system with an interaction that is longitudinal in respect to the TLS.
To this end, we diagonalise the TLS-degrees of freedom via $U= \exp(-\mi \theta(x) \sigma_y/2)$, where $x = a +a^\dag$ and $\tan \theta = 2 g x / \omega_\q$, 
\begin{multline}
    U^\dag H U  
     = \frac{\omega_\m}{4} (   x^2 +   p^2 )
+ \frac{V(  x)}{2} \sigma_z  
\\ 
 + \frac{\omega_\m}{4} \theta'(  x) ^2 -\frac{\omega_\m}{4} \sigma_y ( p \theta'(  x) +\theta'( x) p ) 
 \,, 
\end{multline}
where $p = - \mi ( a - a^\dag) $, $V(x)  = \sqrt{\omega_\q^2 +( 2 g x)^2} $ and $ \theta'(x) 
  =  
 (2 g/\omega_\q ) ( 1 + 4 g^2 x^2 / \omega_\q^2 )^{-1}$. 
Except for the last term, this Hamiltonian has the form of a particle in a  spin-dependent potential $V(x)\sigma_z$.

\begin{figure*}[htbp] 
\begin{overpic}[width=0.7\linewidth]{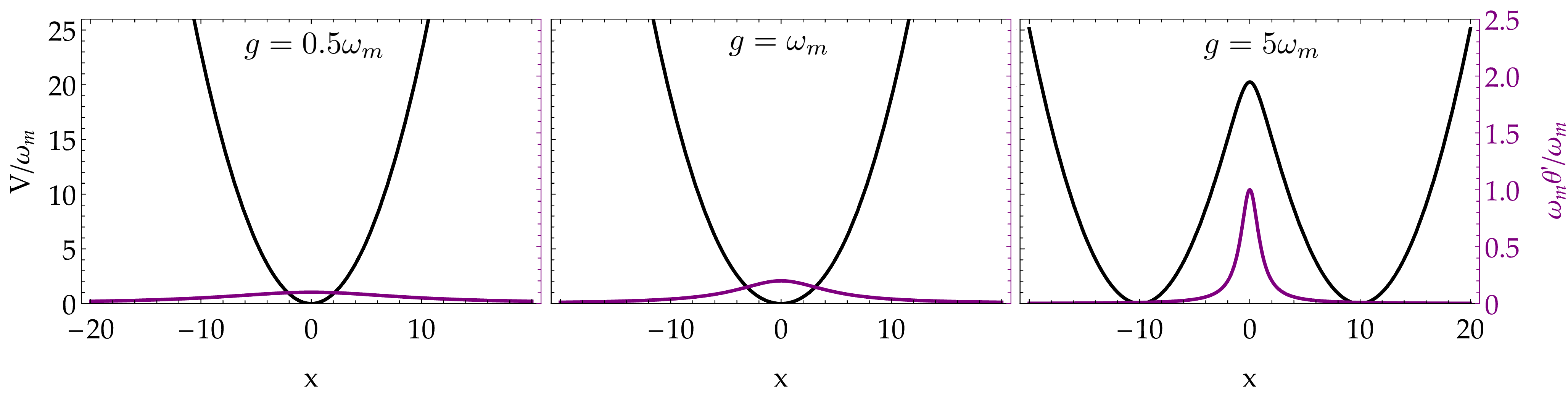}
\put(0,22){a)}
\end{overpic} 
\\ 
\begin{overpic}[width=0.7\linewidth]{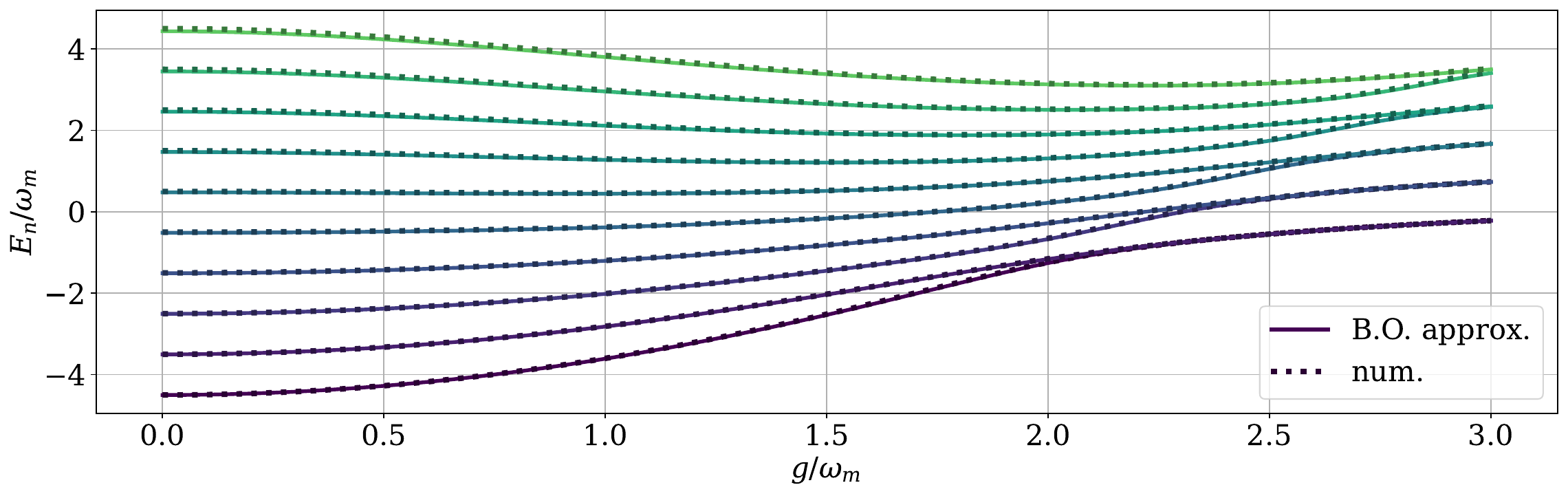}
\put(0,28){b)}
\end{overpic} 
\caption{ 
a)~The potential $V(x)$ (black) in comparison with the energy-term $\omega_\m \theta'(x)$ (purple) for different values of $g$ and $\omega_\q = 10 \omega_\m$. Note, that the scale for $\omega_\m \theta'(x)$ is magnified by 10 to enhance visibility.
b)~Eigenvalues of the Born-Oppenheimer Hamiltonian, Eq.~\eqref{HBO} in comparison with the ones of the full Hamiltonian Eq.~\eqref{Hqrm} as a function of $g$ and for $\omega_\q = 10 \omega_\m$. 
\label{fig:BO}}
\end{figure*}

We can show, that this term however is negligible for all couplings $g$. In the case of $g \ll \omega_\q$, we find $ \omega_\m \theta'(x) \sim \omega_\m g / J \ll 1$ and it is hence negligible. 
For larger values of $g$, the potential $V(x)$ forms a double well at the critical value of $g  =\sqrt{\omega_\m / \omega_\q} \omega_q/2$ with two minima $x_{1/2} \approx \pm 2 g / \omega_\m $ and the wave function in this potential then localizes around these two minima.
In contrast, the term $\theta'(x)$ maximizes at $x =0$ and vanishes for $|x| \gg \omega_\q/g$. 
Hence, if the minima position $|x_{1/2}|$ is much larger than the $\omega_\q/g$, the term $\theta'(x)$ vanishes in the region where the wave function has appreciable support. This condition translates into $g \gg \sqrt{\omega_\m / \omega_\q} \omega_q/ \sqrt{2}$, which coincides with the limit of the double-well formation. 
The remaining coupling regime is then defined by $\omega_\q \ll g \ll \sqrt{\omega_\m / \omega_\q} \omega_q$, which is non-existent for the applied limit of $ \omega_\m \ll \omega_\q$.  
This is shown in Fig.~\ref{fig:BO} for $\omega_\q = 10 \omega_\m$.
We may hence neglect $\theta'(x)$ in the Hamiltonian and approximate
\begin{align}
    U^\dag H U  \approx H_\mathrm{BO} =  \frac{\omega_\m}{4} (   x^2 +   p^2 )
+ \frac{V(  x)}{2} \sigma_z   
\label{HBO}
 \,.
\end{align}
This is equivalent with the Born-Oppenheimer approximation of separating fast (TLS) and slow (oscillator) time-scales of the Hamiltonian. 
Solving the eigen-spectrum of the Born-Oppenheimer approximated Hamiltonian numerically, we may compare it once again with the one obtained from using the full Hamiltonian Eq.~\eqref{Hqrm}.
The comparison is included in Fig.~\ref{fig:BO} and shows excellent agreement for all coupling strengths. 

Although the Born-Oppenheimer approximation does not yield a diagonal form of the Hamiltonian, it is valuable for developing intuition in the deep-strong coupling regime. It reveals a clear crossover in the nature of the lower-band bosonic modes: starting from harmonic oscillator modes at $g=0$, becoming progressively anharmonic for $g < \omega_\m$, and ultimately resembling the modes of a double-well potential when $g > \omega_\m$.  

While the closed quantum Rabi model is well understood and has even been solved analytically for all coupling strengths $g$~\cite{braak_integrability_2011}, the open system in the ultrastrong to deep-strong coupling regime has been considerably less studied.

\section{Lindblad master equation beyond full secular approximation}
\label{sec:new_diss}
In this section, we first briefly recall the Lindblad equation and its relation to the bath power spectrum using the example of an uncoupled TLS and a harmonic oscillator, for which the Lindblad form is well known and can be derived in a standard way.
We then present the Lindblad equation for a general system in a Born-Markov environment obtained under the slowly varying bath spectrum (SVS) approximation~\cite{settineri_dissipation_2018,mccauley_accurate_2020}, in a formulation tailored to situations where the bath spectrum is not explicitly known, and discuss its relation to the constructions of Refs.~\cite{settineri_dissipation_2018,mccauley_accurate_2020}.

Dissipation and decoherence of a quantum system arise from its coupling to an environment, also referred to as a bath. By definition, the environment may comprise arbitrarily many degrees of freedom, making its complete description intractable and its state generally inaccessible to measurement. Instead of describing the full system--bath dynamics explicitly, one considers the unitary evolution of the combined system and subsequently traces out the environmental degrees of freedom. This defines a dynamical map acting on the reduced density matrix $\rho(t)$ of the system. 
Under suitable conditions (namely weak system--environment coupling, a rapidly decaying bath correlation function, and an environment that remains close to its equilibrium state) the environment can be approximated as memoryless. In this case, one can apply the Born--Markov approximation  
and find a time-local dynamical map for the system evolution as 
\begin{align}
\dot{ \rho}  (t)
 &= \mi [ H , \rho]
 - \int_0^\infty \dif \tau 
 \tr_E \left[ H_\mathrm{SE} , \left[  H_\mathrm{SE}(-\tau), \rho(t) \otimes \rho_E \right] \right]
\end{align}
where $\rho_E$ is the bath state which is approximated to be stationary and $H_\mathrm{SE}$ is the coupling Hamiltonian between system and environment. 
The resulting master equation—commonly referred to as the Bloch-Redfield equation~\cite{breuer_theory_2007}—does not, in general, guarantee a physical time evolution of the reduced density matrix, as it may fail to preserve its positivity. By contrast, any linear, time-local master equation generating a trace-preserving and completely positive dynamical evolution can be cast into the so-called Lindblad form, which reads 
\begin{align}
    \dot \rho 
    = 
    - \mi [ H +H_\mathrm{LS}, \rho] + \sum_k \gamma_k \mathcal{D}_{C_k}[\rho] \,,
\end{align}
where $\mathcal{D}_{C_k}[\rho]$ is the Lindblad dissipator 
\begin{align}
    \mathcal{D}_{C_k}[\rho]
    = C_k \rho C_k^\dag - \frac{1}{2} \{C_k^\dag C_k, \rho \}
    \,,
    \label{Diss}
\end{align}
where $C_k$ is called jump or collapse operator acting on $\rho$ and $\gamma_k$ is the rate of the respective dissipation process and
$H_\mathrm{LS}$ is the Lamb shift-Hamiltonian, which is usually neglected in many application since the resulting frequency shifts usually induce a small renormalization of the bare frequencies in the system.  
A big advantage of the Lindblad form is its practicality and simplicity; each rate and its collapse operator correspond to a dissipation or decoherence channel and once the correct collapse operators for the system of interest are found, their rates can be left as free parameters that may be inferred from experimental data. 
Due to this, the Lindblad form is often used as first principle and its collapse operators are inferred from phenomenological arguments. 

The standard approach to obtain a Lindblad form from the Born-Markov equation is the full secular approximation.
It is presented in detail in the appendix, see App.~\ref{app:FSA}. 
Applied to the example of a single TLS $\rho_\mathrm{tls}$ coupled to an environment via 
$H_\mathrm{SE} = \sx \cdot \mathcal{E}_x + \sz \cdot \mathcal{E}_z$, where $\mathcal{E}_k$ are Hermitian and uncorrelated environment operators, the secular approximation yields the Lindblad equation,
\begin{multline}
      \dot \rho_\mathrm{tls}
    = - \mi [H_\mathrm{tls}, \rho_\mathrm{tls}] 
    +  \Gamma_\downarrow \mathcal{D}_{\sm}[\rho_\mathrm{tls}]
     \\
    + \Gamma_\uparrow \mathcal{D}_{\sp}[\rho_\mathrm{tls}]
    + \Gamma_\phi  \mathcal{D}_{\sz}[\rho_\mathrm{tls}]
    \label{LME_tls}
    \,, 
\end{multline}
where each dissipator term can be associated with a dissipation channel; emission, absorption and pure dephasing. 
The rates of each process can be related to the spectrum of the coupled environment operators, $\Gamma_\downarrow = S_x( \omega_\q) $, $\Gamma_\uparrow = S_x( - \omega_\q)$ and $\Gamma_\phi = 2 S_z(0)$, where 
\begin{align}
    S_k(\omega) &=  \int_{-\infty}^\infty  \dif \tau 
    e^{\mi \omega \tau} \braket{\mathcal{E}_k(\tau) \mathcal{E}_k(0) } 
    \label{S(w)}
    \,.
\end{align}
with $k\in \{ x , z\}$ and $S_x(\omega)$ being a thermal bath, we can write
\begin{align}
    S_x(\omega) &= 
    \begin{cases}
        n_\mathrm{th}(| \omega|) J_x(|\omega|) \quad &\text{for } \omega<0 
        \\
        (1 + n_\mathrm{th}( \omega)) J_x(\omega) \quad &\text{for } \omega>0         
    \end{cases}
\\
n_\mathrm{th}(\omega) &= \frac{1}{e^{\omega / k_\mathrm{B}T} - 1}
\end{align}
where $J_x(\omega)$ is the bath spectral density, which we assume to be temperature independent for simplicity and thus determines the emission rate at zero temperature as $\Gamma = J_x( \omega_\q)$.  

Similarly, a single harmonic oscillator coupled to a thermal environment via $H_\mathrm{SE} = ( a + a^\dag) \cdot \mathcal{E}_\m $ yields the Lindblad form 
\begin{multline}
    \dot \rho_\m 
     = - \mi [H_\m, \rho_\m]  
     + ( 1 + n_\mathrm{th}( \omega_\m) )\gamma  \mathcal{D}_a[\rho_\m]
     \\
    + n_\mathrm{th}( \omega_\m) \gamma  \mathcal{D}_a^\dag[\rho_\m]
    \label{LME_m}
\end{multline}
with $( 1 + n_\mathrm{th}( \omega_\m) )\gamma = S_\m(\omega_\m)$, $n_\mathrm{th}( \omega_\m) \gamma = S_\m ( - \omega_\m)$ and $\gamma = J_\m(\omega_\m)$.  

The master equations Eqs.~\eqref{LME_tls} and~\eqref{LME_m} are well known. However, when the two systems are coupled (ultra-)strongly, the dissipative dynamics are modified, making the construction of a valid Lindblad equation for the ultrastrong-coupling regime of the QRM a nontrivial task that has been studied extensively in the literature~\cite{forn-diaz_ultrastrong_2019}.
The often employed local master equation, where the incoherent evolution of the coupled system is simply approximated by the dissipators of the uncoupled case found in Eqs.~\eqref{LME_m} and~\eqref{LME_tls}, 
holds well in the Jaynes-Cummings regime but is known to introduce significant artificial heating into the system in the limit of USC, 
while the alternative standard approach of performing a full secular approximation on the system-bath coupling only holds for well isolated systems, \textit{i.e.}~if the rates of any incoherent dynamics is much smaller than the induced anharmonicity.

Here, we present a Lindblad equation which holds beyond the regime of full secular approximation and relies on a slowly varying bath spectrum (SVS)~\cite{settineri_dissipation_2018,mccauley_accurate_2020}; our formulation differs from those of Refs.~\cite{settineri_dissipation_2018,mccauley_accurate_2020} in the parametrization of the rates and thermal weights, as discussed below.
Its derivation is given in the appendix, see App.~\ref{app:our_LME}. 

We first introduce the approach for a general system-environment coupling given by 
\begin{align}
    H_\mathrm{full} &= H  + H_\mathrm{SE} + H_\mathrm{E}
    \\
    H_\mathrm{SE} &=   A \cdot  \mathcal{E}
\end{align}
where $H$ is the system Hamiltonian, $H_\mathrm{E}$ is the environment Hamiltonian and $\mathcal{E}$ is a Hermitian environment operator coupling to some Hermitian system operator $A$.
Assuming a Born-Markov environment and a slowly varying bath power spectrum $| S(\omega) -  S(\omega')| \ll S(\omega)$ for all positive environment-induced transition frequencies $\omega$, $\omega' > 0$ and under the assumption that all transition frequencies $\omega$ are much larger than the effective decay rate of the system, 
we can approximate the evolution of the system $\rho(t)$ with the Lindblad master equation -- neglecting the Lamb shift -- given by 
\begin{align}
    \dot \rho = - \mi [H , \rho] + \gamma \mathcal{D}_{A_\downarrow} [ \rho ]
    + \gamma \mathcal{D}_{A_\uparrow} [ \rho ]
    +  \gamma_\phi \mathcal{D}_{A_0} [ \rho ]
    \label{new_diss}
\end{align}
where $A_k$ are given by 
decomposing $A$ into its positive-negative frequency components and using the thermal Bose factors as weights, 
\begin{align}
    A_\downarrow &=
    \sum_{n<m} \sqrt{1 + n_\mathrm{th}(\omega_{nm})}\braket{\Psi_n|A|\Psi_m} \ket{\Psi_n}\bra{\Psi_m}   
   \label{A_dec_first}
    \\
    A_\uparrow &=
    \sum_{n>m} \sqrt{n_\mathrm{th}(\omega_{nm})}
    \braket{\Psi_n|A|\Psi_m} \ket{\Psi_n}\bra{\Psi_m}     
    \\
    A_0 &=
    \sum_{n}
    \braket{\Psi_n|A|\Psi_n} \ket{\Psi_n}\bra{\Psi_n}     
   \label{A_dec_last}
\end{align}
where $\ket{\Psi_n}$ are eigenstates of $H$ with eigenenergies $E_n$ and $\omega_{nm} =E_m - E_n$ are transition frequencies of $A$.
The rates are then given by 
\begin{align}
    \gamma &= J(|\omega_{nm}|) \,, 
    \gamma_\phi =  2 S(0)\,,
    \label{gammas_new_diss}
\end{align}
where we approximated $J(|\omega_{nm}|) = J (|\omega_{kl}|)$ for any transition frequency of interest. 
In the following we refer to the Lindblad equation obtained via this approximation as the SVS Lindblad equation.

The collapse operators constructed in this way ensure that the populations relax to those of the thermal Gibbs state, $\rho_\mathrm{th}  = \exp(- H/k_\mathrm{B}T )/Z$. Both the SVS Lindblad equation and the Bloch–Redfield equation, however, allow couplings between populations and coherences. These nonsecular contributions are suppressed by $\gamma_\mathrm{eff} /\omega_\m \ll 1$. Consequently, to leading order in $\gamma_\mathrm{eff} /\omega_\m$, the population dynamics obey detailed balance and the stationary state coincides with the Gibbs state, with corrections of order $\mathcal{O}(\gamma_\mathrm{eff} /\omega_\m)$. 
In many applications that are concerned with short time behavior or correlations at low thermal occupation numbers $n_\mathrm{th} \ll 1$, 
it is sufficient to approximate $n_\mathrm{th}(\omega) = n_\mathrm{th}(\omega')$ for the transition frequencies appearing in the sum. 
The thermal weights can then be taken out of the sum and be redefined into the rates, such that the definition of rates and respective collapse operators take on a more familiar form, 
$\gamma_\downarrow = ( 1+ n_\mathrm{th}) \gamma$, $\gamma_\uparrow= n_\mathrm{th} \gamma$. 
 This more severe approximation is analogue to the one presented in Ref.~\cite{settineri_dissipation_2018} and it approximates the steady state in thermal equilibrium as the one obtained from a purely harmonic system.  
The collapse operators in this approximation are given by 
\begin{align}
A^{(a)}_\downarrow = ( A^{(a)}_\uparrow)^\dag = \sum_{n <m} \braket{\Psi_n|A|\Psi_m} \ket{\Psi_n}\bra{\Psi_m} 
\end{align} 
and $A = A^{(a)}_\downarrow + A^{(a)}_\uparrow + A_0$.
On the other hand, an even less restrictive approximation can be obtained by introducing a multitude of rates $\gamma_{nm} = S( \omega_{nm})$ and integrate them into the definition of the collapse operators, \textit{i.e.}~
\begin{align}
    {A^{(b)}_\downarrow = \sum_{n <m } \sqrt{ \gamma_{nm} ( 1 + n_\mathrm{th}(\omega_{nm} )) } \braket{\Psi_n|A|\Psi_m} \ket{\Psi_n}\bra{\Psi_m}}
    \,. 
\end{align}
This corresponds to the collapse-operator construction introduced in Ref.~\cite{mccauley_accurate_2020}.
When the bath spectrum $S(\omega)$ is known  explicitly, this formulation is advantageous, as it
is valid under the less restrictive condition $|S'(\omega) |\ll 1$ for all relevant transition frequencies $\omega$ , whereas our formulation requires $| ( \omega - \omega') S'(\omega)| \ll S(\omega)$ for all relevant transition frequencies $\omega$ and $\omega'$. 
Moreover, McCauley \emph{et al.} further demonstrated that, at zero temperature, the condition $|S'(\omega) |\ll 1$ is equivalent with the Born-Markov approximation. 
However, if the bath spectrum is not known explicitly, this formulation requires introducing an independent parameter $\gamma_{nm}$ for each transition or assuming a specific model for $S(\omega)$, neither of which is necessarily practical in realistic applications.

In contrast, the Lindblad master equation obtained with the present parametrization, Eqs.~\eqref{A_dec_first}--\eqref{A_dec_last} introduces a minimal set of unknown environment parameters, namely the rate $\gamma$, the temperature $T$ and possibly a pure dephasing rate $\gamma_\phi$.
The derivation is given in the appendix, see App.~\ref{app:our_LME} with a comparison to the usual secular approximation, which is derived in App.~\ref{app:FSA} and a detailed discussion about its limitations. 

\section{Lindblad master equation of the quantum Rabi model}
\label{sec:diss_QRM}
In this section, we apply the SVS approximation to the system of interest to obtain a Lindblad equation that is valid beyond the conventional full secular approximation (FSA). 
Within the perturbative Kerr regime, we obtain analytical expressions for the dissipator and compare it to other standard Lindblad equations, showing that the obtained master equation reduces to known ones in the respective limits. 

We consider the quantum Rabi model~\eqref{Hqrm} with coupling to the environment as 
\begin{align}
    H_\mathrm{full} &= H + H_\mathrm{SE} + H_E
    \\
    H_\mathrm{SE} &=   x \cdot \mathcal{E}_\m
    +  X \cdot \mathcal{E}_x
    +  Z \cdot\mathcal{E}_z
    \,,
    \\
    x &= a + a^\dag \,, \quad X = \sigma_x \,,\quad Z = \sigma_z \,.
\end{align}

By constructing the dissipator defined in Sec.~\ref{sec:new_diss} for each system-environment coupling, we 
get the SVS Lindblad master equation 
\begin{align}
    \dot \rho = - \mi [H , \rho] 
    & + \gamma \mathcal{D}_{x_\downarrow} [ \rho ]
    + \gamma  \mathcal{D}_{x_\uparrow} [ \rho ]
    \nn
    &+ \Gamma \mathcal{D}_{X_\downarrow} [ \rho ]
    + \Gamma \mathcal{D}_{X_\uparrow} [ \rho ]
    \nn
    & + \Gamma^{(z)}  \mathcal{D}_{Z_\downarrow} [ \rho ]
    + \Gamma^{(z)}  \mathcal{D}_{Z_\uparrow} [ \rho ]
    +  \Gamma_{\phi} \mathcal{D}_{Z_0} [ \rho ]\,,
\end{align}
with $H$ being the quantum Rabi Hamiltonian Eq.~\eqref{Hqrm} and 
\begin{align}
    C_\downarrow &= \sum_{n < m} \sqrt{1 + n_\mathrm{th}(\omega_{nm})} \braket{\Psi_n| C| \Psi_m} \ket{\Psi_n}\bra{\Psi_m} \,, 
    \\
    C_\uparrow &= \sum_{n > m} \sqrt{ n_\mathrm{th}(\omega_{nm})} \braket{\Psi_n| C| \Psi_m} \ket{\Psi_n}\bra{\Psi_m} \,, 
    \\ 
    Z_0 &= \sum_{n} \braket{\Psi_n| \sz | \Psi_n} \ket{\Psi_n}\bra{\Psi_n}  \,, 
\end{align}
where $C \in \{x, X, Z\}$,  $\ket{\Psi_n}$ are the eigenstates of the QRM, 
and $x_0 = X_0 = 0$ due to the eigenstate structure. In principle, the rates are defined by Eq.~\eqref{gammas_new_diss} for each system-environment coupling but in practice, they can be treated as free parameters. 
The approximation is valid for $|\omega_{nm}| \gg \gamma_\mathrm{eff}$, where $\gamma_\mathrm{eff}$ is the effective decay of the system resulting from the Lindblad equation and for $|S(\omega_{nm}) - 
S_k(\omega_{kl})| \approx | ( \omega_{nm} - \omega_{kl}) S'_k(\omega_{nm})| \ll S_k(\omega_{nm})$ for all $k \in \{ \m,x,z\}$ and $\omega_{nm}$ relevant for the definition of the respective $C_{\downarrow}$.

In the perturbative Kerr regime, we can write $H$ as $H_\mathrm{Kerr}$, see Eq.~\eqref{HKerr} for the lower band of the Hilbert space. We can then analytically calculate the collapse operators for the Lindblad master equation. 
In the coupled eigenstate basis of the lower band $\ket{\Psi_{n -}}$ we can approximate in leading order
\begin{align}
    \sx  & \approx   \frac{2 g \omega_\q }{\omega_\q^2 - \omega^2 _\m}
    ( b  +b^\dag )  + \mathcal{O}(g^3) \label{map_sx}
    \\
    \sz  & \approx      
   \frac{2  g^2  \omega_\q}{ (\omega_\q^2 -\omega_\m^2)^2 } 
   b^\dag b  
+
\frac{g^2  }{\omega_\q^2-\omega_\m^2}( (b^\dag)^2+ b^2 )
+ \mathcal{O}(g^4)
\label{map_sz}
\end{align}
while $a$ and $a^\dag$ map to $b$ and $b^\dag$ in leading order, respectively. 
At zero temperature, the Bose factors in the definition of the collapse operators vanish and the Lindblad master equation in the perturbative Kerr regime takes on a familiar form, 
\begin{multline}
    \dot \rho \, \big|_\mathrm{PKR} 
    = - \mi [H_\mathrm{Kerr} , \rho] + \gamma  \mathcal{D}_{b} [ \rho ] 
    + \gamma_\mathrm{eff} \mathcal{D}_{b} [ \rho ] 
    \\
    + \gamma_\mathrm{eff}^{(z)} \mathcal{D}_{b^2} [ \rho ] 
    + 2 \gamma_{\mathrm{eff},\phi} \mathcal{D}_{b^\dag b} [ \rho ]\,,
\end{multline}
where we redefined the coefficients of the mapping in Eq.~\eqref{map_sx} and Eq.~\eqref{map_sz} into effective rates as 
\begin{align}
    \gamma_\mathrm{eff} =  \frac{4 g^2 \omega_\q^2 }{( \omega_\q^2 - \omega^2 _\m)^2} \Gamma 
    \label{gamma_eff}
    \,,
    \\
    \gamma_\mathrm{eff}^{(z)} =  \frac{ g^4 }{( \omega_\q^2 - \omega^2 _\m)^2} \Gamma^{(z)} 
    \,,
    \\
    \gamma_\mathrm{eff, \phi}
    = \frac{4 g^4 \omega_\q^2}{ (\omega_\q^2 -\omega_\m^2)^4 } \Gamma_\phi \,.
    \label{gamma_phi_eff}
\end{align}
These effective rates coincide with the ones found in the conventional secular approximation~\cite{pistolesi_proposal_2021}.

Beyond the perturbative regime the decomposition of each environment-coupling operator can be easily carried out numerically. We can then define the effective parameters in an analogous way, such that \textit{e.g.}
\begin{align}
    \gamma_\mathrm{eff} 
    &= |\braket{\Psi_1| \sx | \Psi_0}|^2 \Gamma 
    \,,
    \label{gamma_eff_num}
\end{align}
which converges to its analytical approximation Eq.~\eqref{gamma_eff} for $g/\omega_\m \ll 1$. 
The environment-coupling with $\sigma_z$ leads to a highly suppressed dephasing rate $\gamma_\mathrm{eff,\phi} $ and decay rate $\gamma_\mathrm{eff}^{(z)}$,
each suppressed by $g^4/\omega_\q^4$, while the decay from the $\sx$-coupling is only suppressed by $g^2/\omega_\q^2$ and we assume the intrinsic decay of the mechanical oscillator to be in comparison negligibly small, \textit{i.e.}~$\gamma \ll \omega_\m$ by several orders of magnitude.

We hence focus on the dominant dissipation channel, $H_\mathrm{SE} = \sigma_x \cdot \mathcal{E}$, yielding 
\begin{align}
    \dot{\rho}
    = - \mi [ H, \rho]
    + \Gamma \mathcal{D}_{X_\downarrow}[\rho]
    + \Gamma \mathcal{D}_{X_\uparrow}[\rho]
    \,,
    \label{SVS_sx}
\end{align}
and 
\begin{align}
    \dot \rho \big|_\mathrm{PKR}
    = - \mi [H_\mathrm{Kerr} , \rho]  
    + \gamma_\mathrm{eff} \mathcal{D}_{b} [ \rho ] \,,
\end{align}
in the PKR limit at zero temperature. 
As mentioned previously, this master equation neglects the Lamb shift resulting from the system-environment coupling. A discussion on the Lamb shift's influence on the anharmonicity of the system is given in App.~\ref{app:Lamb}, where we explicitly show that Lamb shift-induced anharmonicity is usually negligible. 

In the uncoupled case $g=0$, we simply find  
${X_\downarrow = \sqrt{ 1 + n_\mathrm{th}(\omega_\q)}\sm}$, 
such that we recover the known master equation for a bare harmonic oscillator and a TLS, 
\begin{multline}
    \dot \rho \,  \big|_{g = 0}
     = - \mi [H_\m + H_\mathrm{
    tls}, \rho ]  
    \\
    +  \Gamma  (1 + n_\mathrm{th}( \omega_\q )) \mathcal{D}_{\sm}[\rho ]
    +  \Gamma   n_\mathrm{th}( \omega_\q )  \mathcal{D}_{\sp}[\rho ] 
    \label{ME_g=0}
    \,.
\end{multline}
Similarly, in the Jaynes-Cummings regime where 
${H_\mathrm{JC} = H_\m + H_\mathrm{tls} + g ( a \sp + a^\dag \sm)}$ and ${g,\,|\omega_\q - \omega_\m | \ll \omega_\q + \omega_\m }$, the coupling Hamiltonian does not reorder the
negative-positive frequency decomposition in comparison to the uncoupled case. Approximating then additionally $n_\mathrm{th}(\omega_{nm})$ inside the definition of the collapse operators Eqs.~\eqref{A_dec_first}--\eqref{A_dec_last} with the thermal occupation evaluated at the bare eigen-frequencies $\omega_\m$ and $\omega_\q$, recovers the local master equation, where the dissipator coincides with the one of the uncoupled system, Eq.~\eqref{ME_g=0}. 

In the perturbative Kerr regime, we can readily show that the local master equation predicts significant artificial heating once the rotating-wave approximation breaks down and counter-rotating terms become relevant. 
At zero temperature and with only the dominant dissipation channel the local master equation reads
\begin{align}
    \dot \rho_\mathrm{lme}= - \mi [H , \rho_\mathrm{lme}]  
    + \Gamma \mathcal{D}_{\sm} [ \rho_\mathrm{lme} ] \,,
\end{align}
which can be written in the eigenbasis by approximating $\sm$ in leading order
\begin{align}
    \sm \approx \frac{g}{\omega_\q - \omega_\m} b + \frac{g}{\omega_\q + \omega_\m} b^\dag \,.
\end{align}
For $\omega_\q \gg \omega_\m$, this describes an absorption process with $b^\dag$ approximately as much as an emission process with $b$, while both processes are weighted with the same rate $\Gamma$ independently of the actual bath temperature, hence introducing artificial heating with a temperature of $T_\mathrm{art} = \omega_\q / \ln [ (\omega_\q + \omega_\m) / ( (\omega_\q - \omega_\m)] $, which diverges for $\omega_\m / \omega_\q \rightarrow 0$.  

This remains true even in the regime $g \ll \Gamma $, \emph{i.e.}~in the weak coupling limit. 
In this limit, one conventionally uses the local Lindblad equation to approximate system dynamics. However, here the problem is more subtle: since the mechanical oscillator is coupled to the environment only via the TLS, the TLS–oscillator interaction is essential for correctly describing the oscillator's coupling to the environment, and a dissipator constructed from the uncoupled system is therefore not a valid approximation.
A proof that the SVS Lindblad equation remains consistent with the Born-Markov description in this regime is given in App.~\ref{app:TLS_as_env}.

The standard approach to avoid artificial heating effects like this is given by the full secular approximation, which leads to a Lindblad master equation for the perturbative Kerr regime at zero-temperature given by 
\begin{gather}
    \dot \rho_\mathrm{fsa} 
     = - \mi [H , \rho_\mathrm{fsa} ]  
    +\sum_{n} \gamma_\mathrm{eff}^\mathrm{(fsa)}(\omega_{n,n-1})   \mathcal{D}_{c_n} [ \rho_\mathrm{fsa}  ] \,,
    \label{rho_dot_kerr_fsa}
    \\
    c_n 
     = \sqrt{n} \ket{n-1}\bra{n}
    \label{c_n}
    \,,
    \\
    \gamma_\mathrm{eff}^\mathrm{(fsa)} (\omega )  
     = \gamma_\mathrm{eff} \frac{J(\omega )}{\Gamma}
    \,, 
\end{gather}
which within the SVS approximation, \textit{i.e.}~$J(\omega_{n,n-1}) = \Gamma$ for all $n$, gives the constant effective rate $\gamma_\mathrm{eff}$ defined by Eq.~\eqref{gamma_eff}.
This is closely related to the SVS Lindblad master equation but neglects, in comparison, additional interference terms
which can only be done if $\gamma_\mathrm{eff} \ll \chi$, see App.~\ref{app:FSA}.
This condition can be related to the original system parameters by using Eqs.~\eqref{chi} and~\eqref{gamma_eff}, such that the FSA limit in the perturbative Kerr regime is given by
\begin{align} 
    \Gamma 
\ll 
\frac{g^2 \left(\omega_\m^2+3 \omega_\q^2\right)}{\omega_\q \left(\omega_\q^2-\omega_\m^2\right)}
\approx \frac{3 g^2}{\omega_\q}\,,\quad \omega_\m \ll \omega_\q
 \,.
 \label{FSA_cond}
\end{align}
The difference in the state evolution between the SVS Lindblad master equation and the one obtained from FSA is then the occurrence of transfer of coherence: when using the SVS Lindblad master equation the coherences $\rho_{nm} = \braket{n| \rho|m}$ evolve in the perturbative Kerr regime as 
\begin{multline}
    \dot \rho_{nm}
     = - \mi \omega_{nm} \rho_{nm}
    - \frac{\gamma_\mathrm{eff}}{2} (n +m ) \rho_{nm}  
    \\
    + \gamma_\mathrm{eff} \sqrt{(n+1)(m+1)} \rho_{n+1,m+1}
    \label{toc}
\end{multline}
where the last term describes the transfer of coherence and it decouples effectively if $\rho_{n+1,m+1}(t)$ oscillates at frequencies $\omega_{n+1,m+1} = E_{n+1} - E_{m+1} = \omega_{nm} + (m -n) \chi$ that are sufficiently different from $\omega_{nm}$ in comparison to the effective decay, \textit{i.e.}~for $\gamma_\mathrm{eff} \ll \chi$. 
In the FSA on the other hand, this effective decoupling is applied manually, 
such that the coherences evolve as 
\begin{align}
    [ \dot \rho_\mathrm{fsa}]_{nm}
    &= - \mi \omega_{nm} [\rho_\mathrm{fsa}]_{nm}
    - \frac{\gamma_\mathrm{eff}}{2} (n +m ) [\rho_\mathrm{fsa}]_{nm}
    \,.
    \label{no_toc}
\end{align}

Finally, adding an additional perturbation to the system, such as a drive can in principle make it necessary to redefine the dissipator. 
Only for a sufficiently small perturbation the unperturbed dissipator can be kept. 
The limit for this perturbation is again set by different scales in the FSA and SVS Lindblad equation; in the secular approximation the perturbation must be much smaller than the anharmonicity $\chi \ll \omega_\m$ while the SVS Lindblad dissipator holds for any perturbation much smaller than the transition frequencies itself $\omega_{nm} \sim \omega_\m$. 
This is discussed in more detail in Sec.~\ref{sec:drive} and shown explicitly for the driven Kerr oscillator. 
 
In summary, we showed that even for arbitrarily small coupling $g$, the local master equation introduces a significant artificial temperature $T_\mathrm{art}$ outside of the Jaynes-Cummings regime, while the full secular approximation neglects interference effects that are only negligible for $\Gamma \ll g^2 / \omega_\q$. 
In contrast, the SVS Lindblad master equation is valid beyond this limit and avoids unphysical heating effects. It reduces to the local master equation in the Jaynes-Cummings regime as well as to the secular approximated master equation for sufficiently small decay rates and finally the SVS Lindblad dissipator is applicable without further adjustments to a wider range of driving amplitudes.

\section{Phonon-blockade and intermediate bunching regime}
\label{sec:g2}
With the help of the SVS Lindblad master equation we may now study the dynamics of the QRM in the ultrastrong and up to deep-strong coupling limit for parameter regimes inaccessible with the standard approach of secular approximation. It furthermore provides a unified description across the full parameter regime, interpolating continuously between the known limiting cases.
This includes the regime beyond complete anti-bunching, \textit{i.e.}~full phonon blockade where the hybrid system can be addressed as a two-level system by a drive. 
To quantify the bunching behavior, we study the two-phonon-correlation function given by the equal-time second-order correlation function $g^{(2)}(0)$ in the infinitesimal-drive limit, where the drive amplitude is taken to be the smallest energy scale in the system and the response is governed by the lowest-order excitation processes~\cite{liu_qubit-induced_2010, shi_tunable_2018}. 
The function indicates anti-bunching for $g^{(2)}(0) <1$, bunching for $g^{(2)}(0) > 1$ and a complete phonon blockade at $g^{(2)}(0) \rightarrow 0$ where the anharmonicity induced by the TLS-coupling prevents multi-phonon excitations in close analogy to photon blockade~\cite{tian_quantum_1992, imamoglu_strongly_1997, birnbaum_photon_2005,rabl_photon_2011}. 
Complete phonon blockade is of particular interest as it indicates the realization of a mechanical qubit. 
As we demonstrate explicitly, however, the phonon-blockade regime coincides with the validity limit of the secular approximation, leaving a large relevant parameter space beyond its applicability, where the SVS Lindblad approach provides a consistent description.

As introduced in the previous section, we keep the focus of the discussion on the dominant dissipation channel $H_\mathrm{SE} = \sigma_x \mathcal{E}$, and assume zero temperature $k_\mathrm{B} T \ll \omega_\m$.
To induce dynamics into the system, we consider an infinitesimal drive via the TLS and close to resonance of the mechanical mode, 
\begin{align}
    H_D(t)  = \varepsilon_D \cos( \omega_D t) \sx \,,
    \label{H_D}
\end{align}
with $\omega_D \sim  \omega_\m$.  
For an infinitesimal drive, where $\varepsilon_D \ll \chi ,\, \gamma_\mathrm{eff}$ and all other relevant system scales, the drive-induced modifications of the transition frequencies can be neglected. 
Consequently, the dissipator can be constructed from the unperturbed Hamiltonian. 
This approximation and its validity is discussed in detail in Sec.~\ref{sec:drive}.
At zero temperature the master equation then reads
\begin{align}
    \dot \rho 
    &= - \mi [ H +H_D(t), \rho ] 
    + \Gamma \mathcal{D}_{X_\downarrow}[ \rho ] 
    \,.
    \label{dot-rho_driven}
\end{align}
\begin{figure}
\centering
\includegraphics[width=   \linewidth]{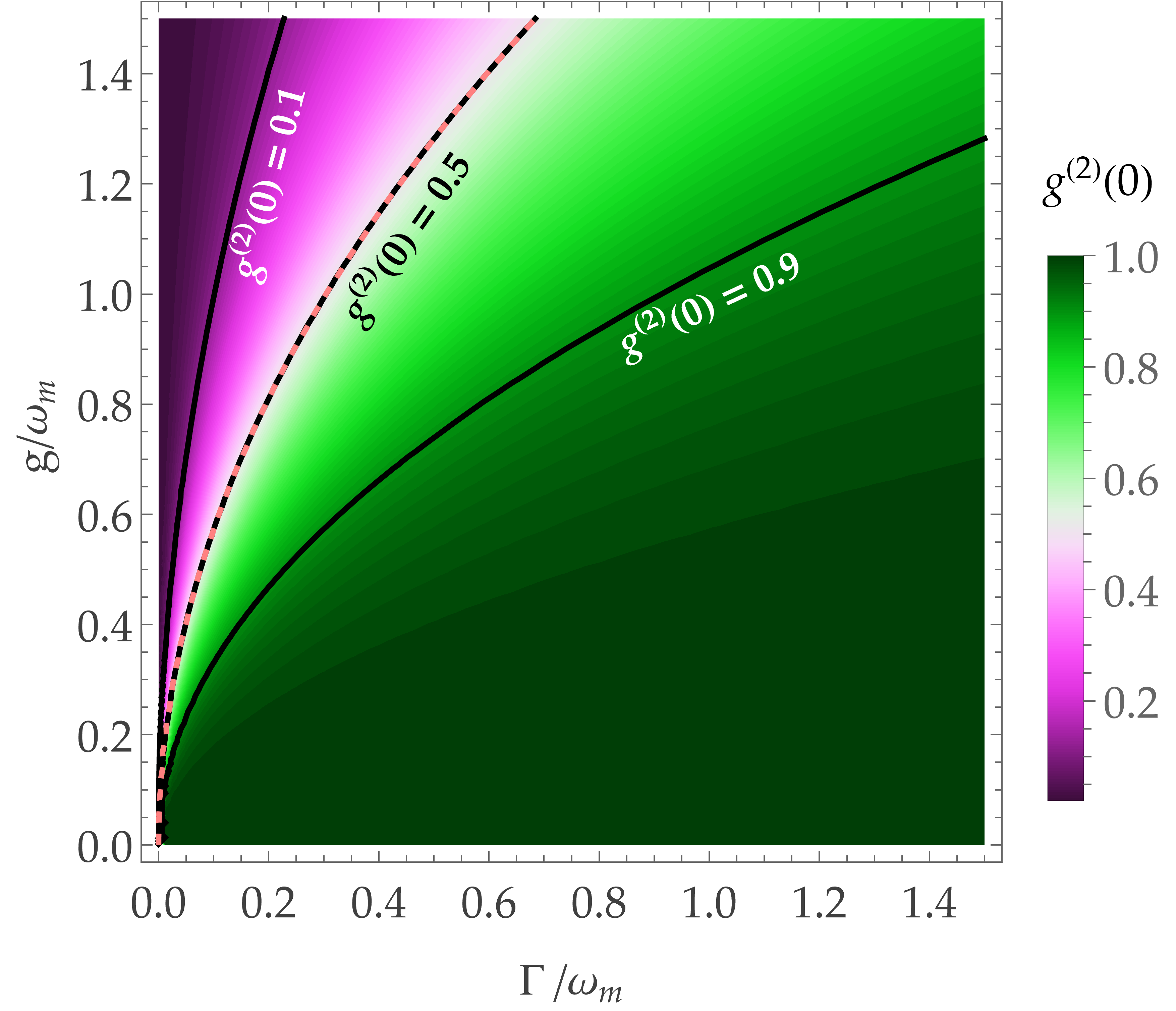}
\caption{
The two-phonon correlation function given by Eq.~\eqref{g2_Kerr} for an infinitesimal drive on resonance with the lowest transition in the effective Kerr oscillator as a function of the oscillator-TLS coupling $g$ and the TLS decay rate at zero-temperature $\Gamma$ for $\omega_\q = 10 \omega_\m$. 
The FSA limit is additionally plotted as a pink dashed line and coincides exactly with the value of $g^{(2)}(0) = 0.5$. 
\label{fig:g2_ana}}
\end{figure}
We start by discussing the dynamics in the perturbative Kerr regime where the system Hamiltonian is approximated by $H_\mathrm{Kerr}$, Eq.~\eqref{HKerr}. 
The drive then maps to $b$ and $b^\dag$ in leading order as 
\begin{align}
    H_D &= \tilde\varepsilon_D \cos( \omega_D t)  ( b+ b^\dag) + \mathcal{O}(g^3/\omega_\q^3 (b^3 + b^{\dag 3})) 
    \,,
    \label{PKR_drive}
    \\
    \tilde \varepsilon_D &=  \frac{2 g \omega_\q \varepsilon_D}{\omega_\q^2 - \omega_\m^2}
    \label{eps_D_eff}
    \,,
\end{align} 
where $\tilde \varepsilon_D$ is the effective drive amplitude. 
With the drive only considered in leading order, we may perform a rotating wave approximation for a weak drive $\tilde \varepsilon_D \ll \omega_D$. 
The dissipator is invariant under the transformation within the rotating wave approximation and we find for the master equation at zero temperature
\begin{align}
    \dot \rho 
    &= - \mi [ H_\mathrm{rwa},\rho] 
    + \gamma_\mathrm{eff}  \mathcal{D}_{b}[ \rho ] 
    \label{rho_rwa}
    \,,
    \\
    H_\mathrm{rwa}
    &= - \Delta b^\dag b + \chi/2 b^\dag b^\dag b b  + \frac{\tilde{\varepsilon}_D}{2} ( b + b^\dag) 
    \label{H_rwa}
    \,,
\end{align}
where $\Delta = \omega_D - \tilde \omega_\m$ is the detuning from the lowest transition frequency and the effective rate in the perturbative limit is given by Eq.~\eqref{gamma_eff}. 

The second-order normalized correlation function is given by 
\begin{align}
    {g^{(2)}(t,\tau) = \frac{\braket{b^\dag (t) b^\dag (t + \tau) b(t + \tau) b(t)}}
    {\braket{b^\dag (t)  b(t)}\braket{b^\dag (t + \tau)  b(t + \tau)} }} \,,    
\end{align}
which at $\tau =0$  and for $t \rightarrow \infty$ characterizes the bunching behavior of the phonons in the steady state and we hence refer to it then as the two-phonon correlation,  
\begin{align}
    g^{(2)}(0) =
    \lim_{t \rightarrow \infty} \frac{\braket{b^\dag(t) b^\dag (t)b(t) b(t)}}{\braket{b^\dag(t) b(t)}^2}
    \,,
    \label{g2}
\end{align}
where 
the expectation value is taken in the rotating-frame steady state of the driven system.

For infinitesimal drives, the two-phonon-correlation function is independent of $\varepsilon_D$ and instead depends only on properties of the undriven system. 
We can then calculate $g^{(2)}(0)$ from Eq.~\eqref{rho_rwa} analytically and find at zero temperature,
\begin{align}
   g^{(2)}(0)\Big|_{\varepsilon_D\rightarrow 0 }  =  \frac{\gamma_\mathrm{eff} ^2+4 \Delta ^2}{\gamma_\mathrm{eff} ^2+(2 \Delta +\chi )^2}
   \,,
   \label{g2_Kerr}
\end{align}
see the appendix, App.~\ref{app:g2} for the derivation. 
When driving in resonance with the first transition, $\Delta = 0$ the two-phonon correlation function for infinitesimal drives takes a very simple form $g^{(2)}(0)  = ( 1 + \chi^2 / \gamma_\mathrm{eff}^2 )^{-1} $ which increases from zero for $\chi \gg \gamma_\mathrm{eff}$, corresponding to a complete phonon blockade to unity in the opposite limit, corresponding to a coherent state in the driven system. 
We can relate $\gamma_\mathrm{eff}$ and $\chi$ back to the original parameters of the system, see Eqs.~\eqref{chi} and~\eqref{gamma_eff}, expressing $g^{(2)}(0)$ as function of $g$ and $\Gamma$. This is plotted in Fig.~\ref{fig:g2_ana}, 
where one can see that the cross-over between phonon-blockade and coherent state at $g^{(2)}(0) = 0.5$ coincides exactly with the limit for secular approximation, Eq.~\eqref{FSA_cond}.
 
\begin{figure*} 
\centering
\begin{overpic}
   [width= 0.35 \linewidth]{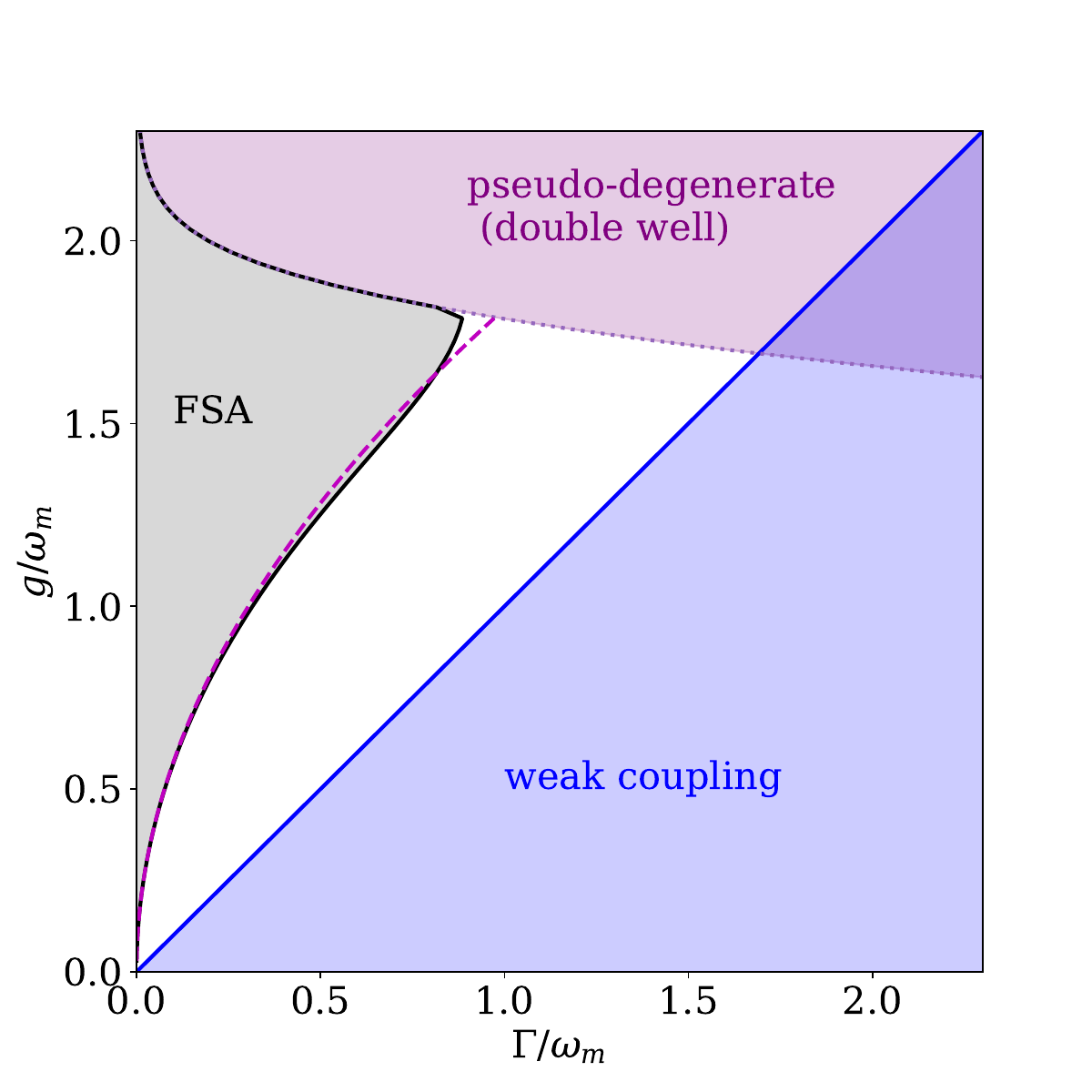}
\put(0,85){a)} 
\end{overpic}
\begin{overpic}
   [width=  0.5 \linewidth]{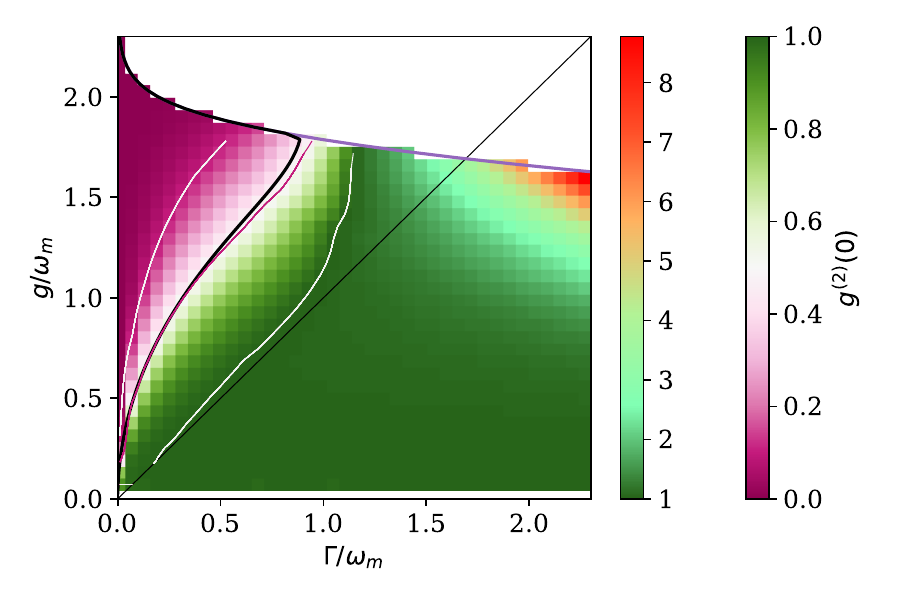}
\put(0,60){b)} 
\end{overpic}
\caption{a) 
Phase diagram of the different regimes found in the quantum Rabi model for $\omega_\q = 10 \omega_\m$ as a function of the TLS decay rate $\Gamma$ at  zero-temperature and the coupling constant $g$. Deep inside the gray area, we find $\gamma_\mathrm{eff} \ll \mathrm{min}\{\chi , 2\omega_{10}\}$, where $\chi = |\omega_{21} - \omega_{10}|$ is the difference of the two lowest lying transition frequencies of the eigenstates and $\omega_{10}$ is the lowest transition frequency. For FSA to be valid the effective decay rate must be lower than each of them. 
The analytical FSA-limit approximated for $g \ll \omega_\m$ is plotted as a dashed pink line. 
The purple area marks the region of pseudo-degeneracy, where $\omega_{10} < \gamma_\mathrm{eff}$ where the interaction forms an effective double well, see Fig.~\ref{fig:BO}. In this region neither the SVS Lindblad master equation nor FSA are valid in the presented form. The blue area marks the weak coupling region where $\Gamma > g$, which is well described by the SVS Lindblad master equation as discussed in App.~\ref{app:TLS_as_env}. 
b)
The two-phonon correlation function found via numerical simulation of the time-dependent Liouville equation Eq.~\eqref{dot-rho_driven} on resonance with $\omega_{10}$. The white contours mark $g^{(2)}(0) = 0.1$ in the pink region and $g^{(2)}(0) = 1$ in the green region, while the pink contour marks $g^{(2)}(0) = 0.5$ which coincides with the FSA-limit (black contour). The effective drive amplitude was chosen as $\tilde \varepsilon_D = 
\chi/50$ for which the simulation results are converged with respect to further reductions of $\varepsilon_D$. 
\label{fig:g2_num}}
\end{figure*}

Outside of the perturbative Kerr regime, the system cannot be approximated by a simple Kerr oscillator and the mapping of the relevant operators into the eigenbasis must be carried out numerically.
The full driven system cannot be cast into a time-independent problem by a rotating wave approximation and we need to integrate the time-dependent master equation, Eq.~\eqref{dot-rho_driven} numerically. 
Additionally, the state then does not converge into a time-independent steady state as it is still oscillating with the drive frequency, we refer to the long-time limit of $\rho(t)$ hence as pseudo steady state. 

Additionally, special care is required when defining the two-phonon correlation function $g^{(2)}(0)$ outside of the perturbative Kerr regime. We follow the solution discussed in Ref.~\cite{ridolfo_photon_2012} on input-output theory for photon detections in an ultrastrongly coupled system 
which is based on the original work of Glauber~\cite{glauber_quantum_1963}. 
Here, we quickly summarize the idea.
Following Glauber~\cite{glauber_quantum_1963} (but translating to phononic fields), the probability of detecting a phonon via an ideal detector for mechanical displacement, is proportional to $\braket{x_+ x_-}$, where $x_\pm$ are the positive/negative frequency components of the displacement operator $x = a^\dag + a$ in the coupled eigenbasis, 
\begin{align}
    x_-& = \sum_{n < m } \braket{\Psi_n| x | \Psi_m} \ket{\Psi_n}\bra{\Psi_m}
    \\
    x_+ &= (x_-)^\dag
\end{align}
such that $x = x_- + x_+$ and $x_- \rightarrow a$ for $g \rightarrow 0$.
Analogously, higher order correlation functions are given by $\braket{x_+(t) x_+(t') x_-(t') x_-(t)}$. 
With this re-definition the energy flux associated with the measured output field is proportional to $\braket{x_+ x_-}$, which yields correctly a zero-energy flux in the ground state $\braket{\Psi_0| x_+ x_-|\Psi_0} =0$, 
while using the original annihilation and creation operators instead predicts a constant energy flux in the ground state $ \braket{\Psi_0|a^\dag a | \Psi_0} > 0$. 
The two-phonon correlation function then is given by 
\begin{align}
    g^{(2)} (0)
    &= \lim_{t \rightarrow \infty} \frac{\braket{x_+(t) x_+(t) x_-(t) x_-(t)} }{ \braket{x_+(t) x_-(t)} ^2 }
    \,. 
\end{align}
 
We can then calculate the infinitesimal-drive correlation function $g^{(2)}(0)$ at zero temperature by numerically integrating the Lindblad equation, Eq.~\eqref{dot-rho_driven} as a function of $g$ and $\Gamma$. 
The result is shown in Fig.~\ref{fig:g2_num}b). 
Again, we find that the transition from anti-bunching to no bunching behavior at $g^{(2)}(0) = 0.5$ coincides with the limit of FSA validity.
The numerical result shows very good agreement with the leading order approximation plotted in Fig.~\eqref{fig:g2_ana} and given by Eq.~\eqref{g2_Kerr} for $g \ll \omega_\m$. 
However, even for small $g \ll \omega_\m$, the numerical result shows slight bunching behavior ( $g^{(2)}(0) >1$) with increasing dissipation rate which is absent in the approximated result.
In the case of $g^{(2)}(0) \sim 1$, the system is excited into higher levels $n>1$ and as shown in Fig.~\ref{fig:4th-order_comp} the lowest order approximation used for the analytical result gets increasingly worse with excitation number $n$ for any given coupling $g$. 
Due to this, even for small couplings $g \ll \omega_\m$, the actual $g^{(2)}(0)$ might differ from the approximated one in the region where $g^{(2)}(0)\sim 1$. 
To obtain the numerical value of $g^{(2)}(0)$ in its pseudo steady state, the state was numerically evolved with the time-dependent driven Hamiltonian. 
We used a simulation time of $t = 10/\gamma_\mathrm{eff}$, assuming that this is sufficient to reach the pseudo steady state and average the resulting expectation values over a period of the drive frequency. The convergence of the evolution at $t=10/\gamma_\mathrm{eff}$ was verified at several representative parameter points.

In Fig.~\ref{fig:g2_num}a), the phase diagram presents the enlarged parameter space that is accessible with the SVS Lindblad equation, while secular approximation is only valid deep inside the marked FSA-area. 
However, as it turns out even outside its regime of validity the full secular approximation yields a qualitatively similar result for $g^{(2)}(0)$ as the one presented here and derived from the valid SVS Lindblad equation. In the perturbative regime the result from both master equations is even exactly the same, as shown in the appendix, see App.~\ref{app:g2}.
As we will show in the next sections, this is partly due to the infinitesimal drive and zero temperature assumption. 

\section{Thermal oscillator spectrum}
\label{sec:Sxx}
\begin{figure*}[htbp] 
\includegraphics[width=   \linewidth]{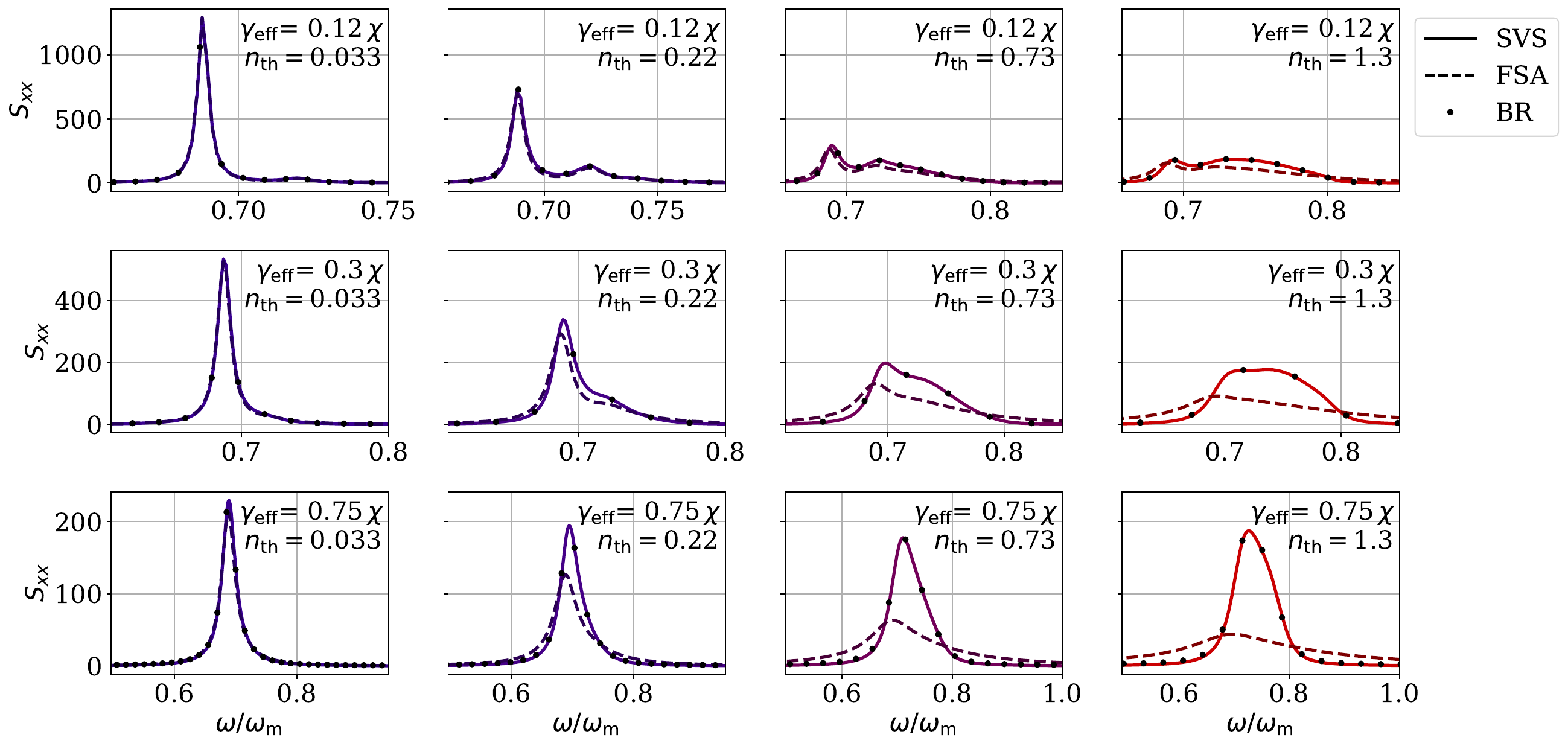} 
\caption{
Thermal spectrum of the oscillator displacement in the deep-strong coupling limit $g = 1.2\omega_\m$ and $\omega_\q = 10 \omega_\m$ obtained from different master equations and for different temperatures and decay rates of an ohmic bath. 
The anharmonicity is then given by $\chi = 0.032 \omega_\m$ and outside of the Kerr-regime defined by the difference of the two lowest transition frequencies $\omega_{21} - \omega_{10}$. 
The full secular approximation breaks down with increasing temperature and decay rate, while the SVS approximation is in excellent agreement with the results from the full Bloch-Redfield equation for all parameters.
\label{fig:thermal_Sxx_DSC}}
\end{figure*}
\begin{figure} 
\includegraphics[width=   \linewidth]{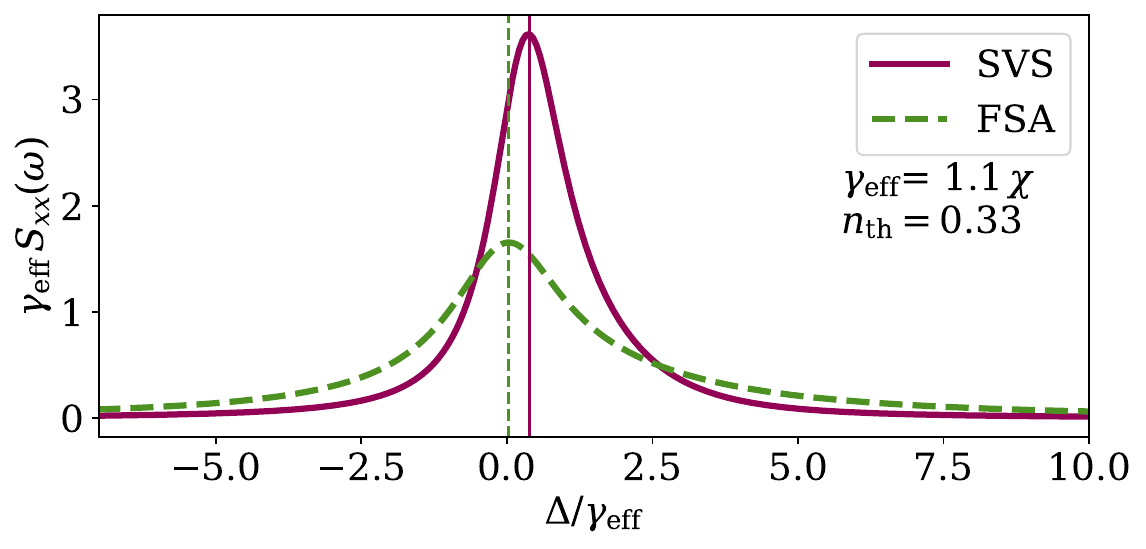} 
\caption{
Thermal spectrum $S_{xx}(\omega)$ of the oscillator displacement in the USC regime $g = 0.3\omega_\m$ and $\omega_\q = 10 \omega_\m$ obtained from FSA and SVS approach as a function of the detuning $\Delta = \omega - \omega_\m$. 
The discrepancy in the linewidth between FSA and SVS approximation observed in the deep strong coupling limit, see Fig.~\ref{fig:thermal_Sxx_DSC} persists in the ultrastrong coupling regime as well as a shift between the resonance peaks. The effective rate $\gamma_\mathrm{eff} = 1.1 \chi$, $\chi = 10^{-4}\omega_\m$ corresponds to a TLS zero-temperature rate of $\Gamma = 0.03 \omega_\m$. 
\label{fig:thermal_Sxx_PKR}
}
\end{figure}
While the two-phonon-correlation function has been measured previously~\cite{cohen_phonon_2015,hong_hanbury_2017}, it is not easily accessible experimentally. 
In contrast, the thermal spectrum of the oscillator $S_{xx}(\omega)$ is easily accessible and has been measured in a variety of set-ups~\cite{ schliesser_resolved-sideband_2008,teufel_sideband_2011,moser_nanotube_2014,de_bonis_ultrasensitive_2018,li_cavity_2021,huang_room-temperature_2024}. 
It is given by 
\begin{align}
    S_{xx}(\omega) 
    = \int \dif \tau e^{- \mi \omega \tau } \tr\left[ x(\tau) x(0) \rho_\mathrm{ss}\right]
    \,,
\end{align}
where $x = a + a^\dag$ and $\rho_\mathrm{ss}$ is the thermal steady state. 

Here, we study this observable beyond the limit of secular approximation and show that the conventional secular approximation 
overestimates the width of $S_{xx}(\omega)$
systematically for a finite anharmonicity and with increasing temperature. 
We provide a numerical comparison between the different approximations and an in depth analytical study of the origin of the observed discrepancy and the temperature dependence of $S_{xx}(\omega)$.

Additionally, we compare the numerical result obtained from the SVS Lindblad equation to the one obtained from using the full Born-Markov equation without any further approximations in the case of an ohmic bath, given by 
\begin{align}
    J(\omega) 
    &= \alpha \omega \quad\text{for } \omega < \omega_\mathrm{c}\, , \ \omega_\mathrm{c} \gg \omega_\q 
    \\
    S( \omega)  
    &= 
    \begin{cases}
        ( 1 + n_\mathrm{th}( \omega ) ) J(\omega) 
        \quad & \text{for } \omega >0 
        \\
        n_\mathrm{th}(| \omega |)  J(|\omega|) 
        \quad & \text{for } \omega <0 
    \end{cases}
    \,.
\end{align}  
In the SVS Lindblad equation~\eqref{SVS_sx} we then introduce a single rate $\Gamma = J(\omega_{01})$, where $\omega_{01} = E_1 - E_0$ is the lowest transition frequency of the QRM, while the full Born-Markov equation -- also referred to as Bloch-Redfield equation -- as well as the full secular approximation evaluate the power spectrum $S(\omega)$ at all occurring transition frequencies $\omega_{nm}$.

The result from numerical simulations is shown in Fig.~\ref{fig:thermal_Sxx_DSC} for different rates $\Gamma$ and increasing temperatures in the DSC regime $g = 1.2 \omega_\m$ and for $\omega_\q = 10 \omega_\m$. We observe that even for $\Gamma < \chi$, the FSA result deviates from the full Born-Markov result with increasing temperature while the SVS Lindblad equation remains in excellent agreement. 
In general, the full secular approximation predicts a broader width of $S_{xx}(\omega)$, which implies a larger effective decay of the correlation function 
$C(\tau) =  \tr\left[ x(\tau) x(0) \rho_\mathrm{ss}\right]$. 
This is consistent with the numerical observation of Ref.~\cite{settineri_dissipation_2018}; 

Here, we identify the underlying mechanism analytically and trace it to the transfer of coherence.
We can analyze the evolution of the coherences further in the perturbative Kerr regime, where the effect persists as shown in Fig.~\ref{fig:thermal_Sxx_PKR} for $g = 0.3 \omega_\m$. 
The correlation function evolves with the respective Lindblad as 
\begin{align}
    C(t)  
    &= 
    \tr \{ x e^{\mathcal{L}t} \cdot ( x \rho_\mathrm{ss} ) \} 
    \\
    &= 
    \tr \{ x \Bar{\rho}(t) \} 
    \,, 
\end{align}
where $e^{\mathcal{L}t}$ is a superoperator in Liouville space acting on $\rho$ defined as a vector in said space and we defined 
$\Bar{\rho}(t) = e^{\mathcal{L}t}\cdot( x \rho_\mathrm{ss}) $ which is not a physical density matrix but evolves with the Lindblad operator of the system. 
The density matrix $\Bar{\rho}(t)$ initially consists only of coherences, \textit{i.e.}~$\Bar{\rho}_{nn}(0) = 0$, since $\rho_\mathrm{ss} = \exp( - H/ kT ) / \tr[ \exp( - H/ kT) ]$ is purely diagonal in the Hamiltonians eigenspace, such that 
\begin{align}
    \rho_\mathrm{ss}
    &= \sum_n p_n \ket{n}\bra{n}
    \,,
    \\
    \Bar{\rho} ( 0 ) &= \sum_n p_n  ( \sqrt{n}\ket{n-1}\bra{n} + \sqrt{n+1}\ket{n+1}\bra{n} ) 
    \label{bro_0}
    \,,
\end{align} 
where $\ket{n}$ is an eigenstate of $H$.  
The difference between the FSA and SVS Lindblad equation in this limit is the occurrence of transfer of coherence, see Eqs.~\eqref{toc} and~\eqref{no_toc}. 
In the following, we assume a small finite temperature, such that $n_\mathrm{th}(\omega_{nm}) \ll 1$ and for simplicity, we additionally approximate $n_\mathrm{th}(\omega_{nm}) \approx n_\mathrm{th} (\omega_{01}) = n_\mathrm{th}$ as constant in the construction of the collapse operators, Eqs.~\eqref{A_dec_first}--\eqref{A_dec_last}. Then the coherences $\rho_{nm} = \braket{\Psi_n|\rho|\Psi_m}$ evolve 
as 
\begin{multline}
    \dot{  {\rho}} _{nm}
    = - \mi \omega_{nm} \rho_{nm}
    \\
    - \frac{\gamma_-}{2} ( n+ m )  \rho_{nm}
    - \frac{\gamma_+}{2} ( n+ m +2 )   \rho_{nm}
    \\
    + \delta_\mathrm{SVS} 
    \big[ 
    \gamma_- \sqrt{(n+1)(m+1)}  \rho_{n+1,m+1}
    \\ 
    +  \gamma_+ \sqrt{n m}  \rho_{n-1,m-1}
    \big] \,, 
    \label{dot_rho_toc}
\end{multline}
\newcommand{\nth}{n_\mathrm{th}}
where $\delta_\mathrm{SVS} = 0$ for FSA and 1 for SVS and with 
\begin{align}
    \gamma_- = ( 1 +\nth ) \gamma  \,,
    \quad
    \gamma_+ =\nth \gamma  \,.
    \label{gamma_pm}
\end{align}
From this we can conclude that the evolution of the coherences can be separated into subspaces which decouple and are defined by equidistant level spacing $\Delta_{nm} = n - m$. 
According to Eq.~\eqref{bro_0}, We are only interested in the evolution of the two subspaces given by $\Bar{\rho}_{n n+1}$ and $\Bar{\rho}_{n+1 n}$. 
Each subspace spanned by $v_n = \Bar{\rho}_{n n+1}$ or $v_n = \Bar{\rho}_{n+1 n }$, 
evolves with the same master equation. 
In Liouville space we can write the Lindblad equation of this subspace as a matrix equation
\begin{align}
    \dot{\vec{v}} = M \cdot \vec{v}
\end{align}
with matrix elements given
\begin{align}
    M_{nn} &=  -\bigg[ \mi \omega_{nm} + \frac{\gamma_-}{2} ( 2 n+ 1 )  
    + \frac{\gamma_+}{2} ( 2 n +3 ) \bigg] 
    \\
    M_{n n+1} &= 
     \delta_\mathrm{SVS}   
    \gamma_- \sqrt{(n+1)(n+2)} 
    \\ 
    M_{n n-1} &= 
     \delta_\mathrm{SVS} 
    \gamma_+ \sqrt{n (n+1) }  
    \,. 
\end{align}
The real part of the eigenvalues of this Liouville operator $M$ yields the decay rates which appear in the evolution of $\rho_{nm}(t)$. 

Within the full secular approximation, \textit{i.e.}~$\delta_\mathrm{SVS} = 0$, $M$ is a diagonal matrix with the real part of its diagonal elements yielding the rates as 
\begin{align}
    \gamma_n^\mathrm{FSA} 
     &= \frac{1}{2} \gamma  (2 n+1)+ 2 \nth  \gamma  ( n + 1)  
    \label{gamma_n_FSA}
    \,,
\end{align}
where we used the explicit expressions for $\gamma_{\pm}$, see Eq.~\eqref{gamma_pm}.
In this case, all $\rho_{nm}(t)$ decouple from each other and their evolution is given by 
\begin{align}
    \rho_{\substack{n+1 n \\ n n +1} } (t) = e^{- \gamma_n^\mathrm{FSA} t} e^{\pm \mi (\omega_m + n \chi) t} \rho_{\substack{n+1 n \\ n n +1} }(0) 
    \,.
\end{align}

While within the SVS approximation, \textit{i.e.}~$\delta_\mathrm{SVS} = 1$ the master equation of each  subspace is given by a non-Hermitian tridiagonal matrix in Liouville space, 
\begin{multline}
    M =  
    \\ 
    \begin{pmatrix} 
  \lambda_0  + \nth \lambda'_0 & (1+ \nth) A_0 & \cdots & 0 \\
\nth  A_0 &\lambda_1  + \nth \lambda'_1 &  \ddots & \vdots \\ 
  \vdots & \ddots &   \ddots & (1+\nth) A_{N-1} \\
0 & \cdots & \nth A_{N-1} &\lambda_N  + \nth \lambda'_N
\end{pmatrix}, 
\label{M_Liouville}
\end{multline}
with 
\begin{align}
    \lambda_n &= \pm \mi  (\tilde \omega_\m + n \chi )   
    - \frac{\gamma }{2} ( 2 n +1 )  
    \label{lambda_n}
    \\
    \lambda_n' &= - \frac{   \gamma }{2} (  n +1)  
    \\ 
    A_n &= \gamma \sqrt{( n+1)(n+2)} 
    \label{An}
    \,,
\end{align}
such that $\gamma_n^\mathrm{FSA} = - \re[ \lambda_n + \nth \lambda_n'] $ and we used that $\omega_{nm} = \pm (\tilde \omega_\m + n \chi ) $ for $(n,m) = (n+1 ,n)$ and $(n,m) = (n ,n +1)$, respectively. 
The eigenvalues of this matrix can then be calculated approximately for small temperatures.
 
\begin{figure}[htbp]
\begin{overpic}
   [width=   \linewidth]{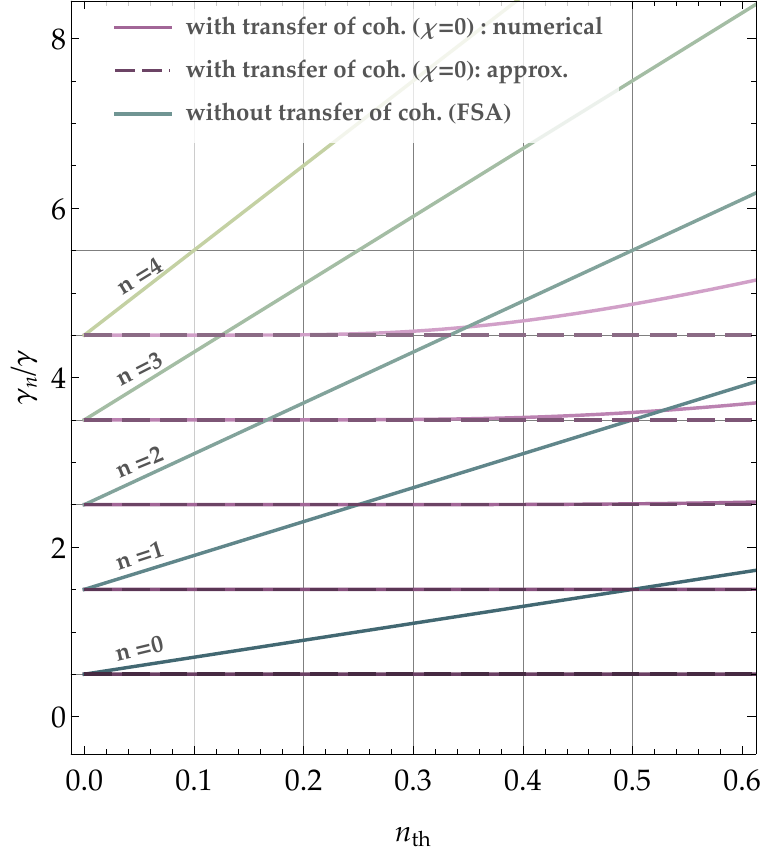}
\put(0,98){a)} 
\end{overpic}
\begin{overpic}
   [width= 0.8  \linewidth]{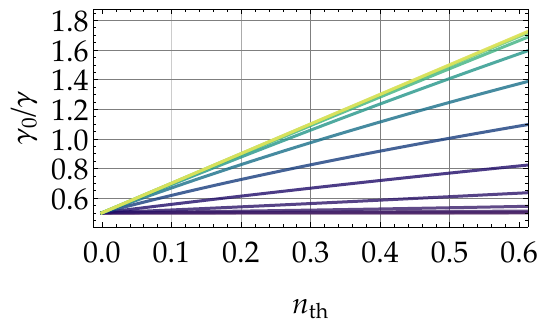}
\put(0,60){b)} 
\end{overpic}
\includegraphics[width= 0.18  \linewidth]{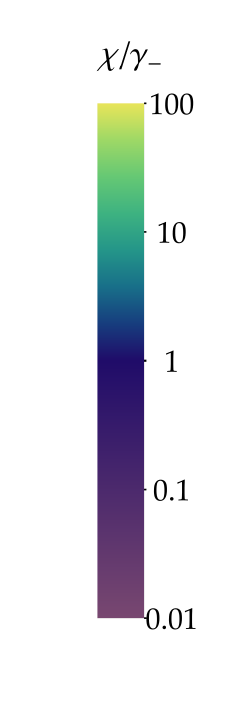}
\caption{
a) Real part $\gamma_n$ of the eigenvalues of the Lindblad operator $M$ in the limit of FSA, \textit{i.e.}~$\chi/\gamma_- \rightarrow \infty$ and via the SVS approximation in the limit of $\chi/\gamma_- \rightarrow 0$. The number $n$ in the figure labels the eigenvalues of the Lindblad master equation. 
b) The lowest rate $\gamma_0$ for different values of $\chi/\gamma_-$ as a function of $\nth$. 
\label{fig:compare_dec_rates}}
\end{figure}

For small temperatures $k_\mathrm{B} T \ll \tilde \omega_\m$, we can treat $\nth$ as a small perturbation and define $M = M_0 + \nth M_1$
and calculate the complex eigenvalues $\Lambda_n$ of $M$ perturbatively in lowest order of $\nth$.
The matrix $M_0 = M\big|_{\nth = 0}$ is a triangular matrix and its eigenvalues $ \Lambda^{(0)}_n$ are directly given by the diagonal entries,
\begin{align}
    \Lambda^{(0)}_n = \lambda_n \,.
\end{align}
The $n$th right eigenvector $\vec{r} ^{(n)}=  (r_0^{(n)} ,\dots )^T$ of $M_0$ is given by the recursive formula
\begin{align}
    r^{(n)}_{k>n} &= 0 
    \,, \quad
    r^{(n)}_{n}  = 1 
    \,, \quad
    r^{(n)}_{k < n }  = - r^{(n)}_{k+1} \frac{A_{k}}{\lambda_{k} - \lambda_{n}  }
    \,. 
    \label{righties}
\end{align}
Similarly, the left eigenvectors can be constructed by 
\begin{align}
    l^{(n)}_{k<n} &= 0 
    \,, \quad
    l^{(n)}_{n}  = 1 
    \,, \quad
    l^{(n)}_{k > n }  = - l^{(n)}_{k -1 }\frac{A_{k-1}}{\lambda_{k} - \lambda_{n}  }
    \label{lefties}
\end{align}
such that $(l^{(n)} )^T \cdot r^{(m)} = \delta_{nm}$. 
The first order correction to the eigenvalues $\Lambda_n = \Lambda_n^{(0)} + \Lambda_n^{(1)}$ can then be calculated by~\cite{kato_perturbation_1995}
\begin{align}
    \Lambda_n^{(1)} = \nth  \left( 
    \left( \vec{l}^{(n)} \right)^T  \cdot M_1 \cdot \vec{r}^{(n)}  \right) 
    \,.
    \label{Lambda_corr_1}
\end{align}
The matrix $M_1$ is tridiagonal with $[ M_1 ]_{kk} \neq 0$ , $[M_1]_{k+1,k} \neq 0$ and $[M_1]_{k,k+1} \neq 0$,  
which by using the definition of the left and right eigenvectors leaves only three terms in Eq.~\eqref{Lambda_corr_1}, 
\begin{align}
    \Lambda_n^{(1)} =
    \nth   \left(  \lambda'_n 
    - \frac{A_{n}^2}{\lambda_{n+1} - \lambda_{n}}  
    - 
    \frac{A_{n-1}^2}{\lambda_{n-1} - \lambda_{n}} 
    \right) 
    \,.
\end{align}
Using the definitions Eqs.~\eqref{lambda_n}--\eqref{An}, the first order correction then reads 
\begin{multline}
    \Lambda_n^{(1)} 
     =   
   \pm \mi 2 \nth (n+1)  \frac{\chi }{1 +( \chi/\gamma)^2} 
    \\
    - 2 \nth \gamma  (n+1)  \frac{\chi ^2}{\gamma ^2+\chi ^2}
\end{multline}
and hence, we get the corrected rates 
$\gamma_n = -\re[\lambda_n + \Lambda_n^{(1)}]$ in first order of $  \nth $ as 
\begin{align}
    \gamma_n =    
     \frac{\gamma}{2} ( 2 n + 1) 
     + 2 \nth  \gamma  (n+1) \frac{\chi ^2}{\gamma ^2+\chi ^2} \le \gamma_n^\mathrm{FSA}\,.
\end{align}
The rate $\gamma_n$ is hence smaller than the one obtained from secular approximation, see Eq.~\eqref{gamma_n_FSA} for finite temperature and anharmonicity. 
It converges to $\gamma_n^\mathrm{FSA}$ in the two different limits of $\chi/\gamma \rightarrow \infty$ and $\nth \rightarrow 0$.
While for vanishing anharmonicity $\chi/ \gamma \rightarrow 0$, i.e.~in the purely harmonic case, the rates become constant in temperature, 
\begin{align}
    \lim_{\chi/\gamma \rightarrow 0 }
    \gamma_n =  \frac{\gamma}{2} ( 2 n +1 ) 
    \,.
\end{align}
This implies that, at finite temperature, the secular approximation systematically overestimates the decay of coherences and, consequently, of the correlation function $C(t)$. Moreover, in the limit $\chi/\gamma_- \rightarrow 0$, the rates $\gamma_n$ become independent of temperature for low temperatures, $n_\mathrm{th}\ll 1$, in contrast to the FSA prediction, see Fig.~\ref{fig:compare_dec_rates}. 

Additionally, we find a temperature-dependent shift $\Delta_n = \im \Lambda^{(1)}_n$ of the Lindblad eigen-frequencies that is absent in the secular approximation, 
\begin{align}
    \Delta_n =\pm 2 \nth (n+1)  \frac{\chi }{1 +( \chi/\gamma)^2} \,.
\end{align}
These shifts of the resonances of the Lindblad eigen-modes imply a resulting shift of the peaks in the correlation function. And consistently,a shift between the FSA and SVS results is also visible in the numerical comparison, see Fig.~\ref{fig:thermal_Sxx_PKR}.
This shift vanishes both in the harmonic-oscillator limit, $\chi/\gamma \rightarrow 0$, and in the FSA limit, $\chi/\gamma \rightarrow\infty$, and is maximal for $\chi=\gamma$. 

The analysis presented above is based on the eigenvalues of the Lindblad operator $M$. The real parts of these eigenvalues determine the decay rates that enter the explicit time evolution of the density-matrix elements $\rho_{nm}(t)$. However, they do not generally correspond to the decay rates of individual matrix elements, since each $\rho_{nm}(t)$ is a superposition of several eigenmodes of $M$, each decaying with its respective rate $\gamma_n$.
The resulting evolution of $\rho_{nm}(t)$ is therefore more involved and, due to the transfer of coherence between different matrix elements, is not state-independent. Consequently, the rates discussed above and their temperature dependence cannot be directly interpreted as the decay rates of individual $\rho_{nm}(t)$. We now discuss some consequences of this distinction.

While at zero temperature, the eigenvalues of $M = M_0$ are given by the diagonal entries in either case, with and without the secular approximation (\emph{i.e.}~with $\delta_\mathrm{SVS} =1$ or $0$), the corresponding eigenmodes differ in the two cases, and therefore the resulting evolution of $\rho_{nm}(t)$ is not identical. 
In case of steady state dynamics, this difference in the evolution cannot be observed as there is no population of higher modes and the transfer of coherence has no effect on the dynamics.

A further point concerns the temperature dependence of the actual decay of the coherences $\rho_{nm}(t)$ in the presence of transfer of coherence.
The temperature independence of the rates $\gamma_n$ discussed above may appear counter-intuitive, since one would generally expect the coherences decay to increase with temperature.
This apparent contradiction is resolved by recalling that the $\gamma_n$ are the real parts of the eigenvalues of $M$, rather than the decay rates of individual coherences. 
Each $\rho_{nm}(t)$ contains several eigenmodes, whose relative weights depend on temperature, \emph{i.e.}~they evolve as 
\begin{align}
    \rho_{n n+1}(t) = \sum_k e^{- \gamma_k t} e^{- \mi \im \Lambda_n  t}  C_{kn}(\nth) \, 
\end{align}
and the coefficients $C_{kn}(\nth)$ depend on the initial state $\rho(0)$ as well as the temperature. 
Hence, the temperature dependence of the decay of $\rho_{nm}(t)$ is not determined by the $\gamma_n$ alone.

In summary, we find that the conventional secular approximation systematically overestimates the decay of coherences $\rho_{nm}(t)$ in the presence of finite anharmonicity and temperature, resulting in an overestimation of the linewidth of $S_{xx}(\omega)$. 
In this regime, an appropriate Lindblad equation, such as the SVS Lindblad equation, is therefore required. 
By mapping the full quantum Rabi model onto a Kerr oscillator in the appropriate coupling regime, we could analyze the origin of this observation.
We trace the overestimation back to the transfer of coherence, which is neglected by the FSA. Importantly, this coherence transfer counteracts the temperature-induced increase of the decay rates.

\section{Driven system with weak but finite drive}
\label{sec:drive}
\begin{figure*}[t] 
\includegraphics[width= 0.8  \linewidth]{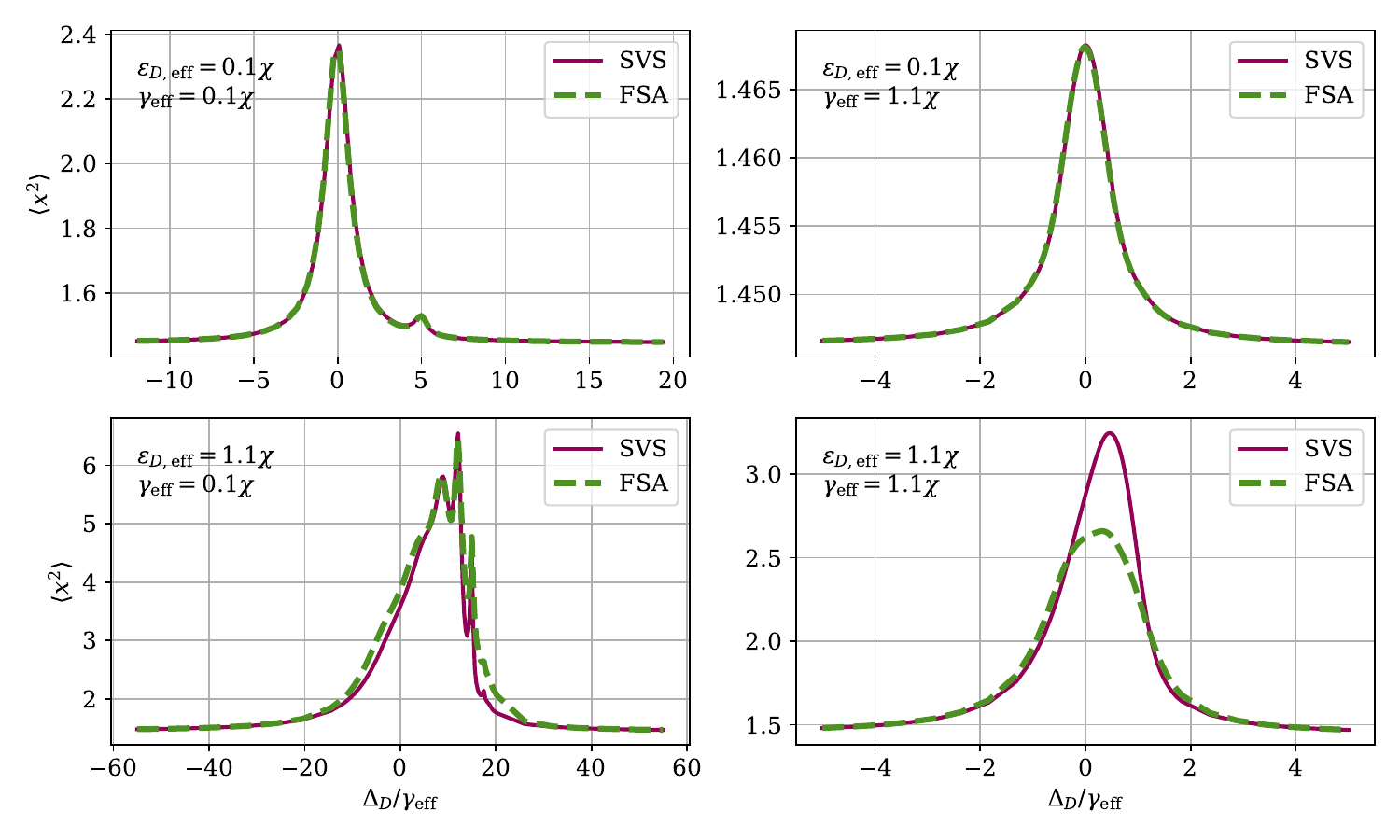} 
\caption{
 Expectation value of $x^2 = ( a + a^\dag)^2$ for $g = 1.2 \, \omega_\m$ and $\omega_\q = 10 \omega_\m$ after a time of $  \gamma_\mathrm{eff} t = 10$ as a function of drive frequency detuning $\Delta_D = \omega_D - \omega_{10}$, where $\omega_{10}$ is the transition frequency of the lowest transition, for different values of drive amplitude $\varepsilon_D$ and decay rate $\Gamma$. The effective values correspond to the following physical values, $\tilde \varepsilon_D = 0.1 \chi ( 1.1 \chi)$ corresponds to $\varepsilon_D = 0.012 \omega_\m ( 0.13 \omega_\m )$ and $\gamma_\mathrm{eff}  = 0.1 \chi ( 1.1 \chi)$ corresponds to $\Gamma = 0.04 \omega_\m ( 0.5 \omega_\m )$. Note that $x$ is defined dimensionless and is given in terms of the zero-point motion of the uncoupled mechanical oscillator $x_\mathrm{zpm}^{(0)} = \sqrt{1/(2 m \omega_\m)}$. 
\label{fig:x_squ_wD}}
\end{figure*}
Another example of an experimentally accessible observable is the mean-square displacement $\braket{x^2}$ in the presence of a drive. 
We consider the drive again applied via the quantum dot, see Eq.~\eqref{H_D}.
The drive acts as a perturbation to the quantum Rabi Hamiltonian. 
As its strength increases, the construction of the dissipator must be reconsidered as the perturbation modifies the system eigenstates and transition frequencies on which the dissipator is based.
In the eigenbasis of the QRM Hamiltonian $\ket{\Psi_n}$, the effective drive amplitude is given by 
\begin{align}
    \tilde \varepsilon_D = | \braket{\Psi_1|\sx | \Psi_0 }| \varepsilon_D 
\end{align}
which in the perturbative Kerr limit yields Eq.~\eqref{eps_D_eff}. 
In the secular approximated master equation, the decomposition of the collapse operators resolves transitions on the scale of the anharmonicity $\chi$, see App.~\ref{app:FSA}. Consequently, the drive strength must satisfy $\tilde \varepsilon_D \ll \chi$ in order for the perturbative treatment underlying the dissipator construction to remain valid.
In contrast, the SVS dissipators are constructed by separating only the positive- and negative-frequency components of the system operators, corresponding to the much larger scale $\omega_{nm} \sim \omega_\mathrm{m}$. 
Their validity therefore requires only $\tilde \varepsilon_D \ll \omega_\m$, with $\omega_\m \gg \chi$.

We may show this explicitly in the perturbative Kerr regime, where we can cast the problem into a time-independent one via the rotating wave approximation, see Eq.~\eqref{H_rwa}.
The collapse operators $c_n$ of the secular approximation are given by Eq.~\eqref{c_n} and result from a spectral decomposition of the system-operator $\sx$ in the system-environment coupling $H_\mathrm{SE}$, such that $\sx = \sum_n ( c_n + c_n^\dag)$. 
The secular approximation is then based on effectively decoupling highly oscillating terms in $\dot \rho$ that appear in terms that have products with two system operators in the interaction picture, e.g.~terms that are proportional to $
    \sx(t) \rho  \sx(t)$, 
where $\sx(t)$ evolves with the quantum Rabi Hamiltonian, see App.~\ref{app:FSA},  include terms such as
\begin{align} 
\sum_{ij} e^{\mi (\omega_i - \omega_j) t } c_i^\dag \rho   c_j
    \underset{\mathrm{FSA}}{\rightarrow} \sum_i c_i^\dag \rho  c_i 
    \,.
\end{align} 
As a result, each $c_n$ oscillates necessarily with a unique frequency $\omega_n$.  
However, in the presence of a drive and in the rotating frame the FSA collapse operators evolve in interaction picture as
\begin{align}
    c_n(t)  &= e^{-\mi \omega_D t } U^\dag c_n U 
    \\
    U(t) &= \exp\left( - \mi  H_\mathrm{rwa}  t \right)
    \,. 
\end{align}
and in lowest order of $\varepsilon_D$ we can express them in the eigen state basis $\ket{\tilde n }$ of $ H_\mathrm{rwa}$ as 
\begin{multline}
    c_n 
    \approx  
    \sqrt{n} 
    {\ket{ \tilde n} } 
    {\bra{ \tilde n +1} } 
    -   
    \frac{\varepsilon_D}{2 \chi } 
    \bigg[ 
    \frac{\sqrt{n+1}}{   n } \ket{\tilde n }\bra{\tilde n }
    \\
    -
    \frac{\sqrt{n+1}}{  n   } \ket{\tilde n +1}\bra{\tilde n + 1 }
    + 
    \frac{\sqrt{n }}{  (n-1)  } \ket{\tilde n -1 }\bra{\tilde n }
    \\
    -
    \frac{\sqrt{n+2}}{  (n+1)   }  \ket{\tilde n }\bra{\tilde n +2  }
    \bigg] \,,
\end{multline} 
where we assumed zero detuning $\Delta = 0$ for simplicity.
From this expression it is immediately apparent that, under the time evolution generated by $U(t)$, the operator $c_n$ acquires several frequency components, 
$\omega \in \{ \omega_{n n+1}, \omega_{n-1,n}, \omega_{n,n+2}, 0\}$.
 Consequently, $c_n$ contains components oscillating at the same frequencies as, for example, $c_{n+1}$, while all collapse operators acquire a zero-frequency component. The set of operators $\{ c_n\}$ therefore no longer constitutes the spectral decomposition of the system coupling operator, in contradiction to the assumption underlying the secular approximation. Only in the limit of a negligible perturbation, $\varepsilon_D / \chi \rightarrow 0$, is the original spectral decomposition used to construct the FSA collapse operators recovered.

 On the other hand,
 the collapse operators $b$ of the SVS approach result from decomposing the system-operator $\sx$ simply into its negative and positive frequency components. Hence, at zero temperature $b$ stays a valid choice for the collapse operator as long as it oscillates with negative frequency in the system-environment interaction picture. 
 In the rotating frame and in this interaction picture, the time-evolution is again given by
 \begin{align}
     b(t) = e^{- \mi \omega_D t} U^\dag a U \,,
 \end{align}
which remains oscillating with exclusively negative frequencies, provided that $U = \exp( - \mi H_\mathrm{rwa} t )$ does not generate oscillation frequencies of the order of $\omega_D$ or larger, \textit{i.e.}~for $\varepsilon_D \ll \omega_D$.

Outside of this regime, we may not cast the driven system into a time-independent problem. Instead, we have to rely on numerical simulations. 
In Fig.~\ref{fig:x_squ_wD}, we compare the results obtained by numerically integrating the SVS Lindblad master equation with the additional drive of Eq.~\eqref{H_D} to those obtained using the FSA dissipator. All simulations are performed at zero temperature for different decay rates $\Gamma$ and drive amplitudes $\varepsilon_D$.
For weak drives, $\tilde \varepsilon_D \ll \chi$, the FSA and SVS predictions are in excellent agreement, even when the effective decay rate lies outside the nominal secular approximation regime, $\gamma_\mathrm{eff} \gtrsim \chi$. In this case, the weak drive and the absence of thermal excitations suppress the mechanisms responsible for the breakdown of the full secular approximation.
As the drive strength increases, the FSA and SVS results begin to deviate from each other, even for $\gamma_\mathrm{eff} \ll \chi$, consistent with the requirement that the drive remain perturbative on the scale of the anharmonicity. When both the effective decay rate and the effective drive strength exceed the FSA validity regime, the discrepancy becomes pronounced.
 
\begin{figure*}
\begin{overpic}
   [width=  0.45 \linewidth]{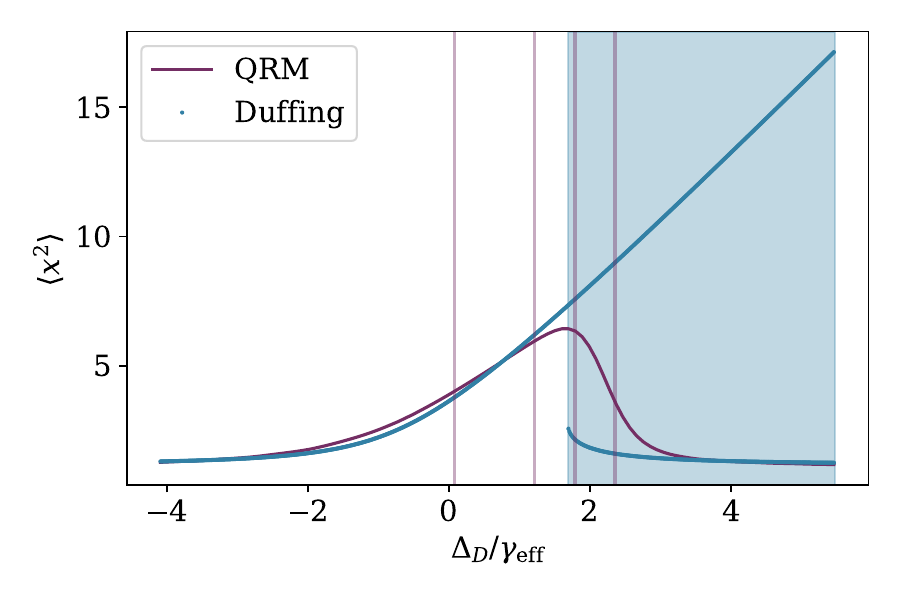} 
\put(0,70){a)} 
\end{overpic}
\begin{overpic}
   [width=   0.45 \linewidth]{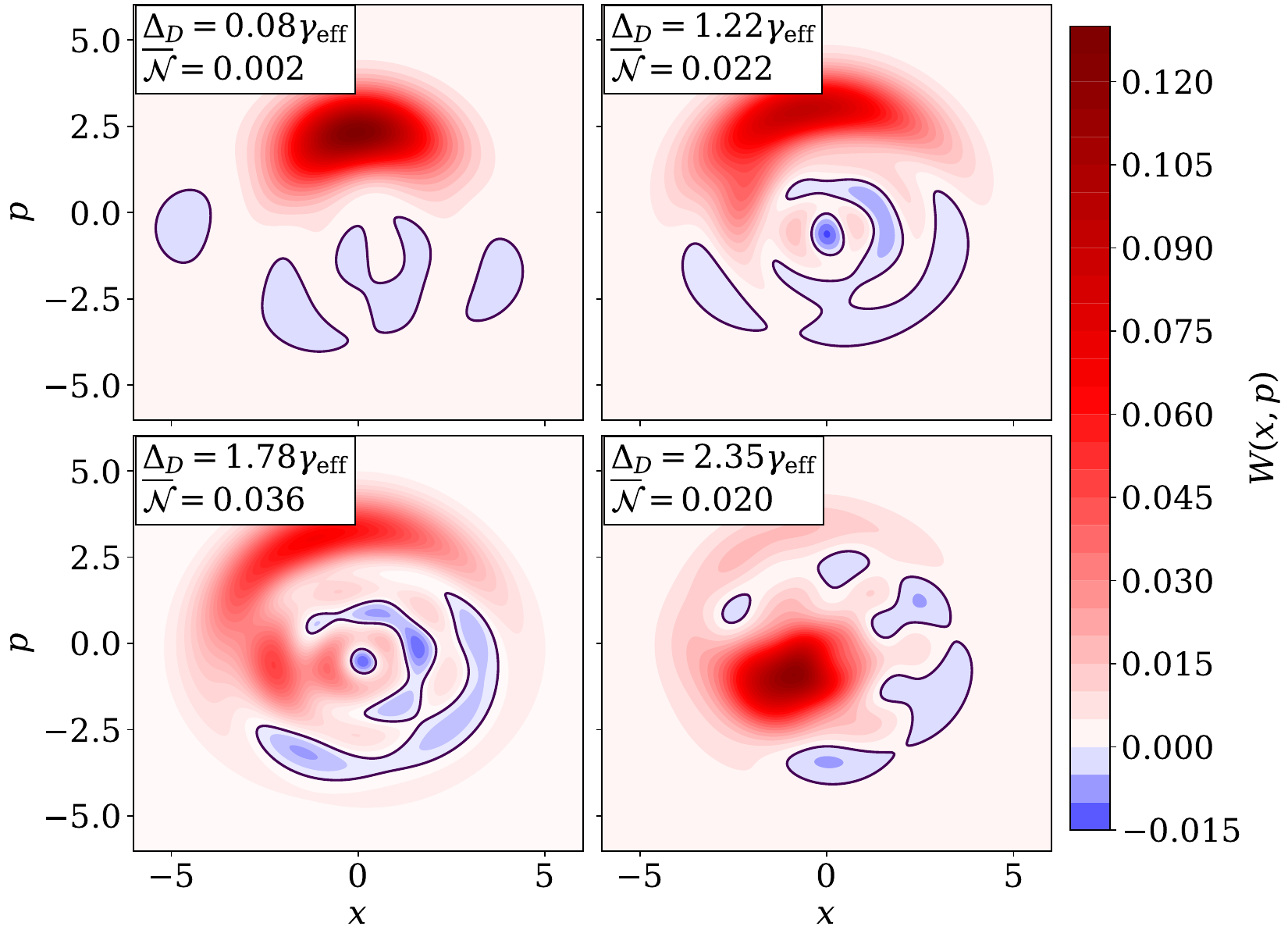}
   \put(0,70){b)} 
\end{overpic}
\caption{a) Expectation value $\braket{x^2}$ as a function of the drive frequency detuning $\Delta_D = \omega_D - \omega_{10}$ in the perturbative Kerr regime $g=0.5 \omega_\m$, $\omega_\q = 10 \omega_\m$, yielding $\chi = 8\times 10^{-4} \omega_\m$ and at a driving amplitude of $\varepsilon_D = 1.5\times10^{-3} \omega_\m$, which maps to an effective drive amplitude of $\tilde \varepsilon_D =  3 \chi$ compared to the bistable amplitude squared $A^2$ of a classical Duffing oscillator, where the blue area marks the bistable region. The classical analogue neglects the zero-point fluctuations $\braket{x_\mathrm{zpm}}^2 = \omega_\m / \tilde \omega_\m$, which we added to the Duffing result manually. Note, that $x = a + a^\dag$ is defined dimensionless, \textit{i.e.}~normalized to the uncoupled zero-point motion $x^{(0)}_\mathrm{zpm} = \sqrt{ 1/( 2m \omega_\m)}$ of the mechanical oscillator. 
b) The Wigner function $W(x,p,t)$ of the driven state for different driving frequencies and $t = 10 \gamma_\mathrm{eff}$ and their respective averaged negativity $\overline{\mathcal{N}} = \int_t^{t+ T_D} \dif \tau \mathcal{N}(\tau) / T_D$ with $T_D = 2 \pi / \omega_D$. The black contour marks $W =0$, where the Wigner function becomes negative. 
\label{fig:Duffing}}
\end{figure*}

In contrast, the SVS approximation remains valid in this regime, and the resulting response resembles that of a classical anharmonic (Duffing) oscillator.
We can compare the result from the quantum simulation to the classical Duffing response to a drive.   
To this end, we use the Born-Oppenheimer approximation, Eq.~\eqref{HBO}, where we add the drive Eq.~\eqref{H_D} into the Hamiltonian, such that the potential reads
\begin{align}
    V(x) = \sqrt{\omega_\q^2  + ( 2 \varepsilon_D \cos( \omega_D t ) + 2 g x)^2}
    \,. 
\end{align}
Expanding $V(x)$ then for $g \ll \omega_\q$ and only keeping the lowest order in the drive amplitude, one obtains the driven Hamiltonian  
\begin{multline}
    H_\mathrm{BO, D}
    =\frac{\wm}{4} ( x^2 + p^2) + \frac{ \omega_\q}{2} \sz + \frac{g^2 }{\omega_\q} x^2 \sz 
    \\
    - \frac{ g^4}{\omega_\q^3} x^4 \sz - \frac{g}{ \omega_\q} \varepsilon_D x 
    \,,
\end{multline}
which for the lower band, $\braket{\sz} = -1$ corresponds to a classical Duffing oscillator with the equation of motion, 
\begin{align}
    f \cos( \omega_\D t ) &=  \ddot x +  \gamma_\mathrm{eff}  \dot x + \tilde \omega_\m^2  x + \alpha x^3 
    \,,
    \\
    f &= \frac{2 \omega_\m g}{\omega_\q} \varepsilon_D \,,
    \\
    \tilde \omega_\m^2 &=  \omega_\m^2 - \frac{4 g^2 }{\omega_\q} \wm \,,
    \\ 
    \alpha &= \frac{8 \wm g^4}{\omega_\q^3} 
    \,.
\end{align}
Applying the harmonic balance approximation and retaining only the fundamental harmonic, we use the ansatz ${x(t) = A \cos( \omega_D t + \phi )}$ and solve for the amplitude $A$, we then obtain the well known Duffing amplitude equation~\cite{nayfeh_nonlinear_2008,dykman_fluctuating_2012}
\begin{align}
    f^2 = A^2 \left[ \left(  \tilde \omega_\m^2 - \omega_D^2 + \frac{3 \alpha}{4} A^2 \right)^2 + \left( \gamma_\mathrm{eff} \omega_D \right)^2\right]
    \,,
\end{align}
which we numerically solve for $A^2$ to obtain the Duffing amplitude comparison in Fig.~\ref{fig:Duffing}. 
The shape of the response is in good agreement to the Duffing amplitude prediction, where the bistable region is expected to show an expectation value for the amplitude which is averaged between the two stable amplitudes. 

Despite this classical-looking behavior of the observable, the steady state itself remains genuinely quantum, as evidenced by the negativity of its Wigner function, see Fig.~\ref{fig:Duffing}b). 
The Wigner function shows negative values with overall negativities 
\begin{align}
    \mathcal{N}(t) = \frac{1}{2} \left( \int \dif x \dif p \left|  W(x,p,t)\right|  -1 \right)
\end{align}
averaged over one drive period $T_D = 2 \pi / \omega_D$ of the order of $10^{-2}$.

We have shown that the SVS Lindblad dissipator is significantly more robust against the application of an additional perturbation, such as a coherent drive, than the master equation obtained within the conventional secular approximation. The latter ceases to provide a valid description of dissipation once the drive amplitude exceeds the system's anharmonicity. Such driving strengths are, however, particularly relevant in experiments operating with limited detector sensitivity or with reduced anharmonicity. We further demonstrated that the observable $\braket{x^2}$ as a function of driving frequency loses its quantum signature when the dissipation rate exceeds the anharmonicity, even though the underlying quantum state remains non-classical.

\section{Introducing detuning of the double quantum dot}
\begin{figure}[htbp] 
\includegraphics[width=  \linewidth]{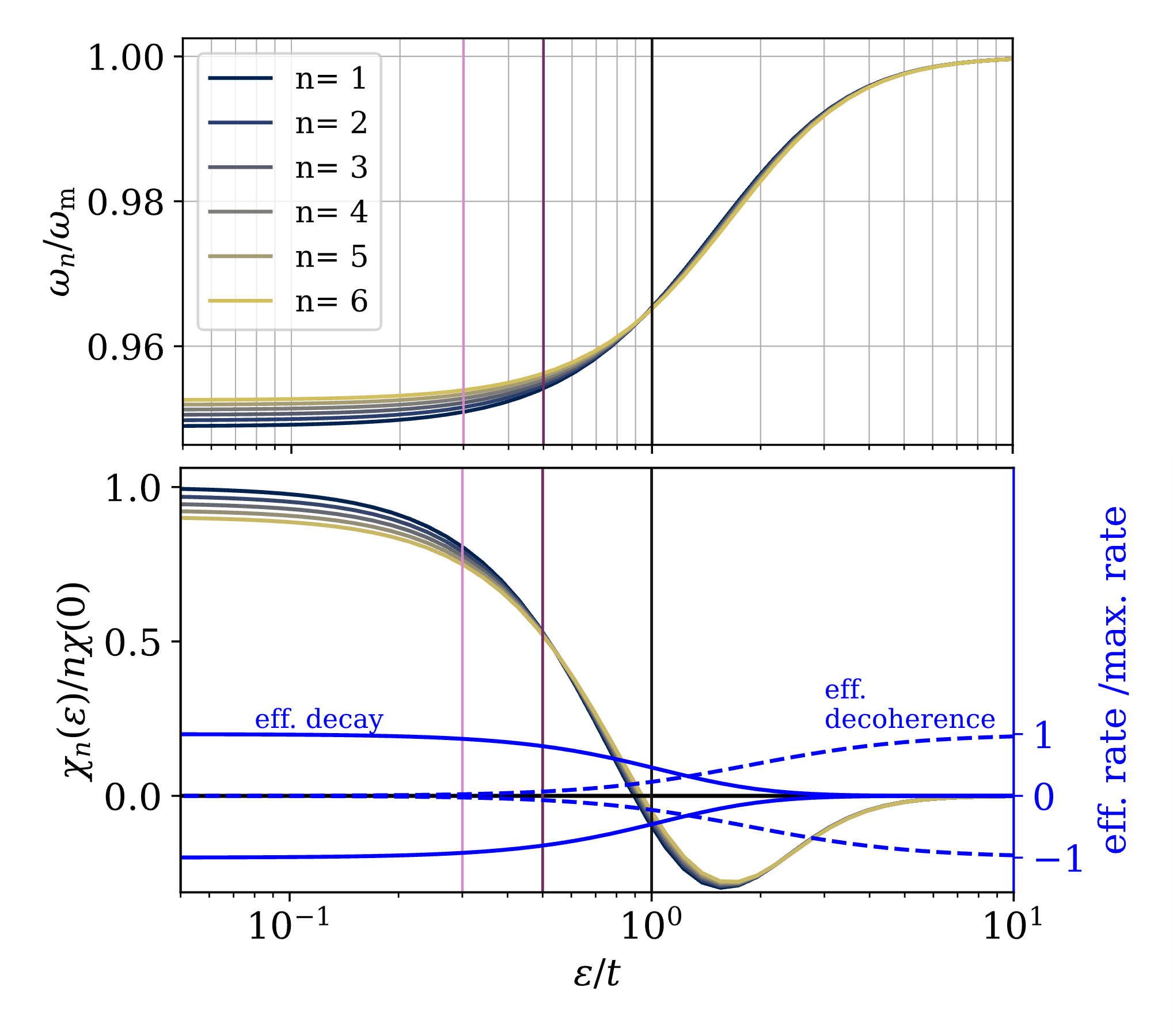} 
\caption{
The first 5 transition frequencies $\omega_{n} = E_n - E_{n-1}$ of the biased QRM, Eq~\eqref{QRM_eps} as a function of the double-dot detuning $\varepsilon$ and 
the related anharmonicity $\chi_n(\varepsilon) = \omega_{n-1,n} - \omega_{10}$ normalized to the zero-detuning Kerr-oscillator value $n \chi(0)$. 
We chose $2t = 10 \omega_\m$ and $g = 0.5 \omega_\m$. 
In blue solid and dashed line, the effective decay rate $\gamma_\mathrm{eff}$ and the effective dephasing rate $\gamma^\mathrm{eff}_\phi$, respectively are plotted normalized to their respective maximum value. 
The chosen values of $\varepsilon$ for Fig.~\ref{fig:x_squ_eps} are marked by vertical lines. 
The anharmonicity defined in this way differs slightly from that obtained from the quartic coefficient in Eq.~\eqref{chi_eps}, due to corrections arising from the cubic term at finite $\varepsilon$. As a result, the point at which the anharmonicity vanishes is shifted away from $\varepsilon=t$.
\label{fig:spec_finite_eps}}
\end{figure}
In the considered set-up of a nanomechanical oscillator, in form of a carbon nanotube, coupled to a double quantum dot, the voltage gates can be tuned to introduce a bias between the left and right potential well of the double dot. This detuning $\varepsilon$ introduces a bias to the QRM considered so far, 
\begin{align}
    H = t \sz + \frac{\varepsilon }{2}\sx + \omega_\m a^\dag a + g \sx ( a + a^\dag) 
    \label{QRM_eps}
    \,,
\end{align}
where $t$ is the tunneling strength, see Fig.~\ref{fig:set-up} and the bare TLS energy splitting is given by $\omega_\q = \sqrt{(2t)^2 + \varepsilon^2}$. 
The simplest way to understand the influence of the detuning is to consider the Born-Oppenheimer-approximated Hamiltonian, see Eq.~\eqref{HBO}, 
\begin{align}
    H_\mathrm{BO} 
    = \frac{\omega_\m}{4} ( x^2 + p^2) + \frac{V_\varepsilon(x)}{2} \sz
\end{align}
with $V_\varepsilon(x) = \sqrt{(2t)^2 + ( \varepsilon + 2 g x)^2}$. 
For $g \ll 2t$, we can expand the potential as 
\begin{multline}
    V_\varepsilon (x) =  \omega _q  
    +   \varepsilon \frac{2  g }{\omega _q} x
    + \frac{8 g^2 t^2}{\omega _q^3} x^2
    \\
    -  \varepsilon   \frac{16 g^3 t^2}{\omega _q^5}  x^3
    -  
    \left(t^2- \varepsilon ^2\right) \frac{32 g^4 t^2 }{\omega _q^7} x^4
    \,.
\end{multline}
From this we find that the detuning introduces asymmetry into the potential, \textit{i.e.}~an effective displacement term linear in $x$ and a cubic non-linearity and finally the detuning tunes the anharmonicity as  
\begin{align}
\frac{\chi_\varepsilon}{6} = \left(t^2- \varepsilon ^2\right) \frac{32 g^4 t^2 }{\omega _q^7}
\label{chi_eps}\,,
\end{align}
where the conversion factor between the quartic coefficient and the anharmonicity comes from normal ordering $x^4$.
The anharmonicity vanishes independently of coupling $g$ for $\varepsilon = t$ and recovers the zero-detuning anharmonicity,~Eq.~\eqref{chi} in the limit of $\omega_\m \ll \omega_\q = 2t$ for $\varepsilon = 0$. 
Note that, at finite $\varepsilon$, the cubic term gives rise to corrections to the anharmonicity obtained from the quartic coefficient. The exact anharmonicity, defined as the difference between the first two transition frequencies, therefore differs slightly from Eq.~\eqref{chi_eps}.

Additionally, the system operator $\sx$ of the system-environment coupling $H_\mathrm{SE}$ gains contributions diagonal in the eigenbasis $\ket{\Psi_n}$ of the system Hamiltonian, such that 
\begin{align}
    | \braket{\Psi_n | \sx | \Psi_n} |^2 > 0
\end{align}
for $\varepsilon \neq 0$. 
Following the SVS collapse operator construction, see Sec.~\ref{sec:new_diss}, this leads to a finite decoherence channel with collapse operator
\begin{align}
    X_0 = \sum_{n}\braket{\Psi_n | \sx | \Psi_n} 
    \ket{\Psi_n} \bra{\Psi_n}
    \,,
\end{align}  
and an effective decoherence rate of 
\begin{align}
    \gamma_\phi^\mathrm{eff}(\varepsilon) 
    = 2 |\braket{\Psi_0 | \sx | \Psi_0} |^2 S(0)
    \,.
\end{align}
The effective decoherence rate vanishes for $\varepsilon/t \rightarrow 0$ and converges to its maximum value $\Gamma_\phi = 2 S(0)$ for $\varepsilon/t \rightarrow \infty$. 
The effective decay rate $\gamma_\mathrm{eff}(\varepsilon) = |\braket{\Psi_1| \sx | \Psi_0}|^2 \Gamma$ shows an opposite behavior to the decoherence rate, where for $\varepsilon \rightarrow 0$ it takes its maximum value and vanishes with $\varepsilon \rightarrow \infty$. 
The behavior of these rates as a function of detuning $\varepsilon$ is shown in Fig.~\ref{fig:spec_finite_eps}. 

Hence, even in the regime $\gamma_\mathrm{eff}(\varepsilon=0)\ll\chi$, where the full secular approximation is applicable at zero detuning, increasing the double-quantum-dot detuning $\varepsilon$ continuously reduces the induced anharmonicity $\chi(\varepsilon)$ and can drive the system into a regime where $\chi$ becomes arbitrarily small. The full secular approximation then ceases to be valid, whereas the SVS Lindblad master equation remains applicable throughout the entire detuning range.

\begin{figure}[htbp] 
\includegraphics[width=  \linewidth]{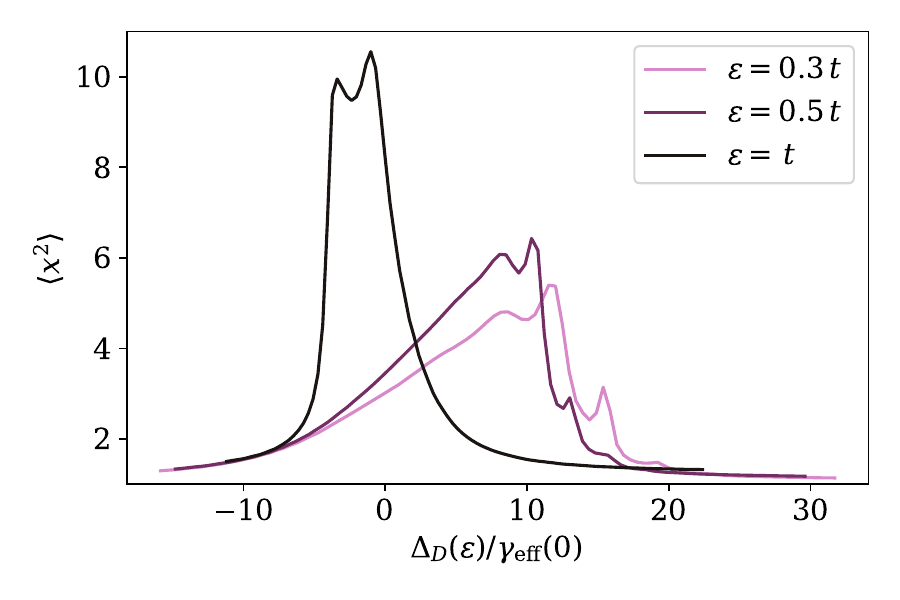} 
\caption{
Expectation value $\braket{x^2}$ for finite quantum dot detuning $\varepsilon$ as a function of drive detuning $\Delta (\varepsilon) = \omega_D - \omega_{1}(\varepsilon)$. 
The transition frequency $\omega_1$ changes as a function of $\varepsilon$, see Fig.~\ref{fig:spec_finite_eps}.
We chose $g = 0.5 \, \omega_\m$, $2t = 10 \omega_\m$, $\varepsilon_D = 8.6 \times 10^{-3}\,\omega_\m$, $\Gamma = \Gamma_\phi = 7.6\times10^{-3}\, \omega_\m$. This yields the effective parameters for $\varepsilon=0$ as $\tilde \varepsilon_D (0) = 1.1 \, \chi(0)$, $\gamma_\mathrm{eff}(0) = 0.1 \chi (0)$ and $\chi(0) = 8\times10^{-4} \, \omega_\m$. 
In the presence of finite detuning $\varepsilon$, the effective parameters change. 
For the three chosen values plotted here, one obtains, $\varepsilon= 0.3 t$: $\chi(\varepsilon) = 8.7 \gamma_\mathrm{eff}(\varepsilon) = 3.2 \gamma^\mathrm{eff}_\phi(\varepsilon)$;
 $\varepsilon= 0.5 t$: $\chi(\varepsilon) = 6.6  \gamma_\mathrm{eff}(\varepsilon) = 0.8 \gamma^\mathrm{eff}_\phi(\varepsilon)$;
 $\varepsilon=  t$: $\chi(\varepsilon) = -2.1 \gamma_\mathrm{eff}(\varepsilon) = -0.05 \gamma^\mathrm{eff}_\phi(\varepsilon)$. 
Note that $x = a + a^\dag$ is the dimensionless oscillator displacement normalized to the zero-point motion $x_\mathrm{zpm}^{(0)} = ( 2 m \omega_\m )^{-1/2}$ of the uncoupled mechanical oscillator. The effective parameters are determined by $\tilde \varepsilon_D = |\braket{1| \sx|0}| \varepsilon_D$, $\chi = \omega_{2} - \omega_{1}$ and $\gamma_\mathrm{eff} = |\braket{0|\sx|1}|^2 \Gamma$, where $\ket{n}$ is an eigenstate of the Hamiltonian. 
\label{fig:x_squ_eps}} 
\end{figure}

Using the SVS master equation, we numerically integrate the driven, time-dependent dynamics for a weak drive, $\varepsilon_D/\omega_\mathrm{m}=8.6\times10^{-3}$, which, for the chosen parameters of $g = 0.5 \omega_\m$ and $\omega_\m = \omega_\q /10$, is larger than the anharmonicity at zero-detuning $\chi(0) = 8 \times 10^{-4} \omega_\m$,
and compute the experimentally relevant response $\langle x^2\rangle(\omega_D)$ as a function of the drive frequency. The resulting spectra for different detunings are shown in Fig.~\ref{fig:x_squ_eps}. As the detuning approaches $\varepsilon=t$, the spectral response becomes progressively more harmonic, reflecting the reduction of the induced anharmonicity, while retaining a finite nonlinear signature. In this regime, the effective anharmonicity satisfies $\chi\ll\gamma_\phi^\mathrm{eff}$, where the full secular approximation is no longer justified, yet the SVS approach continues to provide a consistent description. The calculated response is directly related to the experimentally accessible signal in the corresponding nanomechanical setup~\cite{moller_tunable_2026}.
This provides direct theoretical access to experimentally measurable spectra throughout the full detuning range, including regimes in which conventional Lindblad treatments fail.

\section{Conclusion}
This work has been motivated by recent proposals and experiments that showed how the flexural modes of suspended carbon nanotubes can be coupled to a double quantum dot embedded in the nanotube itself in the ultrastrong regime~\cite{pistolesi_proposal_2021, moller_tunable_2026}. The system is well described by the open quantum Rabi model. We have investigated its behavior in the experimentally relevant regime where the mechanical frequency is much smaller than the TLS frequency, assuming a Markovian environment and hence working within the Born--Markov approximation.

For this purpose, we employed a Lindblad master equation based on the slowly varying bath spectrum (SVS) approximation~\cite{settineri_dissipation_2018,mccauley_accurate_2020}, in a formulation combining a minimal parameter set with correct finite-temperature thermalization.
It accurately describes the system across a large and experimentally relevant parameter space within a unified approach. Using this consistent dissipative description, we studied experimentally accessible observables in the presence of finite temperature, coherent driving, and finite detuning of the double quantum dot.

We showed that the conventional full secular approximation ceases to provide a valid description in these regimes and can lead to qualitatively incorrect predictions. We found that the temperature dependence of the coherence and hence correlation decay rates is governed by the coupling-induced anharmonicity. 
We analyzed the eigenvalues of the Lindblad operator in Liouville space as a function of induced anharmonicity and temperature.
From this we found that depending on the ratio between anharmonicity and population decay rate, the temperature-induced increase of the coherence decay is suppressed due to the transfer of coherence.  
We further identified a small anharmonicity-induced shift of the peak in the mechanical correlation spectrum at finite temperature, which vanishes in both the large- and small-anharmonicity limits.

We showed that the dissipator obtained within the full secular approximation is highly sensitive to perturbations on the scale of the anharmonicity, implying that it must, in principle, be reconstructed whenever such perturbations are present. In contrast, the SVS Lindblad dissipator remains valid in its unperturbed form for perturbations that are small compared to the mechanical transition frequency, making it considerably more robust for the description of driven systems. We also showed that when the driving strength and dissipation become comparable to the induced anharmonicity, observable response functions, specifically the average displacement squared, may already exhibit an essentially classical appearance while the underlying steady state remains non-classical. This demonstrates that observable signatures of non-classicality may disappear well before the quantum character of the state itself is lost, highlighting the importance of a dissipative description that remains reliable beyond the regime of validity of the conventional secular approximation.

Finally, we studied the driven system in the presence of a finite double-dot detuning, where the induced anharmonicity is strongly reduced while an additional decoherence channel emerges. Using the SVS Lindblad master equation, we predicted experimentally accessible observables in this regime of reduced anharmonicity.

We expect these results to provide a reliable theoretical framework for interpreting current experiments and to facilitate the exploration of dissipative phenomena in nanomechanical quantum Rabi systems beyond the regime where conventional secular approaches remain applicable.

\begin{acknowledgments}
We thank A.~Bachtold, R.~Tormo-Queralt, and C.~B.~M{\o}ller for comments and discussions.
We acknowledge financial support from the French Agence Nationale de la Recherche through contract ANR MORETOME ANR–22–CE24–0020–03, from the Conseil R\'egionale de Nouvelle-Aquitaine contract UFOs, from the French government in the framework of the University of Bordeaux’s France 2030 program/GPR LIGHT, and from the European Union Horizon Europe research and innovation program under grant agreement n. 101257982 MECH-QUBIT. 
\end{acknowledgments}
\appendix
\section{Perturbation coefficients and energy terms}
\label{app:4th_order_H}
The energy terms in Eq.~\eqref{H_4th} from fourth-order time-independent perturbation theory are given by:
\begin{align}
\tilde \omega_\q 
&= \omega_\q + \frac{g^2 \omega_\q}{\omega_\q^2-\omega_\m^2}-\frac{g^4 \omega_\q \left(\omega_\m^2+3 \omega_\q^2\right)}{\left(\omega_\q^2-\omega_\m^2\right){}^3}
\,,
\\ 
\Delta \omega_\m 
&= 
\frac{4 g^4 \omega_\q \left(\omega_\m^2+3 \omega_\q^2\right)}{\left(\omega_\q^2-\omega_\m^2\right){}^3}
\,,
\\
    \tilde{\omega}_\m 
    &= \omega_\m
    -\frac{2 g^2 \omega_\q}{\omega_\q^2-\omega_\m^2}
    \nn
    & \quad 
    +
    \frac{2 g^4 \omega_\q \left(3 \omega_\m^2 \omega_\q-6 \omega_\m \omega_\q^2-2 \omega_\m^3+\omega_\q^3\right)}{\omega_\m \left(\omega_\m^2-\omega_\q^2\right){}^3}
\\
\chi 
&= 
\frac{4 g^4 \omega_\q \left(\omega_\m^2+3 \omega_\q^2\right)}{\left(\omega_\q^2-\omega_\m^2\right){}^3} 
\,.
\end{align}

The coefficients for the 4th-order approximated eigenstates $\ket{\Psi_{n-}} = \sum_{m \sigma}  \braket{ \Psi_{n - } |m \sigma } \ket{m \sigma}$, where $\ket{n \pm}$ is an eigenstate of the bare Hamiltonian $H_0 = H_\m + H_\mathrm{tls}$, are given by
\begin{widetext}
\begin{multline}
    \braket{ \Psi_{n -} | n  - }
    = 
    1 
    - 
    \frac{1}{2} g^2 \left(\frac{n}{\left(\omega_\q-\omega_\m\right){}^2}+\frac{n+1}{\left(\omega_\m+\omega_\q\right){}^2}\right)
    \\
    +
    \frac{g^4 }{8 \omega_\m^2 \left(\omega_\m^2-\omega_\q^2\right){}^4}
    \bigg[
    6 (9 n (n+1)+2) 
    \omega_\m^4 \omega_\q^2
    -
    28 (2 n+1) \omega_\m^3 \omega_\q^3
    +
    3 (14 n (n+1)+9) \omega_\m^2 \omega_\q^4
    \\ 
    -
    12 (2 n+1) \omega_\m \omega_\q^5
    -
    (18 n (n+1)+1) \omega_\m^6
    +2 \left(n^2+n+1\right) \omega_\q^6
    \bigg] 
\,,
\end{multline}
\begin{multline}
    \braket{ \Psi_{n -} | n-1 , + } 
    = 
    -
    \frac{g \sqrt{n}}{\omega_\m-\omega_\q}
    +
    \frac{g^3
    \sqrt{n} }{2 \omega_\m \left(\omega_\q-\omega_\m\right){}^3 \left(\omega_\m+\omega_\q\right){}^2}
    \big[3 (n+1) \omega_\m^2 \omega_\q
    \\
    +3 n \omega_\m \omega_\q^2
    -\left((3 n+2) \omega_\m^3\right)
    +(n-1) \omega_\q^3\big] 
\,,
\end{multline}
\begin{multline}
    \braket{ \Psi_{n -} | n+1 ,  + } 
    =\frac{g \sqrt{n+1}}{\omega_\m+\omega_\q}
    +
    \frac{g^3 \sqrt{n+1}}{2 \omega_\m \left(\omega_\m-\omega_\q\right){}^2 \left(\omega_\m+\omega_\q\right){}^3} \bigg[
    -3 n \omega_\m^2 \omega_\q
    +3 (n+1) \omega_\m \omega_\q^2
    \\
    -\left((3 n+1) \omega_\m^3\right)
    -(n+2) \omega_\q^3
    \bigg]
\,,
\end{multline}

\begin{multline}
    \braket{ \Psi_{n -} | n-2 ,  - }  
    = 
    \frac{g^2 \sqrt{(n-1) n}}{2 \omega_\m \left(\omega_\m-\omega_\q\right)}
    +
    \frac{g^4 \sqrt{(n-1) n}}{4 \omega_\m \left(\omega_\q-3 \omega_\m\right) \left(\omega_\q-\omega_\m\right){}^3 \left(\omega_\m+\omega_\q\right){}^2}
    \bigg[
    (11-14 n) \omega_\m^2 \omega_\q 
    \\
    -
    9 (2 n+1) \omega_\m \omega_\q^2
    +
    (1-2 n) \omega_\m^3+(10 n-3) \omega_\q^3
    \bigg] 
\,,
\end{multline}

\begin{multline}
    \braket{ \Psi_{n -} | n+2 ,  - }  
    = 
    \frac{g^2  \sqrt{(n+1) (n+2)}}{2 \omega_\m \left(\omega_\m+\omega_\q\right)}
    +
    \frac{g^4 \sqrt{(n+1) (n+2)} }{4 \omega_\m \left(\omega_\m-\omega_\q\right){}^2 \left(\omega_\m+\omega_\q\right){}^3 \left(3 \omega_\m+\omega_\q\right)}
    \bigg[(14 n+25) \omega_\m^2 \omega_\q
    \\
    -9 (2 n+1) \omega_\m \omega_\q^2
    -\left((2 n+3) \omega_\m^3\right)
    -(10 n+13) \omega_\q^3
    \bigg] 
\,,
\end{multline}

\begin{multline}
    \braket{ \Psi_{n -} | n-3 ,  + }  
    = 
    \frac{g^3 \sqrt{(n-2) (n-1) n}}{2 \omega_\m \left(\omega_\q-\omega_\m\right) \left(\omega_\q-3 \omega_\m\right)} 
\,,
    \hfill
\end{multline}
\begin{multline}
    \braket{ \Psi_{n -} | n+3 ,  + }  
    =
    -\frac{g^3 \sqrt{(n+1) (n+2) (n+3)}}{2 \omega_\m \left(\omega_\m+\omega_\q\right) \left(3 \omega_\m+\omega_\q\right)}
\,,
    \hfill
\end{multline}
\begin{multline}
    \braket{ \Psi_{n -} | n-4 ,  - }  
    = 
    \frac{g^4 \sqrt{( n-3)(n-2) (n-1) n}}{8 \omega_\m^2 \left(\omega_\q-3 \omega_\m\right) \left(\omega_\q-\omega_\m\right)}
\,,
    \hfill 
\end{multline}
\begin{multline}
    \braket{ \Psi_{n -} | n+4 ,  - }  
    = 
    \frac{g^4 \sqrt{(n+1) (n+2) (n+3) (n+4)}}{8 \omega_\m^2 \left(\omega_\m+\omega_\q\right) \left(3 \omega_\m+\omega_\q\right)}
\,,
    \hfill
\end{multline}
\end{widetext}
\section{Derivation of the Lindblad master equation from secular approximation}
\label{app:FSA}
In this section, we derive the Lindblad master equation from the conventional secular approximation and show its validity regime~\cite{breuer_theory_2007}.
Applying the Born-Markov approximation to a system-environment Hamiltonian 
\begin{align}
    H = H_\mathrm{S} + H_\mathrm{E} + H_\mathrm{SE}
\end{align}
to find the master equation of the reduced system density matrix $\rho(t) $ 
yields 
\begin{align}
\dot{\tilde \rho}  (t)
 &= 
 - \int_0^\infty \dif \tau 
 \tr_E \left[\tilde H_\mathrm{SE}(t), \left[ \tilde H_\mathrm{SE}(t-\tau), \tilde \rho  \otimes \rho_E \right] \right]
\end{align}
where $\tilde \rho(t)$ is the system density matrix in interaction picture and $\tilde H_\mathrm{SE}(t)$ is the system-environment interaction in interaction picture and we trace over the bath's degrees of freedom. 
 With $\tilde H_\mathrm{SE}(t) = \lambda  \tilde A(t) \cdot \tilde{\mathcal{E}}(t)  $, where $A$ is a Hermitian system operator and $\mathcal{E}$ is a Hermitian bath operator, we can write the Born-Markov equation as
\begin{align}
\dot{\tilde \rho}  = 
  - \lambda ^2 \int_0^\infty \dif \tau   
 \left[ A(t),  A(t-\tau)  \tilde \rho    \right]   C(\tau) 
 + \mathrm{h.c.}
 \label{tilde_rho_app-fsa}
\end{align}
where $C(\tau) = \braket{\tilde{\mathcal{E}}(\tau)\tilde{\mathcal{E}}(0)} = C^*(-\tau)$ is the correlation function of the bath operator. 
The Born-Markov equation can be rearranged into a form that closely resembles a Lindblad form by introducing the spectral decomposition of the system operator as 
\begin{align}
    \tilde A(t) &= \sum_i \tilde A_i(t)
    \\
   \tilde  A_i(t) &= \sum_{\omega_i = \omega_{nm} } e^{- \mi \omega_i t} A_i 
   \label{spec_dec_A}
   \\
   A_i &= A_{nm} \ket{n}\bra{m}
\end{align}
where $A_{nm} = \braket{n | A | m}$, $\ket{n}$ is an eigenstate of $H_\mathrm{S}$ and $\omega_i = \omega_{nm} = E_m - E_n$ is the transition frequency between eigenstates $m$ and $n$.
Using the spectral decomposition we may write Eq.~\eqref{tilde_rho_app-fsa} as 
\begin{multline}
        \dot{\tilde \rho} (t)
    = \lambda ^2 
    \sum_{i,j} 
     \Lambda(\omega_i) 
     \Big( 
    \tilde A_i(t) \tilde \rho  \tilde A_j^\dag(t) 
    \\ 
    - \tilde A_j^\dag(t) \tilde A_i(t) \rho 
    \Big)
    + 
    \mathrm{h.c.}
    \label{BM-eq}
\end{multline}
where we used that $\tilde A_i^\dagger (t) = \tilde A_k (t)$ with $\omega_k = - \omega_i$ to rearrange terms and 
where $\Lambda(\omega_i ) = \int_0^\infty \dif \tau  \exp(- \mi \omega_i \tau ) C(\tau) $, or alternatively with the explicit time-dependency in the interaction picture, we may write
\begin{multline}
        \dot{\tilde \rho} (t)
    = \lambda ^2 
    \sum_{i,j} \bigg \{ 
     e^{- \mi ( \omega_i - \omega_j)t} \Lambda(\omega_i) \left( 
    A_i \tilde \rho  A_j^\dag  - A_j^\dag A_i \rho 
    \right) 
    \\
    + 
    e^{- \mi ( \omega_i - \omega_j )t }  
    \Lambda^*(\omega_j) 
    \left( 
    A_i \tilde \rho  A_j^\dag  -\rho  A_j^\dag A_i 
    \right)  
    \bigg\} \,,
\end{multline}
which would be directly in Lindblad form if $\Lambda(\omega_i ) = \Lambda^*(\omega_j)$. 

In the full secular approximation, one assumes that $|\omega_i - \omega_j|$ with $i \neq j$ oscillates fast relative to the resulting relaxation time $1/\gamma_\mathrm{eff}$ of the system, such that they effectively decouple, \textit{i.e.}~any term with $i \neq j$ can be neglected in the sum and we may approximate
\begin{multline}
        \dot{\tilde \rho}  (t)
    = \lambda ^2 
    \sum_{i } \bigg \{ 
       \Lambda(\omega_i) \left( 
    A_i \tilde \rho  A_i^\dag  - A_i^\dag A_i \rho 
    \right) 
    \\
    +   
    \Lambda^*(\omega_i) 
    \left( 
    A_i \tilde \rho  A_i^\dag  -\rho  A_i^\dag A_i 
    \right)  
    \bigg\} \,.
\end{multline}
The real part of $\Lambda(\omega_i)$ can now be summarised in Lindblad form, while the imaginary part can be written in terms of a commutator with an additional Hamiltonian contribution that leads to a Lamb shift, which is discussed further in App.~\ref{app:Lamb}. Neglecting the Lamb shift, we finally find the Lindblad equation from FSA written back in Schroedinger picture as
\begin{align}
        \dot{  \rho}(t)
    = - \mi [ H, \rho]
    + 
    \sum_{i }
      \lambda ^2  S(\omega_i) \mathcal{D}_{A_i}[ \rho ] \,,
      \label{rho_fsa_gen}
\end{align}
where $S(\omega ) = 2 \re \Lambda( \omega ) = \mathrm{FT}\{ C(t)\}$ is the Fourier-transform of the bath correlation function and $\mathcal{D}$ is the Lindblad dissipator, see Eq.~\eqref{Diss}. 

The FSA Lindblad master equation hence introduces a multitude of dissipation channels with rates $\gamma_i = \lambda^2 S(\omega_i)$. It is only valid if the effective decay fulfills $\gamma_\mathrm{eff} \ll | \omega_i - \omega_j|$, where $\omega_{i}$, $\omega_{j}$ are transition frequencies appearing in the spectral decomposition. Finally, by definition each collapse operator $A_i$ oscillates with a unique frequency $\omega_i$ under the unitary evolution $U(t) = \exp(-\mi H t)$.
\section{Derivation of the SVS Lindblad equation}
\label{app:our_LME}
In this section, we apply the slowly-varying bath spectrum (SVS) approximation, together with a partial secular approximation to obtain the SVS Lindblad equation. 

In the Born-Markov equation given by Eq.~\eqref{BM-eq}, the sum runs over all $\omega_i = \omega_{nm}$ that appear in the spectral decomposition of the system operator $A$, see Eq.~\eqref{spec_dec_A}. This includes negative frequencies. 
\begin{widetext}
Tracking the sign of the transition frequencies explicitly 
we can rewrite  Eq.~\eqref{BM-eq} with a sum over positive transition frequencies as 
\begin{align}
    \dot{\tilde \rho}  (t)
 = 
 \sum_{\substack{i, \omega_i \ge0 \\ j, \omega_j \ge0 }}  
 \bigg\{
 &
 \Lambda(\omega_i)
 \Big( 
 \tilde A_i(t) \tilde\rho \tilde A_j(t)
 +
 \tilde A_i(t) \tilde\rho \tilde A_j^\dag(t)
 -
  \tilde A_j(t) \tilde A_i(t) \tilde\rho
-  
\tilde A_j^\dag (t) \tilde A_i(t) \tilde\rho 
 \Big)
 \nn
  +
 & \Lambda (- \omega_i)
 \Big( 
 \tilde A_i^\dag(t) \tilde\rho \tilde A_j^\dag(t)
 +
 \tilde A_i^\dag(t) \tilde\rho \tilde A_j(t)
 -
  \tilde A_j^\dag(t) \tilde A_i^\dag(t) \tilde\rho
  -
 \tilde  A_j(t) \tilde A_i^\dag(t) \tilde\rho 
 \Big)
 \bigg\}
 + \mathrm{h.c.}  
\end{align}
where we used that for all $A_i$, we find an $A_k$ in the sum, such that $A_k(t) = A^\dag_i (t)$ with $\omega_k = - \omega_i$. 
We further assume for simplicity that the imaginary part of $\Lambda(\omega)$ is negligible as it will lead to small Lamb-shifts and introduce the power spectrum $S(\omega) =2 \re \Lambda(\omega)$. The Lamb shift and its significance is discussed in more detail in App.~\ref{app:Lamb}. 
By explicitly evaluating the Hermitian conjugate we find, 
\begin{align} 
    \dot{\tilde \rho}  (t)
 = 
 \sum_{\substack{i, \omega_i \ge 0 \\ j, \omega_j \ge 0 }}  
 \bigg\{
 &
\frac{S(\omega_i)}{2}
 \Big( 
 2 \tilde A_i(t) \tilde\rho \tilde A_j^\dag(t) 
-  
\tilde A_i^\dag (t) \tilde A_j(t) \tilde\rho 
-  
 \tilde\rho \tilde A_i^\dag  (t) \tilde A_j (t)  
\nn
 & 
\qquad 
+ 
 \tilde A_i(t) \tilde\rho \tilde A_j(t)
 +
 \tilde A_i^\dag(t) \tilde\rho \tilde A_j^\dag(t)
 -
  \tilde A_j(t) \tilde A_i(t) \tilde\rho
 -
  \tilde\rho \tilde A_j^\dag(t) \tilde A_i^\dag(t)
 \Big)
 \nn
  +
 & 
\frac{S(-\omega_i)}{2}
 \Big( 
 2 \tilde A_i^\dag(t) \tilde\rho \tilde A_j (t) 
-  
\tilde A_i (t) \tilde A_j^\dag(t) \tilde\rho 
-  
 \tilde\rho \tilde A_i  (t) \tilde A_j^\dag (t)  
\nn
 & 
\qquad 
+ 
 \tilde A_i(t) \tilde\rho \tilde A_j(t)
 +
 \tilde A_i^\dag(t) \tilde\rho \tilde A_j^\dag(t)
 -
  \tilde A_j^\dag(t) \tilde A_i^\dag(t) \tilde\rho
 -
  \tilde\rho \tilde A_j(t) \tilde A_i (t)
 \Big)
 \bigg\}
 \label{BME_pos_freq}
 \,. 
\end{align} 
Next, we apply a partial secular approximation, neglecting terms that oscillate with $\omega_i + \omega_j$ or $\omega_i$, \textit{i.e.}~that are proportional to $A_i A_j$ or $A^\dag_i A^\dag_j$ or involve a single 0-frequency component. 
This requires that the effective decay of the system is much slower than these oscillations. 
We then find, 
\begin{align} 
    \dot{\tilde \rho}  (t)
 = 
 \sum_{\substack{i, \omega_i > 0 \\ j, \omega_j > 0 }}  
 \bigg\{
 &
\frac{S(\omega_i)}{2}
 \Big( 
 2 \tilde A_i(t) \tilde\rho \tilde A_j^\dag(t) 
-  
\tilde A_i^\dag (t) \tilde A_j(t) \tilde\rho 
-  
 \tilde\rho \tilde A_i^\dag  (t) \tilde A_j (t)  
 \Big)
 \nn
  +
 & 
\frac{S(-\omega_i)}{2}
 \Big( 
 2 \tilde A_i^\dag(t) \tilde\rho \tilde A_j (t) 
-  
\tilde A_i (t) \tilde A_j^\dag(t) \tilde\rho 
-  
 \tilde\rho \tilde A_i  (t) \tilde A_j^\dag (t)  
 \Big)
 \bigg\} 
  \nn
  +
 \sum_{ \omega_i = 0 }
 & 
S(0) 
 \Big( 
 2 \tilde A_i^\dag  \tilde\rho \tilde A_j  
-  
\tilde A_i \tilde A_j^\dag  \tilde\rho 
-  
 \tilde\rho \tilde A_i   \tilde A_j^\dag  
 \Big) 
 \,.
\end{align} 
Under the SVS assumption that $| S(\omega_i) - S(\omega_j)| \ll S(\omega_i)$, we may approximate $S(\omega_i) \approx S(\omega_j)$ for all $\omega_i$ and $\omega_j$.

This approximation grants some freedom in how to apply it practically. In Ref.~\cite{settineri_dissipation_2018}, the authors substitute $S(\omega_i) \rightarrow \gamma_\downarrow$ (and $S(- \omega_i) \rightarrow  \gamma_\uparrow$) as constant to obtain a Lindblad form, while Ref.~\cite{mccauley_accurate_2020} uses the freedom to substitute $S(\omega_i) \rightarrow  \sqrt{S(\omega_i)S(\omega_j)} = \sqrt{\gamma_i \gamma_j}$ inside the sum to obtain their Lindblad form. 
Here, we apply a compromise between these two paths and introduce
\begin{align}
    S(\omega_i) &= J(\omega_i) (\nth (\omega_i) +1)
    \rightarrow 
    \gamma \sqrt{(\nth (\omega_i) +1) (\nth (\omega_j) +1) } \,,
    \\
    S(- \omega_i)  
    &\rightarrow 
    \gamma \sqrt{ \nth (\omega_i) \nth (\omega_j) } 
\end{align}
such that the zero-temperature rate $\gamma = J(\omega)$ is approximated for all frequencies as a constant, while the thermal weight can be kept true to the eigen-spectrum.
With this we can write 
\begin{align} 
    \dot{\tilde \rho}  (t)
 = 
\frac{\gamma}{2}
 \sum_{\substack{i, \omega_i > 0 \\ j, \omega_j > 0 }}  
 \bigg\{
 &
\sqrt{(\nth (\omega_i) +1) (\nth (\omega_j) +1)}
 \Big( 
 2 \tilde A_i(t) \tilde\rho \tilde A_j^\dag(t) 
-  
\tilde A_i^\dag (t) \tilde A_j(t) \tilde\rho 
-  
 \tilde\rho \tilde A_i^\dag  (t) \tilde A_j (t)  
 \Big)
 \nn
  +
\frac{\gamma}{2}
 \sum_{\substack{i, \omega_i > 0 \\ j, \omega_j > 0 }} 
 &  
\sqrt{ \nth (\omega_i) \nth (\omega_j) } 
 \Big( 
 2 \tilde A_i^\dag(t) \tilde\rho \tilde A_j (t) 
-  
\tilde A_i (t) \tilde A_j^\dag(t) \tilde\rho 
-  
 \tilde\rho \tilde A_i  (t) \tilde A_j^\dag (t)  
 \Big)
 \bigg\} 
  \nn
  +
\frac{\gamma_\phi }{2}
 \sum_{ \omega_i = 0 }
 & 
 \Big( 
 2 \tilde A_i^\dag(t) \tilde\rho \tilde A_j (t) 
-  
\tilde A_i  \tilde A_j^\dag  \tilde\rho 
-  
 \tilde\rho \tilde A_i    \tilde A_j^\dag 
 \Big) 
 \,,
\end{align} 
where $\gamma_\phi = 2 S(0)$. 
This equation is now in a Lindblad form with collapse operators defined by Eq.~\eqref{A_dec_first}--\eqref{A_dec_last}. 
\end{widetext}
Compared to the conventional full secular approximation, the SVS Lindblad equation neglects significantly fewer terms. The remaining neglected terms are also discarded within the FSA Lindblad equation, such that the SVS equation can be viewed as an extension of the full secular approximation.
The full secular approximation is then only valid if the effective decay of the system is much smaller than the transition frequency differences $|\omega_{nm} - \omega_{kl}|$, while the SVS Lindblad equation applies a much less restrictive partial secular approximation that is valid as long as the transition frequencies $\omega_{nm}$ themselves are much larger than the effective decay of the system. Note, that also FSA becomes invalid in this regime.
 
\section{Validity in weak coupling regime}
\label{app:TLS_as_env}
Here, we discuss the validity of the SVS Lindblad equation when the decay rate of the TLS exceeds its coupling to the oscillator, $\Gamma > g$. From the perspective of the TLS, this corresponds to the conventional weak-coupling regime, and it is appropriate to describe the TLS dynamics, which are only weakly perturbed by the oscillator, using the bare TLS dissipator $\mathcal{D}{\sigma_-}[\rho]$. From the perspective of the oscillator, however, the situation is different. As discussed in the main text, using the uncoupled Lindblad dissipator for the coupled system yields unphysical results even for arbitrarily small $g$. This can be understood intuitively: since we assume the oscillator to have a negligible intrinsic dissipation rate, its dominant decay channel for $g < \Gamma$ is through the TLS. Consequently, even in the weak-coupling regime, the oscillator dissipation cannot be regarded as independent of the coupling.

Assuming the coupling $g$ is small enough while the TLS decays sufficiently fast, it is appropriate to describe the TLS as part of the oscillator-environment. 
Let us assume an explicit TLS-environment and show that treating the TLS as part of the environment leads to the SVS Lindblad equation. 
We assume the TLS is coupled to a bosonic bath,
\begin{align}
    H_\mathrm{full}
    &= H_\m + g \sx ( a + a^\dag)  + H_E \,, 
    \\
   H_E &=
     H_\mathrm{tls} + \sum_k \lambda_k \sx ( b_k + b_k^\dag) + \sum_k \omega_k b_k^\dag b_k  
    \,,
\end{align}
where we defined the TLS as part of the environment. 

The environment coupling $ H_\mathrm{SE} = g \sx ( a + a^\dag)$ needs to be written in terms of eigenoperators of the environment. 

The spectrum of $H_E$ is that of the bath modes plus the TLS eigenstates with energy-splitting $\omega_\q \gg \omega_\m$ which is dressed by the bath-modes that are energetically close.
As we will later see, the important part of the environment spectrum is around energies given by $\omega_\m \ll \omega_\q$. In this regime, $\omega \ll \omega_q$ and for small couplings $\lambda_k \ll \omega_k, \omega_q$, the density of states $D(\omega)$ of $H_E$ is approximately unchanged by the presence of the TLS. 
On the other hand, we know how $\sx$ transforms when diagonalising $H_E$ in leading order of coupling $\lambda_k$ and reducing the Hilbert-space to eigenenergies much smaller than $\omega_\q$, we can approximate
\begin{align}
    \sx \rightarrow \sum_{k,\omega_k \ll \omega_q} \frac{2 \lambda_k \omega_q}{\omega_\q^2 - \omega_k^2} ( b_k + b_k^\dag  ) \,.
\end{align}  
With this we can approximate the system-environment coupling as
\begin{align}
    g \sx ( a + a^\dag)  &\rightarrow \sum_k g_k  ( b_k+ b_k^\dag  )  ( a + a^\dag) 
    \,,
    \\
    g_k &= \frac{2 g \lambda_k \omega_q}{\omega_\q^2 - \omega_k^2}
    \,,
\end{align}
such that 
\begin{align}
    H_\mathrm{full}
    = H_\m  + \sum g_k ( b_k + b_k^\dag) ( a + a^\dag) + H_E
    \,,
\end{align}
this is the well known case of an oscillator coupled to a bosonic bath. Applying the Born-Markov approximation and using secular approximation it yields the standard Lindblad master equation (see e.g.~Ref.\cite{breuer_theory_2007})
\begin{align}
    \dot \rho_\m = - \mi [ H_\m 
    , \rho_\m] + \gamma_\downarrow  \mathcal{D}_a [ \rho_\m ] + \gamma_\uparrow  \mathcal{D}_{a^\dag} [ \rho_\m ]\,,
\end{align}
with 
\begin{align}
    \gamma_\downarrow &= 2 \pi \sum_k g_k^2 \delta( \omega_\m - \omega_k) \braket{b_k b_k^\dag }_E 
    \nn
    &= 2 \pi D(\omega_\m ) g^2(\omega_\m ) ( 1 + n_\mathrm{th}(\omega_\m )) \,,
    \\
    \gamma_\uparrow &= 2 \pi \sum_k g_k^2 \delta( \omega_\m - \omega_k) \braket{b_k^\dag  b_k}_E \nn
    &=
    2 \pi D(\omega_\m ) g^2(\omega_\m ) (  n_\mathrm{th}(\omega_\m )) \,,
\end{align}
where we assumed a thermal state of the bath with $\braket{b_k^\dag b_k}_E = n_\mathrm{th}(\omega_k)$ and $\braket{b_k b_k^\dag}_E = 1 +  n_\mathrm{th}(\omega_k)$ and defined $\sum_k g_k^2 \delta( \omega - \omega_k) = D(\omega) g^2(\omega)$ for $\omega \ll \omega_\q$ where $D(\omega)$ is the density of original bath modes.

At zero temperature this simplifies to 
\begin{align}
    \gamma_\downarrow 
    &= 2 \pi D(\omega_\m ) g^2(\omega_\m )  \,,
    \\
    \gamma_\uparrow &= 0\,,
\end{align}
and using the definition of $g_k$, we can write
\begin{align}
    \gamma_\downarrow 
    &= \frac{4 g^2 \omega_\q^2}{( \omega_\q^2 - \omega_\m^2)^2}\Gamma\,,
    \\
     \Gamma &= ( 2 \pi D(\omega_\m ) \lambda^2(\omega_\m ) ) \,,
\end{align}
where $\Gamma$ is the zero-temperature rate that is found in the bare TLS decay, assuming that $D(\omega_\m) \lambda^2(\omega_\m ) \approx D(\omega_\q) \lambda^2(\omega_\q) $. This reproduces the dissipator as found for $g \ll \omega_\m$ with the SVS Lindblad master equation. 

We have shown here, that describing the TLS as part of the environment for $g \ll \Gamma$ and $g \ll \omega_\m$, results in the same Lindblad dissipator as proposed via the positive-negative frequency decomposition. 

\section{Derivation of two-phonon correlation function}
\label{app:g2}
The driven Kerr oscillator has been solved exactly previously~\cite{drummond_quantum_1980}.
To focus on the important point, we provide here an elementary derivation of the two-phonon correlation function $g^{(2)}(0)$ in a Kerr oscillator driven with an infinitesimal driving strength and coupled to an environment via $H_\mathrm{SE} = \lambda \mathcal{E} ( b + b^\dag)$.
The master equation is then given by Eq.~\eqref{dot-rho_driven} in the lab frame and after applying the rotation wave approximation and moving into the rotating frame, the Hamiltonian becomes time-independent, given by Eq.~\eqref{H_rwa}. 
As shown in Sec.~\ref{sec:drive}, for a driving strength that is much smaller than any other scale in the system, it is appropriate to use the unperturbed dissipator.

With this, we find the Lindblad master equation given by Eq.~\eqref{rho_rwa} containing the dissipator of the undriven system.
We want to calculate $\braket{b^\dag b}$ and $\braket{b^\dag b^\dag b b}$ in lowest order of the infinitesimal drive $\tilde \varepsilon_D$ in the steady state of the rotating frame. 
From the master equation, we find the equation of motion
\begin{align}
    \frac{\dif }{\dif t}
    \braket{ b}
    &= \mi \Delta \braket{b }- \mi \chi \braket{b^\dag b^2}  
    - \frac{\gamma_\mathrm{eff}}{2} \braket{b}
    - \mi \frac{\tilde \varepsilon_\D }{2} 
    \,,
    \\
    \frac{\dif }{\dif t}\braket{b^\dag b}
    &= \mi \frac{\tilde \varepsilon_\D }{2} \braket{ b - b^\dag }
    - \gamma_\mathrm{eff} \braket{b^\dag b} \,.
\end{align} 
Equating the equation of motions to zero for the steady state, we find 
\begin{align}
    \braket{b^\dag b }
    &= - \frac{\tilde \varepsilon_D}{\gamma_\mathrm{eff}} \im \left[ 
    \braket{b}\right]
    \,,
    \\
    \braket{b} 
    &=  \frac{\tilde \varepsilon_D}{2 \Delta  + \mi \gamma_\mathrm{eff}}
    \,,
\end{align}
such that 
\begin{align}
    \braket{b^\dag b}
    &= \frac{\tilde \varepsilon_D^2}{ \gamma_\mathrm{eff}^2 + 4 \Delta^2}
    \,.
\end{align}
Similarly, we find 
\begin{multline}
     0 = \frac{\dif }{\dif t} \braket{b^\dag b^\dag  b b}
      = - 2 \gamma_\mathrm{eff}  \braket{b^\dag b^\dag  b b}  
     \\
     + 2 \mi \tilde \varepsilon_D \braket{( b^\dag b^2  - [ b^\dag]^2 b )}
     \,.
     \label{adadaa_dot}
\end{multline}
From the structure of the equation motions, with the only source term $\propto \varepsilon_D$ appears in $b$ (and $b^\dag$), we can conclude that the expectation value of $\braket{[ b^\dag]^n b^m}$ depends in lowest order on $\tilde \varepsilon_D^{n+m}$, \textit{i.e.}~$\braket{b^\dag b^\dag  b b} \propto \tilde \varepsilon^4$.
The equation of motions for the remaining expectation values appearing in Eq.~\eqref{adadaa_dot} are given by 
\begin{multline}
    \frac{\dif }{\dif t} \braket{b^\dag  b ^2}     
     = ( \mi ( \Delta - \chi ) - \frac{3}{2} \gamma_\mathrm{eff} )\braket{b^\dag b^2}
     \\
     + \mi \frac{\tilde \varepsilon_D}{2} (\braket{  b^2} - 2 \braket{b^\dag b })     
     - \mi \chi \braket{[b^\dag]^2 b^3}
     \,,
\end{multline} 
and 
\begin{multline}
     \frac{\dif }{\dif t} \braket{b^2}   
      = ( 2 \mi \Delta - \mi \chi - \gamma_\mathrm{eff} ) \braket{b^2} 
      \\
      - \mi \tilde 
      \varepsilon_D \braket{b} - \mi \chi \braket{b^\dag b^2}
      \,,
\end{multline}
where the terms proportional to $\chi$ can be neglected as they depend on higher orders of $\tilde\varepsilon_D$. 
Equating these equation of motions to zero we find for the steady state
\begin{align}
    \braket{b^\dag b^\dag b b }
    &= 
    \frac{\tilde \varepsilon_D ^4}{\left(\gamma_\mathrm{eff}^2+4 \Delta ^2\right) \left(\gamma_\mathrm{eff}^2+(\chi -2 \Delta )^2\right)}
    \,,
\end{align}
and hence for $g^{(2)}(0) =\braket{b^\dag b^\dag b b }/\braket{b^\dag b }^2 $, 
\begin{align}
   g^{(2)}(0) = \frac{\gamma_\mathrm{eff}^2+4 \Delta ^2}{\gamma_\mathrm{eff}^2+(\chi -2 \Delta )^2}
   \,.
\end{align}

Applying instead the full secular approximation to receive a Lindblad equation for the Kerr oscillator, yields a slightly different Lindblad equation, see Eq.~\eqref{rho_dot_kerr_fsa}
\begin{align}
    \dot \rho_\mathrm{fsa}
    = 
    - \mi [ H_\mathrm{rwa}, \rho_\mathrm{fsa}] 
    + \sum_n \gamma_\mathrm{eff} \mathcal{D}_{c_n} [ \rho_\mathrm{fsa}]
    \,,
\end{align}
with $c_n = \sqrt{n} \ket{n-1} \bra{n}$. 
and as discussed in the main text is known to differ from the SVS Lindblad master equation only by neglecting the transfer of coherence, \textit{i.e.}~the two master equations yield the same populations $\rho_{nn} = [\rho_\mathrm{fsa}]_{nn}$. 
It is easy to prove that expectation values of the form $\braket{[b^\dag] ^m b^m}$ depend only on the evolution of $\rho_{nn}$,
\begin{align}
    \frac{\dif}{\dif t}\braket{[b^\dag] ^m b^m} 
    &= \tr \{ [b^\dag] ^m b^m \dot \rho \} 
    \nn 
    &= \sum_{n, n'} 
    \braket{n|[b^\dag] ^m b^m |n'} \dot \rho_{n'n}
    \nn 
    &= \sum_{n, n'} 
    \braket{n|[b^\dag] ^m b^m |n} \dot \rho_{nn}
\end{align}
and hence the values of $g^{(2)}(0)$ obtained with either master equation coincide with each other in the perturbative Kerr regime. 
Outside of the perturbative regime, the values obtained for $g^{(2)}(0)$ from the different master equation may differ from each other, but numerical calculations show no significant qualitative difference. 
\section{Anharmonic Lamb shift}
\label{app:Lamb}
\begin{figure}  
\includegraphics[width= \linewidth]{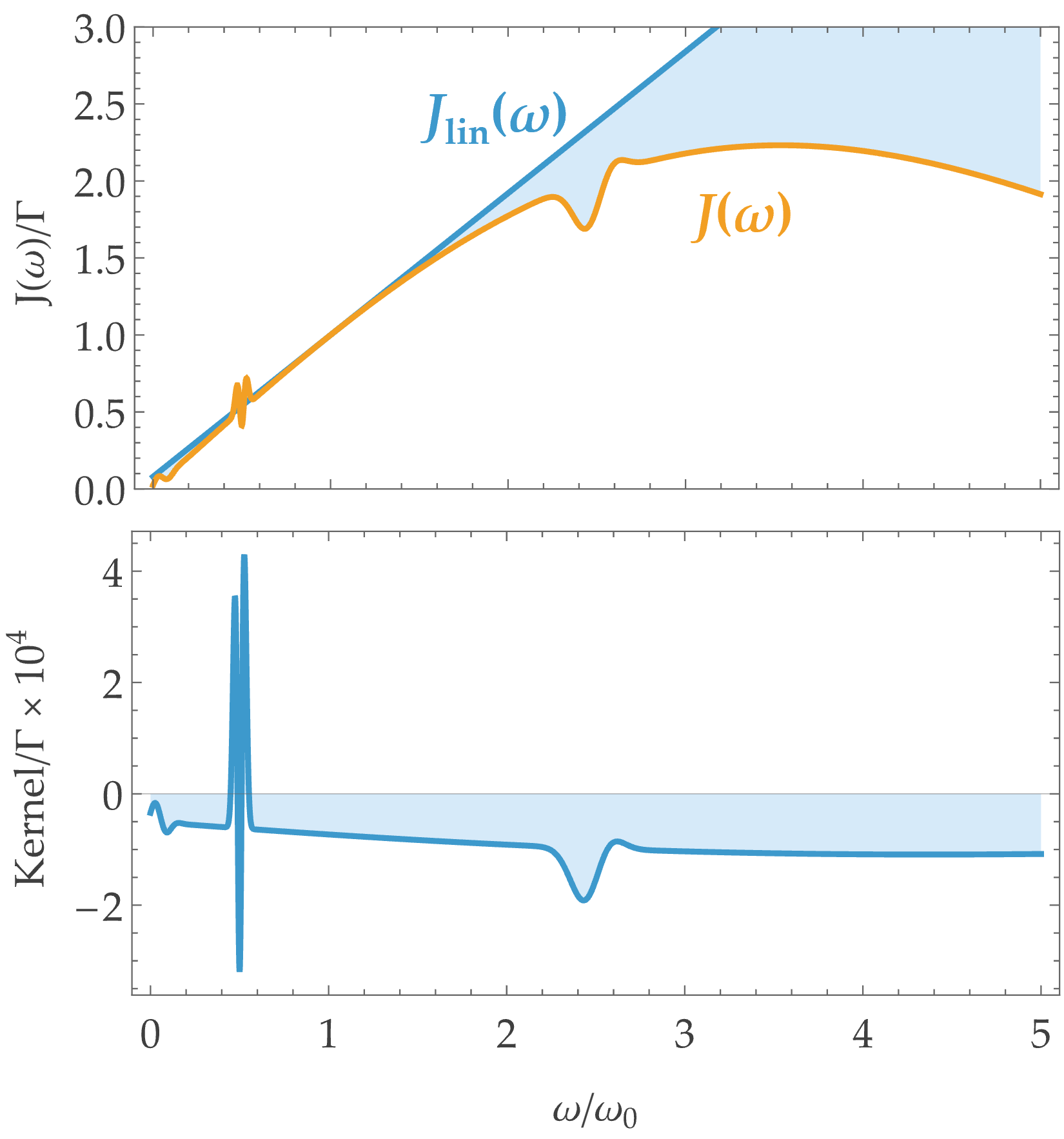} 
\caption{Schematic sketch of the integration Kernel in Eq.~\eqref{chi_LS_gen}. 
The function $J(\omega)$ may vary greatly from its linear approximation $J_\mathrm{lin}(\omega)$ for $|\omega - \omega_0| \gg 0$, however, the denominator of $(\omega - \omega_0)^2$ will suppress the difference then inside the integral. Note that the $y$-axis of the Kernel plot has been scaled by a factor of $10^4$. 
\label{fig:LS_Kernel}
 }
\end{figure}
In this section, we study the additional anharmonicity of the system induced by the Lamb shift, first for a simple ohmic spectrum and then expand the discussion onto a general spectrum. 

In the derivation of the SVS Lindblad master equation in App.~\ref{app:our_LME}, we neglected the imaginary part $\Lambda(\omega)$. Keeping said part, the dissipator gains terms that can be cast into a unitary evolution via a Lamb shift-Hamiltonian. With $A = \sx$, as it is considered in the main text and at zero temperature for simplicity, the Lamb shift Hamiltonian is given by 
\begin{align}
    H_\mathrm{LS}
    &= \sum_{\omega, \omega'} \im \Lambda(\omega) [ X_\downarrow]^\dag (\omega') X_\downarrow(\omega)
    \,,
    \\
    X_\downarrow( \omega)  &= \sum_{n,m:\omega_{nm} = \omega}\braket{m| X_\downarrow |n} \ket{m}\bra{n}  
    \,,
\end{align}
where $X_\downarrow$ is the negative frequency component of $\sx$ and $X_\downarrow(\omega)$ is spectrally decomposed with $H\ket{n} = E_n \ket{n}$ and $\omega_{nm} = E_{m} - E_n$, see Eq.~\eqref{spec_dec_A} with $A = X_\downarrow$ and $A_i = X_\downarrow(\omega_i)$. 
In the perturbative Kerr regime, $X_\downarrow \propto b$ and the Lamb shift Hamiltonian becomes diagonal in the eigenbasis of the QRM Hamiltonian. Outside of this regime, $H_\mathrm{LS}$ can in principle induce transitions but under the approximation of $\omega_{nm} \gg \gamma_\mathrm{eff}$, these terms can be neglected via a rotating wave approximation, such that 
\begin{align}
     H_\mathrm{LS}
    &= \sum_{\omega } \im \Lambda(\omega) [ X_\downarrow]^\dag (\omega ) X_\downarrow(\omega)
    \,. 
\end{align} 
For simplicity, we will focus the remainder of the discussion on the perturbative Kerr regime, where
\begin{align}
     H_\mathrm{LS}
    &=
    |\braket{0| X_\downarrow |1}|^2
    \sum_{\omega } \im \Lambda(\omega) b^\dag (\omega ) b(\omega)
    \,. 
\end{align} 
Using the Kramers-Kronig relation we can then calculate the Lamb shift of a transition frequency $\omega_{i} = \omega_{i+1} - \omega_i$ as 
\begin{align}
    \Delta_{i} &= 
    |\braket{0| X_\downarrow |1}|^2 \im \Lambda(\omega_{i}) 
    \\
    \im \Lambda(\omega_{i})
    &= \frac{1}{2 \pi }\mathcal{P}\!\!\int_0^\Omega \dif \omega 
    \frac{J(\omega)}{ \omega - \omega_{i}}
    \,,
\end{align}
where $\Omega \gg \omega_{i}$ is a cut-off frequency.
 
This shift may differ between the two transitions $\omega_{i}$ and $\omega_{i+n}$, which yields an effective anharmonicity $\chi_n^{\mathrm{(LS)}}$. 
We will first derive the explicit Lamb shift-induced anharmonicity from an Ohmic spectrum and show that it can be approximated by $\chi_n^{\mathrm{(LS)}} = \lambda n \chi$ and hence is a constant factor on the intrinsic anharmonicity $n \chi + \chi_n^{\mathrm{(LS)}}= ( 1 + \lambda ) n \chi$.

For an ohmic spectrum $J(\omega) = \alpha \omega$ and $\Gamma = \alpha \omega_0$ where $\omega_0$ is the lowest transition frequency, we find 
\begin{align}
\chi_n^{\mathrm{(LS)}} &= \Delta_n - \Delta_0 
\\
&=  
    \frac{\gamma_\mathrm{eff} \alpha}{2 \pi \Gamma}  \big[
    \left(\omega_0 + n \chi \right) 
    \log \left(\frac{\Omega -\omega _\m- n \chi}{\omega _0+ n \chi }
    \right) 
    \nn& \qquad 
    - \omega _\m
 \log \left(\frac{\Omega -\omega_0}{\omega _0}\right) 
    \big] 
    \,,
\end{align}
which we can approximate for $n \chi,\ \omega_\m \ll \Omega$ as 
\begin{align}
\chi_n^{\mathrm{(LS)}} 
&=
\frac{\gamma_\mathrm{eff} n \chi }{2 \pi \omega_0} \left[  \log \left( \frac{\Omega}{\omega_\m } \right) - 1 \right] 
\,,
\end{align}
where we used that $\alpha = \Gamma / \omega_0$. 
Thus the Lamb shift yields a constant off-set to the intrinsic anharmonicity. 

In general, we can linearly approximate the difference between two principal value integrals by introducing the Hadamard finite part integral~\cite{i_m_gelfand_generalized_1964}, 
\begin{align}
\chi_n^{\mathrm{(LS)}} 
&=  \frac{\gamma_\mathrm{eff}n \chi}{2 \pi \Gamma} \, 
\mathcal{H}\!\!\!\int_0^\Omega \dif \omega \frac{J(\omega)}{( \omega - \omega_0)^2}
\label{chi_LS_ohmic}
\,,
\end{align}
where the Hadamard finite part keeps the integral regular. 

For a general spectrum, we can linearly approximate $J(\omega)$ around the pole $\omega_0$ and write 
\begin{multline}
    \chi_n^{\mathrm{(LS)}} 
 = \frac{\gamma_\mathrm{eff}n \chi}{2 \pi \Gamma} \, 
\int_0^\Omega \dif \omega 
\frac{J(\omega) - J_\mathrm{lin}(\omega)}{( \omega - \omega_0)^2} 
\\
+ \frac{\gamma_\mathrm{eff} J(\omega_0) n \chi}{2 \pi \Gamma}    \, 
\mathcal{H}\!\!\!\int_0^\Omega \dif \omega 
\frac{1}{( \omega - \omega_0)^2} 
\\
+  
\frac{\gamma_\mathrm{eff} J'(\omega_0) n \chi}{2 \pi \Gamma}  \,
\mathcal{H}\!\!\!\int_0^\Omega \dif \omega 
\frac{1}{  \omega - \omega_0 } 
\label{chi_LS_Hadamard}
\end{multline}
with 
\begin{align}
    J_\mathrm{lin} ( \omega) = J( \omega_0 ) + J'(\omega_0) ( \omega - \omega_0 ) 
    \,.
\end{align}
The first integral in Eq.~\eqref{chi_LS_Hadamard} is then regular and the poles that need to be treated with extra care
only appear in the last two terms which we can calculate explicitly. 
They yield in the limit of $n \chi , \ \omega_0 \ll \Omega$, 
\begin{multline}
        \chi_n^{\mathrm{(LS)}} 
 = \frac{\gamma_\mathrm{eff}n \chi}{2 \pi \Gamma} \, 
\int_0^\Omega \dif \omega 
\frac{J(\omega) - J_\mathrm{lin}(\omega)}{( \omega - \omega_0)^2} 
\\
+ 
\frac{\gamma_\mathrm{eff} n \chi }{2 \pi \omega_0}
\left[  \frac{J'(\omega_0)  \omega_0 }{\Gamma} 
\log \left( 
\frac{\Omega}{\omega_0}
\right) 
- 1
\right]
\label{chi_LS_gen}
\,,
\end{multline}
where we used $\Gamma = J(\omega_0)$. 
For an ohmic spectrum, we find that $J(\omega)$ coincides with its linear approximation $J(\omega) = J_\mathrm{lin}(\omega)$ and with $J'(\omega_0) = \alpha$ and $\Gamma =  \alpha \omega_0$, and hence Eq.~\eqref{chi_LS_gen} recovers the anharmonicity found in the Ohmic case, see Eq.~\eqref{chi_LS_ohmic}. 

The first term of Eq.~\eqref{chi_LS_gen} consists of a factor $\gamma_\mathrm{eff}/\Gamma \ll 1$ and an integral over a function $J(\omega) - J_\mathrm{lin}(\omega)$ divided by $(\omega - \omega_0)^2$.
By construction $J(\omega) - J_\mathrm{lin}(\omega)$  vanishes for $ \omega \rightarrow \omega_0$, while the denominator strongly suppresses the integration kernel at any frequency different from $\omega_0$. 
This is sketched in Fig.~\ref{fig:LS_Kernel}. Hence, it is reasonable to regard the first term as negligible.
The second term consists of two parts, which are small compared to $\chi$, as the first part is proportional to $J'(\omega_0) \chi \log( \Omega/\omega_0)$ where for $\Omega/\omega_0 = 10^N$, the term $\log(\Omega/\omega_0)$ is of order of $N$, and we assume a slowly varying spectrum which entails $J'(\omega_0) \ll 1$. 
While the second part is proportional to $\gamma_\mathrm{eff}/\omega_0 \chi \ll \chi$. 

With this we find that the Lamb shift-induced anharmonicity is much smaller than the intrinsic one and we may neglect it.

\bibliography{better-bib}  

\end{document}